\documentclass{lmcs}

\usepackage[T1]{fontenc}
\usepackage{amssymb}
\usepackage{mathrsfs}

\makeatletter
\def\endfront@text{}
\def\enddoc@text{}
\def\@settitlecomment{}

\def\ps@firstpage{\let\@mkboth\@gobbletwo\def\@oddhead{\vbox to\headheight{\vss\hbox{\vrule width\hsize depth .4pt}}}%
  \def\@evenhead{}\def\@oddfoot{}\def\@evenfoot{}}
\makeatother

\newtheorem{theorem}{Theorem}[section]
\newtheorem{lemma}[theorem]{Lemma}
\newtheorem{proposition}[theorem]{Proposition}
\newtheorem{corollary}[theorem]{Corollary}
\theoremstyle{definition}
\newtheorem{definition}[theorem]{Definition}
\newtheorem{convention}[theorem]{Convention}
\newtheorem{example}[theorem]{Example}
\newtheorem{openproblem}[theorem]{Open Problem}
\theoremstyle{remark}
\newtheorem{remark}[theorem]{Remark}

\newcommand{\Nn}{\mathbb{N}}

\newcommand{\Ee}{\mathcal{E}}
\newcommand{\Tt}{\mathbb{T}}
\newcommand{\Sh}{\mathbf{Sh}}
\newcommand{\Set}{\mathbf{Set}}
\newcommand{\Op}{\mathcal{O}}

\newcommand{\ab}{\{a,b\}}
\newcommand{\Cube}{\square}
\newcommand{\dimn}[1]{\lvert #1\rvert}
\newcommand{\cid}{\mathsf{id}}
\newcommand{\st}{\mathrm{st}}
\newcommand{\acc}{\mathrm{acc}}

\newcommand{\arrO}[1]{\mathrel{\nearrow^{#1}}}
\newcommand{\arrI}[1]{\mathrel{\searrow_{#1}}}
\newcommand{\starter}[2]{{}_{#2}{\uparrow}{#1}}
\newcommand{\terminator}[2]{{#1}{\downarrow}_{#2}}%
\newcommand{\ip}{\mathsf{ev}}
\newcommand{\Ipom}{\mathrm{iPom}}

\newcommand{\Tr}{\mathrm{tr}}
\newcommand{\prin}[1]{\mathrm{#1}}
\newcommand{\EM}{\prin{LEM}}
\newcommand{\UC}{\prin{UC}}

\newcommand{\EIO}{\Sigma^0_2\text{-}\prin{LEM}^{\neg}}

\newcommand{\CD}{\prin{CD}}
\newcommand{\Der}{\vdash}
\newcommand{\TR}{\Tt_{\Lambda}}
\newcommand{\TRab}{\Tt_{\ab}}
\newcommand{\TRm}{\Tt^{-}_{\Lambda}}

\newcommand{\SigR}{\Sigma_{\Lambda}}
\newcommand{\Csyn}[1]{\mathcal{C}^{\mathrm{ren}}_{#1}}
\newcommand{\Jsyn}[1]{J_{#1}}
\newcommand{\Clf}[1]{\Set[#1]}

\newcommand{\JLW}{J_{\LW}}
\newcommand{\sbt}[1]{\mathbf{1}_{#1}}
\newcommand{\MF}[1]{M_{#1}}
\newcommand{\XF}[1]{X_{#1}}
\newcommand{\pprop}[1]{\ulcorner #1\urcorner}
\newcommand{\Pref}{\mathrm{Pref}}
\newcommand{\ext}{\trianglelefteq}
\newcommand{\JX}[1]{J_{#1}}
\newcommand{\BB}{\mathrm{B}}
\newcommand{\chain}[1]{\mathrm{ch}(#1)}
\newcommand{\fle}[1]{\preceq_{#1}}
\newcommand{\feq}[1]{\equiv_{#1}}
\newcommand{\upc}[1]{{\uparrow}#1}
\newcommand{\FX}[1]{\mathbb{F}_{#1}}
\newcommand{\RX}[1]{\mathbb{R}_{#1}}

\newcommand{\IX}[1]{I_{#1}}

\newcommand{\fat}[1]{\mathsf{F}_{#1}}

\newcommand{\nm}[1]{\mathsf{is}_{#1}}
\newcommand{\lab}{\mathrm{lab}}
\newcommand{\lle}{\leq_{\mathrm{lab}}}

\newcommand{\LW}{\mathrm{LW}}

\newcommand{\eqM}{\simeq_{\mathrm{M}}}

\newcommand{\eqD}{\simeq_{\mathrm{d}}}

\newcommand{\eqLs}{\simeq_{\mathrm{ls}}}

\newcommand{\eqFrF}{\simeq^{\circ}_{\mathrm{fr}}}

\newcommand{\BarS}[1]{\mathsf{Bar}(#1)}
\newcommand{\Ext}[1]{\overline{#1}}

\newcommand{\trim}[1]{#1^{\mathrm{t}}}
\providecommand{\SDNE}{\Sigma^0_2\text{-}\prin{DNE}^{\neg}}
\providecommand{\PLEM}{\Pi^0_2\text{-}\prin{LEM}}
\providecommand{\LPO}{\prin{LPO}}
\providecommand{\evo}[2]{\mathsf{ev}_{#1}(#2)}
\providecommand{\Lang}[2]{L(#1,#2)}
\providecommand{\LB}[1]{L_{\mathrm{B}}(#1)}
\providecommand{\LM}[1]{L_{\mathrm{M}}(#1)}
\providecommand{\LMp}[1]{L^{+}_{\mathrm{M}}(#1)}
\providecommand{\Inf}{\mathsf{Inf}}
\providecommand{\MinB}{\mathsf{MB}}
\providecommand{\BinM}{\mathsf{BM}}
\providecommand{\MinBt}{\mathsf{MB}_{\mathrm{tot}}}
\providecommand{\MinBd}{\mathsf{MB}^{\mathrm{dec}}}
\providecommand{\BinMd}{\mathsf{BM}^{\mathrm{dec}}}
\providecommand{\MinBtd}{\mathsf{MB}^{\mathrm{dec}}_{\mathrm{tot}}}
\providecommand{\Sq}[1]{\mathrm{Sq}(#1)}
\providecommand{\Cu}[1]{\mathrm{Cu}(#1)}
\providecommand{\dsum}{\sqcup}

\providecommand{\KA}{\mathsf{A}}
\providecommand{\KB}[1]{\mathsf{B}_{#1}}

\providecommand{\STa}{\mathsf{ST}}
\providecommand{\stw}{\mathrm{w}}
\providecommand{\LangA}[1]{L^{\mathrm{acc}}(#1)}
\providecommand{\LangPAF}[1]{L^{\mathrm{PAF}}(#1)}
\providecommand{\LangPt}[1]{L^{\mathrm{pt}}(#1)}
\providecommand{\LangLp}[1]{L^{\mathrm{lp}}(#1)}

\providecommand{\CUC}{\prin{CUC}}
\providecommand{\BC}{\prin{BC}}

\providecommand{\tcl}{\mathsf{cl}}
\providecommand{\tder}{\mathsf{der}}

\providecommand{\eqDg}{\approx_{\mathrm{dg}}}

\providecommand{\KK}{\mathsf{K}}
\providecommand{\DD}{\mathsf{D}}

\providecommand{\dgB}{\mathsf{dg}}

\providecommand{\matchB}{\mathsf{match}}
\providecommand{\moB}{\mathsf{mo}}

\providecommand{\tyl}[1]{\sim_{#1}}
\providecommand{\tyeq}{\sim}
\providecommand{\rest}{\mathrm{rest}}
\providecommand{\splitP}{\mathrm{sp}}
\providecommand{\Cls}{\mathcal{K}}

\usepackage{tikz}
\usetikzlibrary{arrows,arrows.meta,calc,decorations.markings}
\tikzset{
  v4/.style={>={Computer Modern Rightarrow[length=1.9pt,width=4.2pt]},line width=0.45pt,
    every node/.style={inner sep=1.6pt}},
  v4obj/.style={font=\small,inner sep=2.4pt},
  v4map/.style={->,shorten <=1.2pt,shorten >=1.2pt},
  v4read/.style={v4map,dash pattern=on 2.2pt off 1.6pt},
  v4eqv/.style={-,shorten <=1.2pt,shorten >=1.2pt,dash pattern=on 2.2pt off 1.6pt},
  v4lab/.style={font=\scriptsize,inner sep=1.4pt},
  v4mono/.style={>->,shorten <=1.2pt,shorten >=1.2pt},
  v4monoread/.style={v4mono,dash pattern=on 2.2pt off 1.6pt},
  v4incl/.style={right hook->,shorten <=1.2pt,shorten >=1.2pt},
  v4inclread/.style={v4incl,dash pattern=on 2.2pt off 1.6pt},
  v4epiread/.style={->>,shorten <=1.2pt,shorten >=1.2pt,dash pattern=on 2.2pt off 1.6pt},
  v4part/.style={-{Computer Modern Rightarrow[length=1.9pt,width=4.2pt,harpoon]},shorten <=1.2pt,shorten >=1.2pt},
  v4cell/.style={font=\small,inner sep=0pt},
}

\providecommand{\wQ}{\mathsf{Q}}
\providecommand{\lab}{\mathrm{lab}}

\definecolor{hdaViolet}{HTML}{1F1E33}

\tikzset{
  hdaSq/.style={fill=black!10},
  hdaV/.style={circle,draw,fill=white,line width=0.4pt,inner sep=0pt,minimum size=1.6mm},
  hdaVgone/.style={hdaV,draw=black!45,densely dashed},
  hdaEdge/.style={->,>=stealth',line width=0.4pt},
  hdaLab/.style={font=\scriptsize,inner sep=1.5pt},
  hdaName/.style={font=\footnotesize,inner sep=1.5pt},
  hdaTrack/.style={->,>=stealth',hdaViolet,line width=1.1pt,shorten <=1.6pt,shorten >=1.6pt},
  hdaBand/.style={hdaViolet!22,line width=5.5pt,line cap=round,line join=round},
  hdaDot/.style={circle,fill=hdaViolet,inner sep=0pt,minimum size=4.2pt},
  cellNode/.style={circle,draw,fill=white,line width=0.4pt,inner sep=0pt,minimum size=17pt,font=\footnotesize},
  faceArr/.style={->,>=stealth',line width=0.4pt,shorten <=1pt,shorten >=1pt},
  faceMiss/.style={faceArr,densely dashed,gray},
  stState/.style={cellNode},
  stArr/.style={faceArr},
  stUsed/.style={faceArr,hdaViolet,line width=1.1pt},
  bridgeLeg/.style={->,>=stealth',hdaViolet,line width=0.6pt,dashed},
  bridgeMap/.style={->,>=stealth',line width=0.4pt,black!75},
  impl/.style={double,double distance=1.6pt,line width=0.4pt,-{Implies},shorten <=2pt,shorten >=2pt},
  nimpl/.style={impl,hdaViolet,postaction={decorate},decoration={markings,
    mark=at position 0.5 with {\draw[hdaViolet,line width=0.6pt,-] (-0.09,-0.16) -- (0.09,0.16);}}},
  oimpl/.style={impl,gray,densely dashed},
  eqNode/.style={inner sep=3pt},
  witLab/.style={font=\scriptsize,inner sep=1.5pt,text=hdaViolet},
  hypLab/.style={font=\scriptsize,inner sep=1.5pt},
}

\usepackage[all]{xy}
\providecommand{\BL}{\mathcal{B}_{\Lambda}}
\providecommand{\BST}{\mathcal{B}_{\mathrm{ST}}}
\providecommand{\SW}{\mathrm{ST}}
\providecommand{\JSW}{J_{\mathrm{ST}}}
\providecommand{\fgt}{\varpi}
\providecommand{\lsts}[1]{\ell^{*}_{\mathrm{ST},#1}}
\providecommand{\JimS}[1]{J^{\mathrm{im}}_{\mathrm{ST},#1}}
\providecommand{\eqB}{\simeq_{\mathcal{B}}}
\providecommand{\lst}[1]{\ell^{*}_{#1}}
\providecommand{\fin}{f^{\ast}}
\providecommand{\fsh}{f_{\ast}}
\providecommand{\Ima}{\mathrm{Im}}
\providecommand{\ImaST}{\mathrm{Im}_{\mathrm{ST}}}
\providecommand{\Jim}[1]{J^{\mathrm{im}}_{#1}}
\providecommand{\dset}[2]{\mathrm{d}^{#1}(#2)}
\providecommand{\grph}[1]{\Gamma(#1)}
\providecommand{\dsum}{\sqcup}
\providecommand{\DC}{\prin{DC}}
\providecommand{\BIb}{\prin{BI}_{\mathrm{br}}}
\providecommand{\CS}{\prin{CS}}
\providecommand{\DCS}{\prin{DCS}}
\providecommand{\CUC}{\prin{CUC}}
\providecommand{\BC}{\prin{BC}}
\providecommand{\SDNE}{\Sigma^0_2\text{-}\prin{DNE}^{\neg}}
\providecommand{\PLEM}{\Pi^0_2\text{-}\prin{LEM}}
\providecommand{\Tacc}{\Tt^{\mathrm{acc}}_{\Lambda}}
\providecommand{\FT}{F_{\mathsf{T}}}
\providecommand{\Lmax}[1]{L^{\max}(#1)}
\providecommand{\TB}[1]{\mathsf{T}^{#1}}
\providecommand{\Wq}[3]{W_{#1}(#2,#3)}
\providecommand{\Cs}[2]{\mathsf{C}_{#1}^{#2}}
\providecommand{\pfB}{\mathsf{pf}}
\providecommand{\fshS}{f^{\mathrm{ST}}_{\ast}}
\providecommand{\fsym}[1]{\delta_{#1}}
\providecommand{\evat}[2]{\ip_{#1}=#2}
\providecommand{\nst}{\overline{\mathsf{st}}}
\providecommand{\nacc}{\overline{\mathsf{acc}}}
\providecommand{\wG}{\mathsf{G}}
\providecommand{\wGt}{\tilde{\mathsf{G}}}
\providecommand{\wQ}{\mathsf{Q}}
\providecommand{\wQt}{\tilde{\mathsf{Q}}}
\providecommand{\wQtt}{\tilde{\mathsf{Q}}_3}
\providecommand{\wC}[1]{\mathsf{C}_{#1}}
\providecommand{\wCh}[1]{\hat{\mathsf{C}}_{#1}}
\providecommand{\wB}{\mathsf{N}}
\providecommand{\wBh}{\hat{\mathsf{N}}}
\begin{document}

\title{Higher-Dimensional Automata in Geometric Logic}
\author{Kenan Oggad}
\address{Universit\'{e} Paris-Saclay}
\email{kenan.oggad@universite-paris-saclay.fr}

\begin{abstract}
What does the topos of a higher-dimensional automaton know about its infinite behaviour that its language does
not? We write partial HDAs with an infinite run as models of a geometric theory over an intuitionistic metatheory
with no choice. The topos of a machine is that of sheaves on its prefixes. The label word of a prefix records the
shape and acceptance bit of each cell with the ipomset of each step. Under unique choice these words give a geometric
morphism to the label topos of sheaves on label words, which thus records the shapes and acceptance bits that the B\"uchi language forgets. Machines sharing their B\"uchi language and their frames can differ over the label topos. The bouquet of two squares shares its frame and its B\"uchi language with the interleaved square; each accepted run of either reads the two events in parallel forever. Only the interleaved square has runs
through cells of shape $a$, and the geometric sequent excluding such runs separates the two over the label topos. Over alphabets with decidable equality, finite rigid machines are
unambiguous exactly when their label maps are inclusions of locales, and their equivalence over the label topos is decided
whenever either is unambiguous. Equivalences of frames over the label topos transport points with their labels, and the invariants of fibres and of the language of points follow from this transport. On finite partial HDAs over alphabets with decidable equality, Muller and B\"uchi acceptance coincide exactly under a weak form of excluded middle.
\end{abstract}

\maketitle

\section{Introduction}\label{sec:intro}

Consider two events $a$ and $b$ of a concurrent program or of an asynchronous circuit which may overlap in time. A
finite automaton records the pair only by its interleavings, $a$ then $b$ beside $b$ then $a$, while a
higher-dimensional automaton (HDA) adds the cell of dimension two in which both run, a square whose boundary holds the
orders in which the two can start and finish; the \emph{shape} of a cell lists the events in progress in it, so that
the square has shape $ab$ and its edges the shapes $a$ and $b$. A \emph{partial} HDA may lack some faces, the cells on
the boundary of a cell, so that a square can be entered from some of its edges and not from others. Each step of a run
starts or finishes events, recorded by its \emph{ipomset}, a finite partial order of labelled events. An infinite run
is \emph{accepted} when it starts at a start cell and meets accepting cells at arbitrarily late times; the
\emph{B\"uchi language} collects the partially ordered events read by the runs that return to one accepting cell
infinitely often, classically the accepted runs of a finite machine. Comparing machines by their B\"uchi languages forgets how
runs pass through cells and branch, while a bisimilarity matches each continuation of a run by one of a related run of
the other machine. The question is what lies between, and what an observer of a running machine can tell apart, since
an infinite computation is never seen whole.

A \emph{prefix} is a finite initial part of a run, and a property of runs is \emph{observable} when each run having it
has a prefix that establishes it. On a machine an \emph{open} is a set of prefixes closed under extension, and a set
$S$ of prefixes \emph{covers} a prefix $p$ when $p\in S$ or, for some $n$, $S$ covers every extension of $p$ meeting
an accepting cell after position $n$ (Proposition~\ref{thm:jx}). The \emph{topos} of the machine is the category of
\emph{sheaves} on its prefixes, assignments of sets to prefixes in which elements agreeing on a cover glue to exactly
one (Theorem~\ref{thm:topos-a}). Its \emph{subterminals}, the parts of the one-element sheaf that are sheaves,
correspond to the opens up to identification of opens covering each other's members, and form the \emph{frame} of the
machine, ordered by covering with intersection as meet and union as join (Theorem~\ref{thm:frame-sub}). Observable
properties are thus objects of the topos, and their finite conjunctions and arbitrary disjunctions are the connectives
of geometric formulas, so that geometric logic is their logic. A \emph{point} is a set of prefixes with at least one
member, closed under initial parts, holding a common extension of any two members and meeting every set that covers
one of them (Theorem~\ref{thm:pts-x}(1)); over an alphabet with decidable equality the prefixes of a run form a point
exactly when the run is accepted (Theorem~\ref{thm:pts-x}(4)).

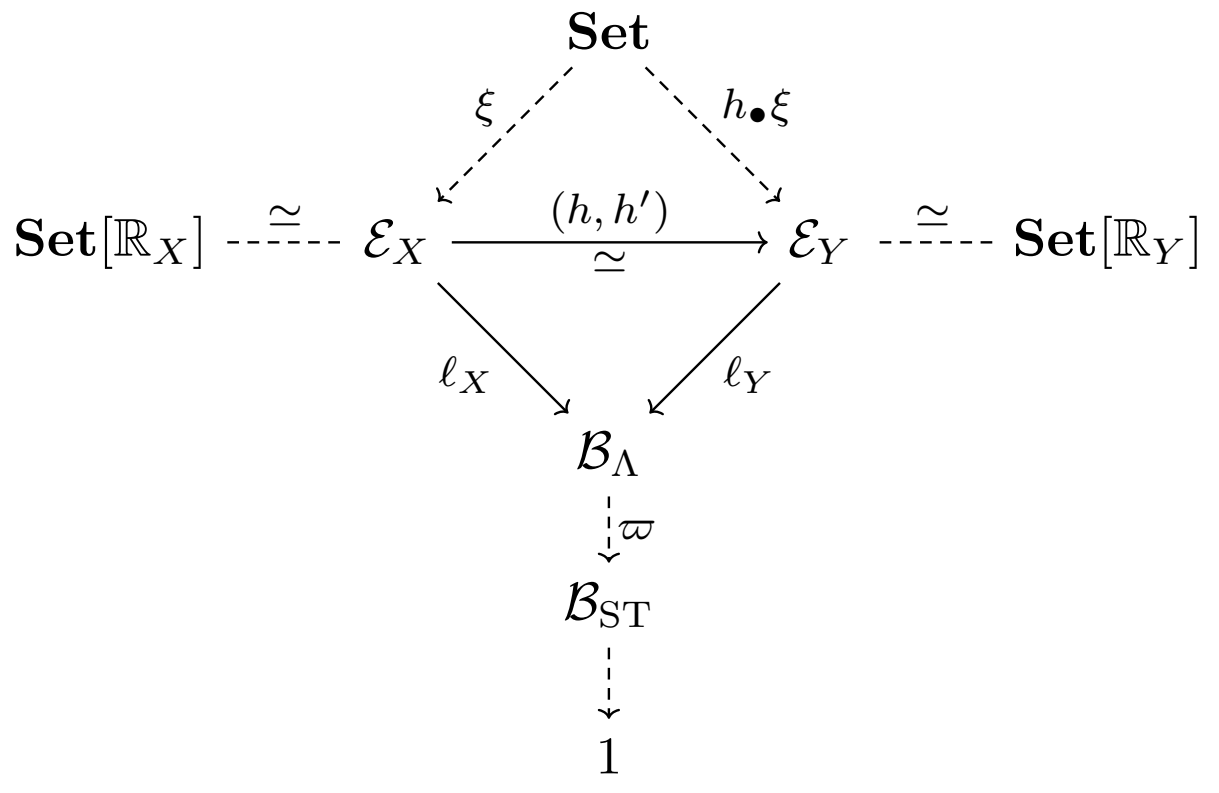
\begin{figure}[htb]
\centering
\begin{tikzpicture}[v4,x=1cm,y=1cm]
  \node[v4obj] (pt) at (0,1.4) {$\Set$};
  \node[v4obj] (ex) at (-1.4,0) {$\Ee_X$};
  \node[v4obj] (ey) at (1.4,0) {$\Ee_Y$};
  \node[v4obj] (bl) at (0,-1.4) {$\BL$};
  \node[v4obj] (bs) at (0,-2.4) {$\BST$};
  \node[v4obj] (one) at (0,-3.4) {$1$};
  \node[v4obj] (rx) at (-3.3,0) {$\Clf{\RX{X}}$};
  \node[v4obj] (ry) at (3.3,0) {$\Clf{\RX{Y}}$};
  \draw[v4read] (pt) -- node[v4lab,above left] {$\xi$} (ex);
  \draw[v4read] (pt) -- node[v4lab,above right] {$h_\bullet\xi$} (ey);
  \draw[v4map] (ex) -- node[v4lab,above] {$(h,h')$} node[v4lab,below] {$\simeq$} (ey);
  \draw[v4map] (ex) -- node[v4lab,below left] {$\ell_X$} (bl);
  \draw[v4map] (ey) -- node[v4lab,below right] {$\ell_Y$} (bl);
  \draw[v4read] (bl) -- node[v4lab,right] {$\fgt$} (bs);
  \draw[v4read] (bs) -- (one);
  \draw[v4eqv] (rx) -- node[v4lab,above=1.5pt] {$\simeq$} (ex);
  \draw[v4eqv] (ry) -- node[v4lab,above=1.5pt] {$\simeq$} (ey);
\end{tikzpicture}
\caption{An equivalence $(h,h')$ witnessing $X\eqB Y$ joins the topos $\Ee_X=\Sh(\Pref(X),\JX{X})$ of $X$ to that
of $Y$ over $\BL$ and carries each point $\xi$ of $X$ with its labels to a point $h_\bullet\xi$ of $Y$
(Theorem~\ref{thm:transport}). Solid arrows are proved, $(h,h')$ as an equivalence of frames of subterminals (Theorem~\ref{thm:frame-sub}) and the label maps as geometric morphisms under unique choice
(Theorem~\ref{thm:prices-uc}). Dashed arrows are readings through cited theorems (Sections~\ref{sec:topos} and~\ref{sec:image}) of points as
geometric morphisms and of $\Ee_X$ as the classifying topos of its run theory $\RX{X}$; the comorphism $\fgt$ to words of step ipomsets
(Theorem~\ref{thm:image-live}) is read as a geometric morphism, and so is the map to the one-point topos $1=\Set$.}
\label{fig:pattern}
\end{figure}

An observer of labels sees only the \emph{label word} of a prefix, the shapes and acceptance bits of its cells with
the ipomsets of its steps. The \emph{label topos} is the category of sheaves on label words, and $X\eqB Y$,
\emph{equivalence over the label topos}, holds when some \emph{equivalence of frames}, order-preserving maps both ways
inverse up to identification, sends, for every open $U$ of label words, the open of the prefixes of $X$ whose label
words lie in $U$ to the open so defined in $Y$ up to identification (Figure~\ref{fig:pattern}).

The bouquet $\wQ^\circ$, two squares at one start vertex, each starting and finishing its two events together, with
only the vertices accepting, and the interleaved square $\mathsf S$, one square which a run may also enter while $a$
runs alone and leave while $b$ runs alone (Example~\ref{exa:dec-witnesses}), have equivalent frames and one B\"uchi language; every accepted run of either reads the two events in parallel
forever, and recognising this language over these two letters takes a cell of dimension two, where both run at once. A \emph{geometric sequent} $\varphi\vdash\psi$
states that every observation of $\varphi$ is one of $\psi$; only $\mathsf S$ has a cell of shape $a$, so
$\mathrm{pass}_a\vdash\bot$, which says that the empty set covers every prefix passing such a cell, holds in
$\wQ^\circ$ and fails in $\mathsf S$, and $\wQ^\circ\eqB\mathsf S$ fails (Theorem~\ref{thm:separation}(1)). A prefix
is \emph{live} when it is an initial part of an accepted run, and machines are \emph{run-tree bisimilar} when some
relation between live prefixes with equal label words, total on both sides, matches each live extension of either by a
related live extension of the other. The bouquet $\wQ$, with the start vertex and one square accepting, and its
\emph{double} $\wQ\dsum\wQ$, two disjoint copies of $\wQ$, read the same sequences of step ipomsets on accepted runs
and are run-tree bisimilar with equivalent frames, while $\wQ\eqB\wQ\dsum\wQ$ fails (Theorem~\ref{thm:separation}(2)).
\emph{Step bisimilarity}, which matches live prefixes one step at a time from length zero, is incomparable with $\eqB$
already on \emph{graphs}, partial HDAs whose cells have dimension at most one (Theorem~\ref{thm:separation}(7)). So $\eqB$ separates pairs that their B\"uchi language and their
frames identify together, and pairs that either bisimilarity identifies.

Classical logic assigns every question about an infinite run an answer, while an observer holds only what finite
stages of the run establish; the metatheory is therefore intuitionistic with no choice, and a \emph{principle}, a classical law such as excluded middle or a form of choice, enters a statement only as a hypothesis, its
\emph{logical strength} when the statement implies it back. Muller acceptance fixes the set of cells met infinitely
often and so asserts of every other cell that it is met only finitely often, an assertion that only the whole infinite run settles. On
finite partial HDAs over alphabets with decidable equality Muller and B\"uchi acceptance recognise the same languages
exactly when $\EIO$ holds, the principle that every Boolean sequence is eventually $0$ or takes the value $1$ at
arbitrarily late times, thus the logical strength of their coincidence (Theorem~\ref{thm:muller-answer}). They
coincide classically (Corollary~\ref{cor:muller-classical}), as conjectured for partial HDAs in the reading of
Remark~\ref{rem:muller-conjecture}.

Deciding $\eqB$ makes the topos over the label topos a computable invariant. The \emph{rank} of a point is the number
of rounds it survives when points alone in some open among those remaining are removed repeatedly, and every
equivalence of frames carries points with their ranks and the number at each rank, and with their labels when it fixes
the opens defined by label words (Theorem~\ref{thm:transport}). A machine is \emph{rigid} when at most one step goes
from any cell to any other and \emph{trimmed} when each prefix is live; it has \emph{closed acceptance} when for some
$\Delta$ every $\Delta$ consecutive steps of a prefix reach an accepting cell (Definition~\ref{def:closed}). Over
alphabets with decidable equality, equivalence of frames alone, \emph{equivalence over} the one-point topos $1$, is
decided on finite rigid machines with closed acceptance, where duality with Boolean algebras reduces it to a
comparison of \emph{clopens}, compact opens with a complement (Theorems~\ref{thm:dec-stone} and~\ref{thm:dec-decide}).
Over the label topos the obstacle is ambiguity, two points that read one infinite sequence of labels; over such
alphabets $\eqB$ is decided on finite rigid machines when one of the two is \emph{unambiguous}, with at most one point over each infinite sequence of labels
(Theorem~\ref{thm:dec-morita-inj}), and on finite rigid trimmed machines over $\ab$ with closed acceptance a decision
exists exactly when one exists on the ambiguous pairs (Theorem~\ref{thm:decide}).
\section{Machines and their theory}\label{sec:setup}

\subsection*{Background}

A \emph{sieve} on an object $c$ of a category is a set of arrows into $c$ closed under composition with arrows into
their domains, and a \emph{Grothendieck topology} chooses for each object a set of \emph{covering} sieves, among them
the maximal sieve, stable under pullback and local in the sense that a sieve covers whenever its
pullbacks along the members of some covering sieve all cover; a category with a Grothendieck topology is a
\emph{site} \cite[Expos\'e~II]{SGA4}. A \emph{presheaf} assigns a set to each
object and a restriction map to each arrow, and it is a \emph{sheaf} when every family of elements on the members of a
covering sieve that agree where they meet comes from exactly one element; a \emph{topos} is a category equivalent to
the sheaves on some site \cite[Expos\'e~IV]{SGA4}\cite{CL25a}.
A \emph{geometric morphism} $f\colon\mathcal F\to\mathcal E$, the analogue for toposes of a continuous map, is a pair
of adjoint functors whose left adjoint $f^\ast$, the \emph{inverse image}, preserves finite limits, and a \emph{point}
of $\mathcal E$ is a geometric morphism $\Set\to\mathcal E$
\cite{SGA4}\cite{CL25a}.
A \emph{geometric theory} is a set of sequents $\varphi\vdash_{\vec x}\psi$ between \emph{geometric formulas}, the
positive existential formulas in which disjunctions of arbitrary families are allowed
\cite{CL25a}, and each geometric theory $\Tt$ has a
\emph{classifying topos} $\Clf{\Tt}$ with a \emph{universal model} $U_{\Tt}$ such that $f\mapsto f^\ast U_{\Tt}$ is
an equivalence from the geometric morphisms $\mathcal E\to\Clf{\Tt}$ to the models of $\Tt$ in $\mathcal E$, so that
the points of $\Clf{\Tt}$ are the models of $\Tt$ in sets
\cite{CL25a}. Two theories whose classifying
toposes are equivalent are \emph{Morita equivalent}
\cite{CL25a}\cite{Caramello19},
and a property of a topos stated invariantly transfers between its presentations by sites and by theories; Caramello
calls such transfers \emph{bridges}, and the bridge that recasts acceptance makes the subtoposes of $\Clf{\Tt}$
correspond both to the Grothendieck topologies refining that of a site presenting it and to the \emph{quotient
theories} of $\Tt$, those obtained by adding axioms in its signature
\cite{CL25a,Caramello09}.
A \emph{comorphism of sites} $F\colon(\mathcal D,K)\to(\mathcal C,J)$ is a functor along which every $J$-covering
sieve on $F(d)$ contains the image of a $K$-covering sieve on $d$, and it induces a geometric morphism from the
sheaves on $(\mathcal D,K)$ to those on $(\mathcal C,J)$ \cite{Caramello19}.

A \emph{frame} is a partially ordered set with finite meets and arbitrary joins in which binary meets distribute over
joins, and a \emph{locale} is a frame regarded as a space known through its opens
\cite{Caramello11}. A \emph{point} of a locale is a frame
homomorphism to the two-element frame, which is the same as a completely prime filter of the frame, the set of opens
in which the point lies \cite{Caramello11}. The
sheaves on a locale form a \emph{localic} topos, whose points are the points of the locale and whose
\emph{subterminal} objects, the subobjects of the terminal sheaf, are the elements of the
frame \cite{Caramello11}. An element of a frame is
\emph{compact} when every family whose join lies above it has a finite subfamily whose join does, and in its localic
form Stone duality makes the Boolean algebras correspond to the locales in which the complemented elements form a
basis; this form uses no choice, while recovering from it a space with enough points uses a form of the axiom of
choice \cite{Caramello11}.

Constructive mathematics classifies the non-constructive principles, statements that classical logic proves and
intuitionistic logic does not, by comparing them up to provability over a base theory such as Heyting arithmetic
\cite{ABHK04}. Unique choice turns a relation holding of exactly one element at each argument into
a function, which sheaf constructions need whenever an amalgamation known to be unique must be given as an operation.
In the arithmetic hierarchy a $\Sigma^0_k$ formula has $k$ alternating quantifiers over the natural numbers, the first
existential, before a quantifier-free matrix, and a $\Pi^0_k$ formula has the first universal; excluded middle and
double-negation elimination $\neg\neg P\to P$ restricted to these classes form a hierarchy of principles in which
double-negation elimination for $\Sigma^0_1$ formulas is Markov's principle
\cite{ABHK04}, and finding the principles that prove a theorem is a
constructive form of reverse mathematics
\cite{ABHK04}.

Higher-dimensional automata, introduced by Pratt and by van Glabbeek
\cite{Pratt91,vG91}, extend finite automata by squares and
cubes of every dimension, where a square stands for two events running concurrently, so that the concurrent execution
$a\parallel b$ differs from the choice between the interleavings $a.b$ and $b.a$
\cite{FJSZ21}. A cell has as \emph{shape} the list of
the events running in it, its \emph{conclist}, and a \emph{face} of a cell is the cell reached by unstarting some of
its events and terminating others \cite{BDFZ26}. In an HDA
every face exists; a \emph{partial} HDA lets faces be undefined and asks only that two defined faces in succession
compose to the face along the composite inclusion of conclists, which makes it a lax functor to sets with partial
maps \cite{BDFZ26}. A run passes between cells by up-steps,
which start events, and down-steps, which terminate them, and each step carries an ipomset, a \emph{starter} for an
up-step and a \emph{terminator} for a down-step
\cite{BDFZ26}\cite{PAF25}.
An \emph{ipomset}, a pomset with interfaces, is a finite set of labelled events with a precedence order, an event
order on the concurrent events and two interfaces holding the events running at its start and at its end, and the
ipomset of a path glues those of its steps along matching interfaces
\cite{BDFZ26}.

An automaton on infinite words accepts an infinite run according to the set of states the run visits infinitely
often; the B\"uchi condition asks that this set contain an accepting state \cite{LP18}
and the Muller condition that it be one of a listed family of sets, and for automata the two conditions recognise the
same languages \cite{PAF25},
while on HDAs with infinite tracks Muller acceptance recognises more languages than B\"uchi acceptance
\cite{PAF25}. The \emph{degree of ambiguity} of an automaton is the
supremum over words of the number of accepting runs on a word
\cite{LP18}.

\subsection{Metatheory}\label{ssec:meta}\label{conv:principles}

The ambient logic is intuitionistic higher-order logic with propositional extensionality and quotients, and
without choice. A \emph{principle} enters a statement only as a hypothesis, and a step of a proof uses none beyond those it
names; when the converse is also proved the statement is \emph{equivalent} to it, and the principle is its
\emph{logical strength}. A result
attributed to a reference is \emph{cited}, taken from it without proof; an identification of a proved object with an
object of the literature rests on the cited result named in its sentence.

\begin{definition}[Principles]\label{def:em-uc}\label{def:eio}\label{def:sdne}\label{def:cuc-bc}\label{def:setup-bibr}\label{def:setup-cd}\label{def:setup-split}
\begin{enumerate}
\item $\EM$ (excluded middle) states $P\vee\neg P$ for every $P$.
\item $\UC$ (unique choice) states that every family of relations $R_i\subseteq A_i$ $(i\in I)$ each holding of exactly one
  element has a function $f\in\prod_{i\in I}A_i$ with $R_i(f(i))$ for every $i$; $\CUC$ (countable unique choice) states it for $I=\Nn$
  and $A_i=A$.
\item $\BC$ (Boolean choice) states that every family $(P_n)_{n\in\Nn}$ of propositions with $P_n\vee\neg P_n$ has some
  $f\colon\Nn\to\{0,1\}$ with $f(n)=1\iff P_n$.
\item $\DC$ (dependent choice) states that every total relation $R$ on a set $A$ and every $x_0\in A$ have some $f$ with $f(0)=x_0$
  and $R(f(n),f(n+1))$ for every $n$.
\item $\EIO$ ($\Sigma^0_2$ law of excluded middle in dual form) states that every $u\colon\Nn\to\{0,1\}$ is eventually $0$
  or satisfies $I(u)$, which is $\forall N\,\exists n\geq N\ u_n=1$. $\SDNE$ ($\Sigma^0_2$ double-negation elimination in dual form) states
  that every $u$ with $\neg I(u)$ is eventually $0$, and $\PLEM$ ($\Pi^0_2$ law of excluded
  middle) states $I(u)\vee\neg I(u)$ for every $u$.
\item A \emph{branch} is a set of finite sequences of naturals containing $\langle\rangle$ and closed under
  initial segments, with exactly one $uk$ for each member $u$. $\BIb$ (monotone bar induction
  with branches in place of sequences) states that if $B$ is closed under
  $u\mapsto uk$ and meets every branch then every $Q\supseteq B$ containing each $u$ whose $uk$ all lie in $Q$
  contains $\langle\rangle$.
\item A \emph{gluing} on a set $X$ gives every family in $X$ constant on an inhabited upward closed subset of $\Nn$ a
  value equal to its members, and $X$ is \emph{complete} when it has one. $\CS$ (complete sets)
  states that every set is complete. A \emph{description operator} on $X$ chooses the
  solution of each predicate on $X$ with exactly one. $\DCS$ (descriptions on complete sets)
  states that every family $(X_i)_{i\in I}$ of sets with a family of gluings has a family of description operators
  on the $X_i$, and $\CD$ (countable descriptions) states $\DCS$ for
  $I=\Nn$.
\end{enumerate}
\end{definition}

For a $\Sigma^0_2$ formula $P$ the dual $P^{\neg}$ exchanges the quantifiers of $P$ and negates its matrix.
$\Phi\text{-}\prin{LEM}$ and $\Phi\text{-}\prin{DNE}$ state $P\vee\neg P$ and $\neg\neg P\to P$ for $P$ in a class
$\Phi$ \cite{ABHK04}; the forms marked ${}^{\neg}$ put $P^{\neg}$ in
place of $\neg P$ for $P$ the statement that $u$ is eventually $0$, whose dual is $I(u)$. Monotone bar induction
on sequences is that of \cite{TvD88}. $\EIO$ follows from $\EM$ and holds exactly when
$\SDNE$ and $\PLEM$ hold; $\UC$ gives $\CUC$ and
$\CD$, and $\CUC$ gives $\BC$.

\subsection{Geometric theories and their points}\label{ssec:geometric}

Grothendieck topologies and their toposes of sheaves are those of \cite[Expos\'es~II and~IV]{SGA4},
and formulas and derivability from theories are those of geometric logic over relational signatures
\cite{CL25a}.
For a proposition $P$ of the metatheory, $\pprop{P}$ is the sentence $\bigvee_{i\in I}\top$ over the set
$I=\{\ast\mid P\}$.

\begin{definition}[Syntactic site on renamings]\label{def:synsite}
For a theory $\Tt$ the category $\Csyn{\Tt}$ has the formulas in context $\{\vec x.\varphi\}$ as objects and as
arrows $\{\vec x.\varphi\}\to\{\vec y.\psi\}$ the renamings $\rho$ of $\vec y$ into $\vec x$ with
$\varphi\Der_{\vec x}\psi[\rho]$ derivable, where $\rho$ and $\rho'$ are identified when
$\varphi\Der_{\vec x}\rho(y)=\rho'(y)$ is derivable for every variable $y$ of $\vec y$. A family
$\rho_i\colon\{\vec x_i.\varphi_i\}\to\{\vec y.\psi\}$ \emph{covers} when
$\psi\Der_{\vec y}\bigvee_i\exists\vec x_i\,(\varphi_i\wedge\vec y=\rho_i(\vec y))$ is derivable, and these
families form $\Jsyn{\Tt}$.
\end{definition}

A \emph{point} is a flat $\Jsyn{\Tt}$-continuous functor $F\colon\Csyn{\Tt}\to\Set$; its \emph{model} $\MF{F}$ has
carrier $F(\{x^s.\top\})$ at $s$, and $R$ holds of the projections of
$F(\{\vec x.R(\vec x)\})$.
On the full syntactic site \cite{CL25a} the sheaves form the classifying topos $\Clf{\Tt}$
\cite{Johnstone02}, whose points are the models of $\Tt$.

\begin{definition}[Generic model]\label{def:generic}
Let $\Tt$ be a theory. A presheaf on $\Csyn{\Tt}$ is a \emph{sheaf} when every matching family on a
$\Jsyn{\Tt}$-covering family has exactly one amalgamation, and the site is \emph{subcanonical} when every
representable presheaf is a sheaf. An element of sort $s$ of the \emph{generic model} $U_{\Tt}$ at an object $c$
is a variable of sort $s$ given locally on $c$; such a variable is a relation between arrows $d\to c$ and variables of $d$
that is stable under composition, local, single-valued up to derivable equality and locally inhabited. Formulas
are interpreted by the sheaf semantics of the site, with $\top,=,\wedge$ stagewise and
$\bot,\vee,\bigvee,\exists$ locally; a relation symbol holds of a tuple when locally it is a tuple of
variables of which the symbol is derivable. A sequent \emph{holds} in $U_{\Tt}$ when at every object its left side
implies its right side.
\end{definition}

The generic model makes derivability in $\Tt$ a matter of truth in one sheaf model on the syntactic site, so that
whether every sequent true in all models in sets is derivable becomes a question about sheaves on that site, which
Theorem~\ref{thm:setup-bar} answers for the covering sequents of a machine.

\begin{theorem}[Generic model]\label{thm:generic}
Let $\Tt$ be a geometric theory.
\begin{enumerate}
\item Every sort of $U_{\Tt}$ is a sheaf on $(\Csyn{\Tt},\Jsyn{\Tt})$.
\item For every renaming $\rho$ of the context of $\varphi$ into that of an object $c=\{\vec x.\chi\}$, the formula
  $\varphi$ holds at $c$ of the tuple of variables of $\rho$ exactly when $\chi\Der_{\vec x}\varphi[\rho]$ is
  derivable.
\item A sequent holds in $U_{\Tt}$ exactly when it is derivable in $\Tt$.
\item $(\Csyn{\Tt},\Jsyn{\Tt})$ is subcanonical exactly when the signature has no
  sort.
\end{enumerate}
\end{theorem}

\begin{proof}
(1) A matching family of local variables is one local variable.
(2) Induction on $\varphi$.
(3) Apply (2) at the identity renaming. (4) On the empty cover of $\{.\bot\}$ an
amalgamation for the representable of $\{x.\top\}$ would rename $x$ into the empty
context; without sorts parallel arrows are equal and a covering family derives the
formula of its target.
\end{proof}

The universal model of $\Clf{\Tt}$ on the full syntactic site is that of
\cite{CL25a}, where syntactic categories are described by
renamings of variables only for theories with no axioms over relational
signatures. Whether the sheaves on
$(\Csyn{\Tt},\Jsyn{\Tt})$ form $\Clf{\Tt}$ with $U_{\Tt}$ as universal model is not claimed; by (4) this site is not
subcanonical once a sort exists, and Section~\ref{sec:discussion} asks two questions about it.

A \emph{preorder site} $C$ is a preorder $\ext$ ($p\ext q$ when $q$ refines $p$) with a Grothendieck topology
on its sieves; a set $V$ \emph{covers} $q$ when a covering sieve on $q$ lies in $V$, and sheaves are as in
Definition~\ref{def:generic}. A \emph{structure in sheaves} on $C$ has a sheaf for each sort and for each relation
symbol a subpresheaf of the product of its sorts containing each tuple whose restrictions to a covering sieve lie
in it; it is a \emph{model} of $\Tt$ when it forces each axiom everywhere by the clauses of
Definition~\ref{def:generic}. A \emph{point of $\Tt$ in sheaves on $C$} is a flat continuous functor $F$ from
$\Csyn{\Tt}$ to sheaves on $C$ whose filtering conditions and $\Jsyn{\Tt}$-covers are witnessed on covering sieves;
$\MF{F}$ has the carriers $F(\{x^s.\top\})$, and $R$ holds of the tuples that are locally projections of
$F(\{\vec x.R(\vec x)\})$.
The \emph{theory of points} of $C$ has an atom $\fat{p}$ for each object and the axioms
$\top\Der\bigvee_p\fat{p}$, $\fat{q}\Der\fat{p}$ for $p\ext q$, $\fat{p}\wedge\fat{q}\Der\bigvee_{p,q\ext r}\fat{r}$
and $\fat{p}\Der\bigvee_{r\in S}\fat{r}$ for $S$ a covering sieve on
$p$. A \emph{subterminal map} $f$ from $C$ to sheaves on a
preorder site $C'$ gives each object $p$ a set $f(p)$ of objects of $C'$ closed under refinement and containing what
it covers, with $f(p')\subseteq f(p)$ for $p\ext p'$; it is \emph{flat and continuous} when $\bigcup_p f(p)$ covers
every object of $C'$, $\bigcup_{p,p'\ext r}f(r)$ every member of $f(p)\cap f(p')$ and $\bigcup_{r\in R}f(r)$ every
member of $f(p)$ for $R$ a covering sieve on $p$.

On a preorder site the classifying property of a geometric theory becomes a statement about sheaves on the site
itself, and in this form it identifies the runs of a partial HDA with the points of the theory of runs $\TR$
(Theorem~\ref{thm:points}) and the models of the theory of points of a preorder site with the flat continuous
subterminal maps out of it, on which the points of a machine rest (Theorem~\ref{thm:pts-x}).

\begin{theorem}[Models in sheaves on a preorder site]\label{thm:classifying}
Let $\Tt$ be a geometric theory over a relational signature and $C$ a preorder site.
\begin{enumerate}
\item Every sequent derivable in $\Tt$ is forced in every model of $\Tt$ in sheaves on $C$,
  and a structure in sheaves on $C$ is a model of $\Tt$ exactly when it is isomorphic to the model of a point of
  $\Tt$ in sheaves on $C$. On the one-object site where only the maximal sieve
  covers, a structure in sets is thus a model of $\Tt$ exactly when it is isomorphic to the model of a point of
  $(\Csyn{\Tt},\Jsyn{\Tt})$.
\item The projections from a point $F$ to the point of $\MF{F}$ are natural and injective, and at every object
  and stage each element of the second is locally an image. They are surjective at every
  object and stage for every point of every theory on every preorder site exactly under
  $\UC$.
\item A structure in sheaves on a preorder site $C'$ is a model of the theory of points of $C$ exactly when the
  sets $f(p)$ of the objects of $C'$ forcing $\fat{p}$ form a flat continuous subterminal map from $C$ to sheaves on
  $C'$.
\end{enumerate}
\end{theorem}

\begin{proof}
(1) Forcing is stable under refinement and local, so soundness is an induction on
derivations. A model gives its functor of
valuations, flat since cones are conjunctions and continuous by soundness; conversely flatness lifts valuations and
continuity projects them by induction on formulas, with forcing for
satisfaction.
At one object local truth is truth.
(2) Injectivity is flatness, and under $\UC$ a sheaf glues sections given locally by a
uniquely satisfied relation. Conversely, for
objects $i$ refining and covering one object $\ast$, the sheaf of the solutions of $R_i$ at $i$ and of their
families at $\ast$ gives a point whose surjectivity at $\ast$ is a choice
function.
(3) By forcing, the axioms of the theory of points are the conditions on $f$.
\end{proof}

The sites are preorders, so Theorem~\ref{thm:classifying} is the localic case of the classifying property;
Diaconescu's equivalence \cite{Diaconescu75,CL25a} identifies the flat
continuous maps of (3) with the geometric morphisms from sheaves on $C'$ to sheaves on $C$.

\subsection{Partial HDAs and their runs}\label{ssec:phda}

An \emph{alphabet} is a set $\Lambda$ whose equality is decidable only where a statement says so; the
alphabet $\ab$ has decidable equality.

\begin{definition}[Conclists and cofaces]\label{def:coface}
A \emph{conclist} is a finite word $U=u_1\cdots u_n$ over $\Lambda$ of \emph{dimension} $\dimn{U}=n$. For
disjoint sets $A$ and $B$ of coordinates of a conclist $V$ the \emph{coface}
$d_{A,B}\colon V\setminus(A\cup B)\to V$ keeps the other coordinates in their order and marks those of $A$
with $0$ and those of $B$ with $1$. It is \emph{lower} ($d^0_A$) when $B=\emptyset$
and \emph{upper} ($d^1_B$) when $A=\emptyset$. A lower or upper coface is \emph{proper}
when its domain has smaller dimension than $V$. Cofaces compose by $d_{A,B}\circ d_{C,D}=d_{A\cup C,\,B\cup D}$, with $\cid_U=d_{\emptyset,\emptyset}$.
\end{definition}

These are the conclist inclusions of \cite{BDFZ26}, with conclists
taken up to isomorphism so that an event is its coordinate; they label
the morphisms of the large precube category of \cite{FJSZ21}.

\begin{definition}[Partial HDA]\label{def:phda}
A \emph{partial HDA} $X$ over $\Lambda$ has a set $X[U]$ of \emph{cells} for each conclist $U$ and a partial
\emph{face map} $\delta_{A,B}=X[d_{A,B}]$ for each coface such that $\delta_{\emptyset,\emptyset}=\mathrm{id}$ and
$\delta_{C,D}\,\delta_{A,B}\subseteq\delta_{A\cup C,\,B\cup D}$ as partial maps; for
$d_{A,B}\colon U\to V$ the map $\delta_{A,B}$ sends each cell of $X[V]$ to a cell of $X[U]$ or to one further value
marking its face undefined, so that whether a face is defined is decided. It carries
bits $\st_X,\acc_X\colon X[U]\to\{0,1\}$ for each conclist $U$ marking the \emph{start} and \emph{accepting} cells.
It is \emph{finite} when it is given a \emph{finiteness structure}, which is a finite list of conclists outside
which every $X[U]$ is empty together with a finite enumeration and decidable equality of each
$X[U]$.
\end{definition}

These are the lax functors $\Cube^{\mathrm{op}}\mathrel{\tilde{\rightarrow}}\Set_\ast$ of
\cite{BDFZ26},
and an HDA in the sense of
\cite{PAF25} is one with every face defined.
The \emph{label} of a cell $x$ of shape $U$ is $(U,\acc_X(x))$; a \emph{graph} has no cell of dimension above $1$.

\begin{definition}[Steps and tracks]\label{def:track}
An \emph{up-step} $x\arrO{A}y$ is a lower coface $d^0_A$ with $\delta^0_A(y)=x$ and ipomset the starter
$\starter{V}{A}$, and a \emph{down-step} $y\arrI{B}x$ an upper coface $d^1_B$ with $\delta^1_B(y)=x$ and
ipomset the terminator $\terminator{V}{B}$, where $V$ is the shape of $y$. A step is \emph{proper} when its
cells differ in dimension. An \emph{$\omega$-track} is a sequence of cells with proper steps
$\beta_0\xrightarrow{s_0}\beta_1\xrightarrow{s_1}\beta_2\cdots$, and $X$ is \emph{rigid} when for all defined cells $c$
and $c'$ at most one proper step goes from $c$ to $c'$.
\end{definition}

Steps are taken in the completion $X_\ast$ of $X$ by a cell $\ast_U$ per shape, the value of every undefined
face, and the cells of $X$ are its \emph{defined} cells. A \emph{concrete ipomset} over $\Lambda$ has events
$0,\dots,n-1$ labelled in $\Lambda$ with a relation of precedence and an event order between them, and two sets of
events as its source and target interfaces; $\Ipom_\Lambda$ is their set. The starter
$\starter{V}{A}$ has the events of $V$ with their labels and the order of the coordinates of $V$ as event order,
with no precedence; its target interface holds every event and its source interface the events outside
$A$. The terminator $\terminator{V}{B}$ has every event in its source interface and
the events outside $B$ in its target interface; these are the starters and
terminators of \cite{FJSZ21,PAF25}.
The \emph{cell graph} of $X$ has the
cells of $X$ as vertices and its proper steps as edges.

\begin{convention}[Acceptance]\label{conv:buchi}
Over any alphabet an $\omega$-track $\beta$ is \emph{recurrently accepting} when for every $n$ some $m>n$ has
$\beta_m$ accepting. Over an alphabet with decidable equality $\beta$ is
\emph{PAF-accepted} (after Passemard, Amrane and Fahrenberg) when it avoids the cells $\ast_U$, $\beta_0$ starts and some accepting cell $c$ has
$\beta_m=c$ for arbitrarily late $m$.
\end{convention}

We call $\beta$ \emph{accepted} when $\beta_0$ starts and $\beta$, avoiding the cells $\ast_U$, is
recurrently accepting. PAF-acceptance is the B\"uchi condition in the form of
\cite{PAF25}; it implies acceptance,
and on finite $X$ the converse holds under $\EIO$.

\begin{definition}[Languages]\label{def:languages}
Let $X$ be a partial HDA over $\Lambda$. The \emph{ST-word} of an $\omega$-track $\beta$ with steps $s_0,s_1,\dots$
is the sequence $\stw(\beta)=(\ip(s_n))_{n\in\Nn}$ of its step
ipomsets. For $\Lambda$ with decidable equality the \emph{accepted language} $\LangA{X}$ is the set of
ST-words of the accepted $\omega$-tracks, and the \emph{PAF-accepted language}
$\LangPAF{X}$ is the set of ST-words of the PAF-accepted $\omega$-tracks.
\end{definition}

Tracks match paths of the ST-automaton $\STa(X)$
\cite{PAF25}, so $\LangPAF{X}=L(\STa(X))$.

\begin{definition}[Runs]\label{def:run}
A \emph{run} of $X$ is a pair of a relation $A\subseteq\Nn\times\textstyle\bigsqcup_U X[U]$ saying which cell is at
time $n$ and a relation $\sigma\subseteq\Nn\times\Ipom_\Lambda$ such that
\begin{enumerate}
\item some cell is at time $0$ and each starts, and a cell at $n+1$ has one at $n$; all
  cells at one time have one shape and are equal;
\item cells $x$ at $n$ and $y$ at $n+1$ have $x=\delta^0_{A'}(y)$ along a proper lower coface or
  $y=\delta^1_{B'}(x)$ along a proper upper coface;
\item at each $n$ the relation $\sigma$ holds of at most one ipomset, of one when cells are at $n$ and $n+1$ and of
  none when no cell is at $n+1$; the ipomset $P$ it holds of at $n$ gives a proper step of ipomset $P$ between the
  cells at $n$ and $n+1$;
\item accepting cells are at arbitrarily late times.
\end{enumerate}
\end{definition}

An accepted track $\beta$ has a run $\Tr(\beta)$ determining it; that every run is one is not claimed.

\subsection{Witnesses}\label{ssec:witnesses}

$X\dsum Y$ is the disjoint sum of $X$ and $Y$, and
$X^{(4)}=(X\dsum X)\dsum(X\dsum X)$.

\begin{example}[Witnesses]\label{exa:dec-witnesses}\label{exa:image-sq}\label{exa:image-pk}\label{exa:pair}\label{exa:image-loops}\label{exa:image-branch}
The machines are over $\ab$ and the edges of every graph but $\mathsf{E}$ have the shape $a$.
\begin{enumerate}
\item The \emph{torus} $\TB{\mathfrak A}$ has one cell in each shape $\varepsilon,a,b,ab$, written $q_0,e_a,e_b,q_{ab}$,
  with every face the cell of its shape \cite{PAF25};
  $q_0$ alone starts and $\mathfrak A$ is the set of accepting cells. Its \emph{square loop} is at $q_0$ at even and
  at $q_{ab}$ at odd times. $\mathsf{T}_0=\TB{\{q_0\}}$ and $\mathsf{T}_1=\TB{\{q_0,e_a,e_b\}}$. The
  \emph{two-edge graph} $\mathsf{E}$ has the cells $v=q_0$, $e_a$ and $e_b$ of the torus with their
  faces and $v$ starting; in $\mathsf{E}_v$ the vertex alone accepts.
\item The \emph{bouquet} $\Wq{1}{k}{\mathfrak A}$ has one start vertex $v$ and $k$ loops $e_0,\dots,e_{k-1}$ at $v$, and
  $\Wq{2}{k}{\mathfrak A}$ has one start vertex $v$ and $k$ squares $s_0,\dots,s_{k-1}$ of shape $ab$, each with the face
  $v$ along every coface from the empty conclist and no face along the cofaces from $a$ and from $b$. Both also
  have $k-1$ further vertices $v_1,\dots,v_{k-1}$ that neither start nor are faces of any
  cell, so that no step meets them; they give a bouquet as many vertices as
  loops or squares, which the suspension of Section~\ref{ssec:image-double} asks. In both $\mathfrak A$ is the set of accepting
  cells among $v$ and the loops or squares.
  We write $\wG=\Wq{1}{2}{\{v,e_1\}}$ and $\wGt=\Wq{1}{2}{\{v\}}$;
  $\wQ=\Wq{2}{2}{\{v,s_0\}}$, $\wQ^\circ=\Wq{2}{2}{\{v\}}$ and
  $\wQt=\Wq{2}{2}{\{s_0,s_1\}}$.
  The acceptance of the further vertices is fixed apart from $\mathfrak A$, which no prefix sees since no step meets them, and $v_1$ accepts in $\wG$, $\wGt$, $\wQ$ and
  $\wQ^\circ$ and in no other bouquet.
  $\mathsf{P}$ is $\Wq{2}{1}{\{v\}}$ with the faces of its square along $d_{a,b}$ and $d_{b,a}$ left
  undefined.
\item The \emph{interleaved square} $\mathsf{S}$ adds to $\mathsf{P}$ an $a$-edge $e_a$ from $v$ and a $b$-edge $e_b$ to
  $v$, the faces of $s_0$ before $b$ starts and after $a$ finishes; the other faces of $s_0$ stay undefined and
  $v$ alone starts and accepts.
\item The \emph{spine} $\Cs{m}{\mathfrak A}$ is the graph with vertices $0$ and $1$, a loop $e_0$ at $0$, a loop at $1$
  and $m$ edges $e_1,\dots,e_m$ from $0$ to $1$; the vertex $0$ starts and $\mathfrak A$ is the set of accepting cells. In
  $\wC{m}$ exactly the vertices accept and in $\wCh{m}$ the edges from $0$ to $1$ also
  accept.
\end{enumerate}
\end{example}

\subsection{The theory of runs and its points}\label{ssec:theory}

\begin{definition}[The theory $\TR$]\label{def:theory}
The signature $\SigR$ has a sort for each conclist $U$ and symbols $\fsym{A,B}$ of arity $(V,U)$ for the
cofaces $d_{A,B}\colon U\to V$ together with unary $\mathsf{st}_U,\nst_U,\mathsf{acc}_U,\nacc_U$;
it has further unary $\mathsf{at}_{n,U}$ for $n\in\Nn$ and nullary $\evat{n}{P}$ for
$n\in\Nn$ and $P\in\Ipom_\Lambda$. The theory $\TR$ has the axioms
\[
\renewcommand{\arraystretch}{1.3}
\begin{array}{@{}l@{\;}r@{\;}l@{\enspace}r@{\;}l@{}}
\textit{faces} & \top & \Der_x\fsym{\cid_U}(x,x) & \fsym{A,B}(y,x)\wedge\fsym{A,B}(y,x') & \Der x=x'\\
 & \fsym{A,B}(z,y)\wedge\fsym{C,D}(y,x) & \Der\fsym{A\cup C,B\cup D}(z,x)\\[2pt]
\textit{bits} & \top & \Der_x\mathsf{st}_U(x)\vee\nst_U(x) & \mathsf{st}_U(x)\wedge\nst_U(x) & \Der\bot\\
 & \top & \Der_x\mathsf{acc}_U(x)\vee\nacc_U(x) & \mathsf{acc}_U(x)\wedge\nacc_U(x) & \Der\bot\\[2pt]
\textit{track} & \top & \Der\textstyle\bigvee_U\exists x\,\mathsf{at}_{0,U}(x) & \mathsf{at}_{0,U}(x) & \Der\mathsf{st}_U(x)\\
 & \mathsf{at}_{n+1,U}(x) & \Der\textstyle\bigvee_V\exists y\,\mathsf{at}_{n,V}(y) & \mathsf{at}_{n,U}(x)\wedge\mathsf{at}_{n,U}(x') & \Der x=x'\\
 & \mathsf{at}_{n,U}(x)\wedge\mathsf{at}_{n,V}(y) & \Der\bot\ (U\neq V)\\
 & \multicolumn{4}{@{}l@{}}{\mathsf{at}_{n,U}(x)\wedge\mathsf{at}_{n+1,V}(y)\,[\wedge\,\evat{n}{P}]\Der
    \textstyle\bigvee_{A}\fsym{A,\emptyset}(y,x)\vee\bigvee_{B}\fsym{\emptyset,B}(x,y)}\\[2pt]
\textit{acceptance} & \multicolumn{4}{@{\quad}l@{}}{\top\Der\textstyle\bigvee_{m\geq n}\bigvee_U\exists x\,(\mathsf{at}_{m,U}(x)\wedge\mathsf{acc}_U(x))}\\[2pt]
\textit{steps} & \mathsf{at}_{n,U}(x)\wedge\mathsf{at}_{n+1,V}(y) & \Der\textstyle\bigvee_P\evat{n}{P} & \evat{n}{P}\wedge\evat{n}{Q} & \Der\pprop{P=Q}\\
 & \evat{n}{P} & \Der\textstyle\bigvee_V\exists y\,\mathsf{at}_{n+1,V}(y) & \mathsf{at}_{n,U}(x)\wedge\mathsf{at}_{n,V}(y) & \Der\pprop{U=V}
\end{array}
\]
where the disjunctions range over proper lower cofaces $d^0_A$ and proper upper cofaces
$d^1_B$, restricted to those whose step has
ipomset $P$ when $\evat{n}{P}$ is present. $\TRm$ has the axioms of the first
four groups without $\evat{n}{P}$. $\Tacc$ replaces in $\TRm$ the downward axiom
$\mathsf{at}_{n+1,U}(x)\Der\bigvee_V\exists y\,\mathsf{at}_{n,V}(y)$ by the \emph{totality} axioms
$\top\Der\bigvee_U\exists x\,\mathsf{at}_{n,U}(x)$ for every $n$.
\end{definition}

The faces and bits axiomatise partial HDAs, and the fourth group is the \emph{acceptance} axiom of index
$n$. For an $\omega$-track $\beta$ of $X$ the structure $X_\beta$ interprets $\mathsf{at}_{n,U}(x)$ as
$\beta_n=x$; over an alphabet with decidable equality $X_\beta\models\Tacc$ exactly
when $\beta$ is accepted.

A \emph{relational form} has sets $X[U]$, relations $\delta_{A',B'}\subseteq X[V]\times X[U]$ and four
predicates satisfying the axioms of faces and bits; it \emph{decides its faces} when $\exists x\,\delta_{A',B'}(y,x)$ is decided
for every coface and every $y$, and its runs are those of Definition~\ref{def:run} with face relations for face
maps.
A partial HDA is one through the graphs of its faces; the relational
HDAs of \cite{BDFZ26} differ by allowing several faces along one
coface. A point $F$ of $\TR$ has the relational form $\XF{F}$ and relations $A_F,\sigma_F$
read off $\MF{F}$; it \emph{has decided faces} when $\XF{F}$ does, and is \emph{carried} by $X$ with $A,\sigma$ when
$\MF{F}$ is isomorphic to the structure they
form.

Reading the points of $\TR$ as runs is what lets a property of its classifying topos speak about the runs of
machines; for relational forms the reading holds with no principle, and for partial HDAs over $\ab$ it holds, under
unique choice, exactly when excluded middle does.

\begin{theorem}[Points]\label{thm:points}\label{thm:decided}\label{thm:price-em}
Let $\Lambda$ be any alphabet.
\begin{enumerate}
\item Each point $F$ of $\TR$ gives the run $(A_F,\sigma_F)$ of $\XF{F}$. The
  structure of each relational form $X$ with a run $(A,\sigma)$ is isomorphic to the model of some point of
  $\TR$, and $(X,A,\sigma)$ is isomorphic to the triple
  $(\XF{F},A_F,\sigma_F)$ of the functor $F$ of valuations in that structure.
\item A point $F$ of $\TR$ carried by a partial HDA $X$ with $A,\sigma$ has decided
  faces, $(A,\sigma)$ is a run of $X$, and
  $\XF{F}$ has a finiteness structure exactly when $X$ has one. For every run
  $(A,\sigma)$ of every partial HDA $X$ some point of $\TR$ with decided faces is carried by $X$ with
  $A,\sigma$. Under $\UC$ a functor $\Csyn{\TR}\to\Set$ is a point with decided
  faces exactly when it is a point carried by some partial HDA with a
  run.
\item $\EM$ holds if every point of $\TRab$ has decided faces, and also if each
  is carried by a partial HDA with relations $A,\sigma$. Under $\EM$ and $\UC$ every point
  of $\TR$ is so carried; so under $\UC$ every point of $\TRab$ is carried by a
  partial HDA exactly when $\EM$ holds.
\end{enumerate}
\end{theorem}

\begin{proof}
(1) The axioms of $\TR$ unfold to the clauses of a relational form with a
run, and
Theorem~\ref{thm:classifying}(1) gives the point.
(2) A face in $X$ is defined or not, and the isomorphism carries the answer and finiteness structures
back;
the second claim is (1), and $\UC$ builds partial face maps
from decided faces.
(3) For a proposition $q$ and either $j\in\{0,1\}$ the \emph{gate} is $\mathsf{T}_j$ with the face $\delta^0_a(e_a)=q_0$ kept only under $q$; its
square loop is a run whose point has
that face exactly under $q$, so each hypothesis of (3) decides
$q$.
Under $\EM$ faces are decided and (2) applies.
\end{proof}

Item (2) constructivises the inclusion of partial into relational HDAs
\cite{BDFZ26}.

\begin{remark}\label{rem:not-presheaf}
For both $j\in\{0,1\}$ the run of the square loop of $\mathsf{T}_j$ models $\TRm$ over $\ab$ while no model presented by a
geometric formula maps there. For if a model $M_0$ of $\TRm$ presented by
$\varphi$ at $a$ had a map $g$ there, then $\varphi(g(a))$ would hold in the run, and by finite support in every model of
$\TRm$ that agrees with it up to some time $N$, among them the run that follows the square
loop up to time $2N$ and is at $e_a$ at time $2N+1$; presentation would also map $M_0$ there, and a cell of $M_0$ at
time $2N+1$ would go to cells of its own shape at that time in both runs, while one run is at $q_{ab}$ and the other at
$e_a$. So $\TRm$ is not of presheaf type over $\ab$ by
Theorems~5.1(i) and~7.7(i) of
\cite{Caramello14}.
\end{remark}
\section{The topos of a machine over the label topos}\label{sec:topos}

\subsection{The site of prefixes}\label{ssec:machine}

\begin{definition}[Prefixes]\label{def:prefixes}
A \emph{prefix} $p$ of a partial HDA $X$, of length $\dimn{p}=n$, is a start cell $p_0$ followed by proper steps
$p_0\xrightarrow{s_0}p_1\cdots\xrightarrow{s_{n-1}}p_n$ whose cells $p_0,\dots,p_n$ are all defined, so that a prefix passes no cell
$\ast_U$. A prefix $q$ \emph{extends} $p$, written $p\ext q$, when
$\dimn{p}\leq\dimn{q}$ and $q$ has the cells and steps of $p$ up to $\dimn{p}$. The \emph{chain} $\chain{\beta}$ of
an $\omega$-track is the set of its initial prefixes.
\end{definition}

Prefixes are the paths of \cite{ZvGZ26}; $\Pref(X)$ is a preorder under
$\ext$. In the spine $\wC{1}$ of Example~\ref{exa:dec-witnesses} the
prefix $\kappa_n$ of length $n$ alternates between $0$ and $e_0$, and $\kappa'_2$ is the prefix
$0,e_1,1$.

\begin{definition}[B\"uchi topologies]\label{def:prefixes-site}
Let $X$ be a partial HDA. A \emph{sieve} on $p$ is a set of extensions of $p$ closed under
extension. Let $\BB(p,n)$ be the set of extensions $r$ of $p$ with $r_i$ accepting for some
$n<i\leq\dimn{r}$. A \emph{Grothendieck topology} on $\Pref(X)$ is a relation ``$S$ covers
$p$'' between prefixes and sieves on them under which the maximal sieve on $p$ covers $p$ and a sieve covering $p$
pulls back to a sieve covering each extension of $p$; a sieve $R$ on $p$ covers $p$ whenever the pullbacks of $R$
cover the members of some sieve covering $p$. A Grothendieck topology on $\Pref(X)$ is
\emph{B\"uchi} when each $\BB(p,n)$ covers $p$. A B\"uchi topology is \emph{least} when each
sieve covering $p$ in it covers $p$ in every B\"uchi topology.
\end{definition}

\begin{proposition}[The B\"uchi topology]\label{thm:jx}
Let $X$ be a partial HDA over any alphabet, and let $S$ cover $p$ in $\JX{X}$ inductively when $p\in S$ or some
$n$ has every $r\in\BB(p,n)$ covered by $S$. Then $\JX{X}$ is a Grothendieck
topology in which each $\BB(p,n)$ covers $p$.
It is a least B\"uchi topology, and a relation between prefixes and sets of prefixes
is a least B\"uchi topology exactly when it lets the same sieves cover as $\JX{X}$.
\end{proposition}

\begin{proof}
Induction on the clauses, with stability and transitivity of the given
topology; the Grothendieck axioms see a topology only through its
sieves.
\end{proof}

\begin{definition}[Closed acceptance]\label{def:closed}
Let $X$ be a partial HDA. A prefix that the empty sieve covers in $\JX{X}$ is \emph{dead}. $X$ has \emph{closed acceptance} with
constant $\Delta$ when every prefix $r$ and every $n$ with $n+\Delta\leq\dimn{r}$ have an accepting $r_i$ with
$n<i\leq\dimn{r}$. A set of prefixes \emph{reads} a sequence $w$ of ipomsets when each step
$s_i$ of its members has $\ip(s_i)=w_i$.
\end{definition}

Over an alphabet with decidable equality an $\omega$-track of a machine with closed acceptance is accepted exactly
when it avoids the cells $\ast_U$ and starts; the acceptance
condition thus rejects no $\omega$-track from a start cell along defined cells, while the acceptance bits still enter
the label words of Section~\ref{ssec:labels}.

\subsection{Sheaves and truth values}\label{ssec:built}

\begin{definition}[Sheaves]\label{def:setup-sheaf}
A \emph{presheaf} $\mathscr{F}$ on $X$ is a set $\mathscr{F}(p)$ for each prefix with functorial restrictions
$\mathscr{F}(p)\to \mathscr{F}(q)$ for $p\ext q$. It is a \emph{sheaf} when every matching family on a covering sieve
has exactly one amalgamation, and a \emph{data sheaf} when a function choosing that amalgamation is given.
$\mathscr{F}^+(p)$ is the set of matching families of $\mathscr{F}$ on covering sieves of $p$ up to agreement on a
covering sieve, and $\mathscr{F}^{++}=(\mathscr{F}^+)^+$.
\end{definition}

\begin{definition}[Subobject classifier]\label{def:setup-omega}
A sieve on $p$ is \emph{closed} when it contains every extension $q$ of $p$ at which its pullback covers $q$ in
$\JX{X}$. A \emph{subpresheaf} $\mathscr{P}$ of a presheaf $\mathscr{F}$ is a choice of
subsets $\mathscr{P}(p)\subseteq\mathscr{F}(p)$ preserved by restriction. For
$x\in\mathscr{F}(p)$ the \emph{sieve} of $x$ in $\mathscr{P}$ is
$\chi_{\mathscr{P}}(x)=\{q : p\ext q,\ x|_q\in \mathscr{P}(q)\}$. The subpresheaf
$\mathscr{P}$ is \emph{closed} when it contains every $x\in \mathscr{F}(p)$ whose sieve $\chi_{\mathscr{P}}(x)$
covers $p$ in $\JX{X}$. Let $\mathbf{1}$ be the presheaf with one
element at every prefix. A \emph{subobject classifier} of $X$ is a presheaf $\Omega$
carrying a structure of data sheaf with a morphism $\top\colon\mathbf{1}\to\Omega$ such that for every presheaf
$\mathscr{F}$ the map sending $\varphi\colon\mathscr{F}\to\Omega$ to $\{x : \varphi(x)=\top\}$ is a bijection onto
the closed subpresheaves of $\mathscr{F}$.
\end{definition}

\begin{proposition}[The topos built]\label{thm:setup-built}
Let $X$ be a partial HDA over any alphabet.
\begin{enumerate}
\item The presheaf $\Omega_X$ of closed sieves restricted by pullback carries a structure of data
  sheaf. With $\top$ the maximal sieves it is a subobject classifier of
  $X$, and the inverse of its bijection is $\mathscr{P}\mapsto\chi_{\mathscr{P}}$.
  Two subobject classifiers of $X$ are isomorphic by an isomorphism preserving $\top$. A subpresheaf of a sheaf is closed exactly when it is a sheaf.
\item The presheaf $\mathbf{1}$, the product of two data sheaves and the equaliser of two morphisms from a data sheaf
  to a sheaf carry structures of data sheaf;
  they are terminal, product and equaliser among
  presheaves.
\item For a data sheaf $\mathscr{E}$ and any presheaf $\mathscr{F}$ the presheaf $\mathscr{E}^{\mathscr{F}}$ carries
  a structure of data sheaf; for all presheaves $\mathscr{H}$, $\mathscr{F}$ and
  $\mathscr{E}$ there is a bijection
  $\mathrm{Hom}(\mathscr{H}\times \mathscr{F},\mathscr{E})\cong\mathrm{Hom}(\mathscr{H},\mathscr{E}^{\mathscr{F}})$,
  natural in $\mathscr{H}$.
\item Morphisms from a presheaf $\mathscr{F}$ to a data sheaf factor uniquely through $\mathscr{F}\to
  \mathscr{F}^{++}$.
\item A data sheaf is a sheaf, and a presheaf carries at most one structure of
  data sheaf.
\end{enumerate}
\end{proposition}

\begin{proof}
Closed sieves glue to the closure of their union; limits are formed componentwise and exponentials
pointwise.
A morphism to a data sheaf lifts to $\mathscr{F}^+$ by gluing.
\end{proof}

The data sheaves on $X$ stand in for the \emph{topos} $\Sh(\Pref(X),\JX{X})$ of $X$, and under $\UC$ every sheaf
carries a structure of data sheaf (Theorem~\ref{thm:prices}(1)).

\subsection{The frame and the points}\label{ssec:points}

\begin{definition}[The frame]\label{def:frame}
An \emph{open} of $X$ is a set of prefixes closed under extension, such as the \emph{cylinder} $\upc{p}$ of the extensions of $p$
and the top open $\Pref(X)$.
$\Op(X)$ is the set of opens with $A\fle{X}B$ when $B$ covers every member of $A$, and $\feq{X}$ both ways; a
\emph{least upper bound} of a family is an upper bound below every upper bound. An \emph{equivalence}
$\Op(X)\eqFrF\Op(Y)$ is a pair of monotone maps $h$ and $h'$ back with $h'(h(A))\feq{X}A$ and
$h(h'(B))\feq{Y}B$. An open $W$ is \emph{compact} when every family of opens whose union lies above $W$ has a finite
subfamily whose union lies above $W$, and $\Op(X)$ is compact when $\Pref(X)$ is. $W$ is \emph{complemented} when
some open $W'$, a \emph{complement} of $W$, has $W\cap W'\fle{X}\emptyset$ and $\Pref(X)\fle{X}W\cup W'$. A
\emph{completely prime filter} of $\Op(X)$ is an inhabited set of opens closed upward under $\fle{X}$ that has a
member below any two members and a member of each family whose least upper bound it contains.
\end{definition}

Union is a least upper bound.

Subterminals and global elements of $\Omega_X$ are objects and arrows of the topos of $X$, so that identifying
them with opens of prefixes up to $\feq{X}$ makes every statement about the frame of $X$ a statement about its topos
that is checked by covering alone.

\begin{theorem}[The frame of subterminals]\label{thm:frame-sub}
Let $X$ be a partial HDA over any alphabet. A \emph{subterminal} of $X$ is a closed subpresheaf of $\mathbf{1}$, ordered
by inclusion, and for an open $A$ the subterminal $\sbt{A}$ has its element at the prefixes that $A$ covers.
\begin{enumerate}
\item The intersection of two subterminals is their greatest lower bound, and the closure of the union of a family
  is its least upper bound.
\item $A\mapsto\sbt{A}$ induces a bijection from $\Op(X)/{\feq{X}}$ onto the subterminals that preserves and
  reflects the order and sends $A\cap B$ and $\bigcup_i A_i$ to the meet and the join of their
  images. Each subterminal is $\sbt{A}$ for the open $A$ of the prefixes
  at which it has its element, and an open $A$ is that open of $\sbt{A}$ exactly when it
  contains every prefix it covers.
\item Composed with $\mathscr{P}\mapsto\chi_{\mathscr{P}}$, the bijection of (2) is a bijection from
  $\Op(X)/{\feq{X}}$ onto the morphisms $\mathbf{1}\to\Omega_X$ under which $\fle{X}$ is the pointwise inclusion of
  closed sieves.
\end{enumerate}
\end{theorem}

\begin{proof}
Covering is stable and transitive, so $\sbt{A}$ is closed and $A\fle{X}B$ exactly when
$\sbt{A}\subseteq\sbt{B}$;
Proposition~\ref{thm:setup-built}(1) gives (3).
\end{proof}

A preorder site $(\mathcal{C},J)$ has a localic topos \cite[Example~A4.6.2(e)]{Johnstone02} whose frame of
subterminals is that of $J$-ideals by Theorem~3.1 of
\cite{Caramello11}. Theorem~\ref{thm:frame-sub} is
its instance at $(\Pref(X),\JX{X})$ with no principle; its ideals are the opens containing all they cover.

The quotient $\Op(X)/{\feq{X}}$ is the \emph{frame} of $X$, and $X$ and $Y$ are \emph{equivalent over $1$} when
$\Op(X)\eqFrF\Op(Y)$. A subterminal is a sheaf (Proposition~\ref{thm:setup-built}(1)) with a monomorphism into
$\mathbf{1}$; that every such sheaf is isomorphic to
$\sbt{A}$ for the open $A$ of the prefixes at which it has an element is proved under $\UC$ and open with no
principle.

\begin{definition}[Points]\label{def:points-theory}\label{def:points}
Let $X$ be a partial HDA. Its prefixes under extension with the topology $\JX{X}$ form the preorder site
$(\Pref(X),\JX{X})$. The \emph{theory of points} $\FX{X}$ of $X$ is the theory of points of
this site (Section~\ref{ssec:geometric}), with an atom $\fat{p}$ for each prefix $p$. A \emph{point} of $X$ is a model of
$\FX{X}$ in sets. The \emph{members} of a point are the prefixes whose atoms it makes
true.
\end{definition}

The points of $X$ are where its topos meets its runs, since by item~(4), over an alphabet with decidable equality, the chains of
accepted $\omega$-tracks are points, so that an invariant of the topos computed on points constrains the accepted behaviour of $X$.

\begin{theorem}[Points of a machine]\label{thm:pts-x}\label{prop:ct-points}
Let $X$ be a partial HDA over any alphabet $\Lambda$.
\begin{enumerate}
\item A structure in sets for $\FX{X}$ is determined by the set of prefixes whose atoms it makes
  true, and it is a model exactly when that set is inhabited, closed under initial
  segments and directed and meets every covering sieve on each of its
  members.
\item For a point $\xi$ the opens containing a member of $\xi$ form a completely prime filter of
  $\Op(X)$, and for a completely prime filter $P$ the prefixes $p$ with
  $\upc{p}\in P$ form a point. The point formed from the filter of a point $\xi$
  is $\xi$, and the filter formed from the point of a completely prime filter $P$
  has the members of $P$.
\item The models of $\FX{X}$ in sheaves on a preorder site $C'$ are the flat continuous subterminal maps of
  Theorem~\ref{thm:classifying}(3) from $(\Pref(X),\JX{X})$ to sheaves on $C'$.
\item Over $\Lambda$ with decidable equality the chain of an $\omega$-track $\beta$ is a point
  exactly when $X_\beta\models\Tacc$.
\end{enumerate}
\end{theorem}

\begin{proof}
(1) Read at a set of prefixes the axioms of $\FX{X}$ are the four
clauses. (2) A point meets every
cover of its members by an open, and a completely prime filter contains $\upc{p}$ for some member $p$ of each of its
opens.
(3) is Theorem~\ref{thm:classifying}(3) at
$(\Pref(X),\JX{X})$. (4) A point meets each $\BB(p,n)$, and induction on
$\JX{X}$ gives the converse.
\end{proof}

By (1) we identify a point with the set of its members; the two maps of (2) are then mutually inverse. Item~(2) is the
instance at $(\Pref(X),\JX{X})$ of Theorem~7.3 of \cite{Caramello11}
on the completely prime filters of the frame of $J$-ideals; here no principle is used. Applied to (3),
Diaconescu's equivalence \cite{Diaconescu75,CL25a} identifies the topos of
$X$ with the classifying topos $\Clf{\FX{X}}$. The \emph{language of the points} $\LangPt{X}$ is the set of
sequences read by some point of $X$.

\begin{definition}[Derived sets and derived opens]\label{def:derived}\label{def:dec-derived}
A point \emph{lies in} an open when one of its members does. A point $\xi$ is
\emph{isolated} in a set $D$ of points when some member $p$ of $\xi$ has every point of $D$ that contains $p$
contained in $\xi$. $\dset{0}{X}$ is the set of points and $\dset{n+1}{X}$ the set of points
of $\dset{n}{X}$ not isolated in it. A point has \emph{rank} $n$ when it belongs to
$\dset{n}{X}$ and is isolated in it. A \emph{$0$-th derived open} is an open below every open
of $X$ under $\fle{X}$. An \emph{$(n+1)$-th derived open} is a least upper
bound of an $n$-th derived open $D$ and of every open $A$ such that exactly one completely prime filter of $\Op(X)$
contains $A$ and not $D$.
\end{definition}

By Theorem~\ref{thm:pts-x}(2) the opens $A$ of the last kind are those in which exactly one point outside $D$ lies.
For every $n$ some $n$-th derived open $D$ of $X$ has $\dset{n}{X}$ as the points of $X$ outside
$D$, and we call the closed
complement of $D$ the $n$-th derived subtopos, whose points are those of $\dset{n}{X}$
\cite{CL25b}. The empty open is a $0$-th derived
open, and the union of an $n$-th derived open $D$ with every open in which exactly one point outside $D$ lies is an
$(n+1)$-th one. By induction on $n$ a completely prime filter
misses an $n$-th derived open exactly when it survives $n$ rounds in which each filter that some open holds alone among
the survivors is removed, and these survivors are the points of $\dset{n}{X}$, since
an open that a point alone among the survivors lies in contains the cylinder $\upc{p}$ of one of its members $p$,
which then isolates the point in $\dset{n}{X}$, while a member $p$ isolating it makes $\upc{p}$ such an
open.

$\Omega_X$ is \emph{Boolean} at $p$ when $S\vee\neg S=\top$ for all $S$ in $\Omega_X(p)$ and \emph{De Morgan} when
$\neg S\vee\neg\neg S=\top$, unqualified at every
prefix.
$X$ is \emph{discrete} when each point $\xi$ has a member whose extensions that are not dead lie in
$\xi$. The \emph{stalk} at $\xi$ is the set of \emph{germs} at $\xi$, closed sieves on members of $\xi$
up to agreement on a common member; a germ \emph{holds} when it contains a member of
$\xi$, which is when it is the germ of
$\top$.

\begin{proposition}[Truth at points]\label{thm:setup-truth}
Let $X$ be a partial HDA over any alphabet. For each point $\xi$ every element of the stalk at $\xi$ holds or not
exactly when $\EM$ holds, and Booleanness at a member of a point gives
$\EM$. If $\Omega_X$ is Boolean then $X$ is
discrete. Under $\EM$ and $\DC$ the presheaf $\Omega_X$ is Boolean exactly
when $X$ is discrete, and it is De Morgan exactly when it is Boolean and exactly
when $X$ is discrete.
\end{proposition}

\begin{proof}
The closure of $\{q : p\ext q,\ R\}$ holds exactly when $R$ does, and a point
meets $S\cup\neg S$. Booleanness at a member of a point gives $\EM$, under which a
point without discreteness branches and refutes Booleanness. Under $\EM$ and $\DC$ an uncovered prefix lies on a point avoiding
the cover, and De Morgan rules out a branching
point.
\end{proof}

\subsection{The run theory}\label{ssec:theories}

\begin{definition}[Axiom sieve]\label{def:setup-axsieve}
Let $M_q$ be $X$ without the atoms $\evat{n}{P}$ with $\mathsf{at}_{n,U}(x)$ true when $n\leq\dimn{q}$ and
$q_n=x$. The \emph{sieve} of $\top\Der\psi$ on $p$ is
$S_\psi(p)=\{\,q : p\ext q,\ M_q\models\psi\,\}$.
\end{definition}

\begin{definition}[Acceptance theory and run theory]\label{def:machine-theories}
For a family $K$ of sieves on prefixes the theory $\FX{X}^{K}$ has the axioms of $\FX{X}$
(Definition~\ref{def:points}) other than its covering axioms, and in their place the sequents
$\fat{p}\Der\bigvee_{r\in S}\fat{r}$ for each sieve $S$ of $K$ on a prefix $p$. The theory
$\FX{X}^{\mathrm{acc}}$ is $\FX{X}^{K}$ for the family $K$ whose sieves on each prefix $p$ are the sieves on $p$ of
the acceptance and totality axioms (Definitions~\ref{def:theory} and~\ref{def:setup-axsieve}).
The \emph{run theory} $\RX{X}$ is over the signature with the symbols of $\SigR$ and a unary name $\nm{c}$ on $U$
for each cell $c$ of shape $U$. It has exactly these axioms.
\begin{enumerate}
\item The axioms of $\Tacc$ (Definition~\ref{def:theory}) and those of the diagram of $X$
  below.
\item The step axioms of $\TR$.
\item In place of the axiom of the track of $\TR$ containing $\evat{n}{P}$, the sequents
  \[
    \mathsf{at}_{n,U}(x)\wedge\nm{c}(x)\wedge\mathsf{at}_{n+1,V}(y)\wedge\nm{d}(y)\wedge\evat{n}{P}\Der\textstyle\bigvee_s\top,
    \qquad \nm{c}(x)\wedge\nm{d}(x)\Der\pprop{c=d}.
  \]
  In the first sequent $s$ ranges over the proper steps from $c$ to $d$ with $\ip(s)=P$.
\end{enumerate}
The \emph{diagram} of $X$ has the sequents
\[
  \top\Der\exists x\,\nm{c}(x),\quad \nm{c}(x)\wedge\nm{c}(x')\Der x=x',\quad
  \top\Der_x\textstyle\bigvee_{c\in X[U]}\nm{c}(x),\quad \nm{c}(x)\wedge\nm{d}(x)\Der\bot\ (c\neq d)
\]
and the sequents that give each named element the faces and bits of its cell. The
interpretation $\IX{X}$ of $\FX{X}$ in $\RX{X}$ sends $\fat{p}$ to the conjunction of the formulas
$\exists x\,(\mathsf{at}_{i,U}(x)\wedge\nm{p_i}(x))$ for $i\leq\dimn{p}$ and $U$ the shape of $p_i$ with the atoms
$\evat{i}{\ip(s_i)}$ for $i<\dimn{p}$.
\end{definition}

Theorem~\ref{thm:generic} applies to $\FX{X}$ and to $\RX{X}$ for every machine in every
dimension; the site of $\FX{X}$ is subcanonical since $\FX{X}$ has no sort, and that of
$\RX{X}$ is not.

\subsection{The label topos}\label{ssec:labels}

\begin{definition}[Label words and the label map]\label{def:labels}\label{def:setup-base}
A \emph{label word} over $\Lambda$ has a length $n$, a shape and an acceptance bit at every time and an ipomset at
every time below $n$; $\LW_\Lambda$ is their set. The label word $\lab(p)$ of a prefix $p$
records $\dimn{p}$, the shape and acceptance bit of each $p_i$ and the ipomset $\ip(s_i)$ of each
step. Let $u\lle w$ when $u$ is not longer than $w$ and has its shapes, acceptance bits and
ipomsets up to its length. The relation $\lle$ is reflexive and
transitive. In $\JLW$ a set $S$ covers $w$ when $w\in S$ or when for some $n$ every
extension of $w$ longer than $n$ is covered by $S$. An \emph{open} of label words is a set
closed under extension. For a label word $u$ the open $\lst{X}(u)$ holds the prefixes $q$ of
$X$ with $u\lle\lab(q)$. For a set $U$ of label words $\lst{X}(U)$ is
the union of the opens $\lst{X}(u)$ for $u\in U$. For an open $U$ of label words
the prefixes $p$ with $\lab(p)\in U$ form an open of $X$. The \emph{label open} in $Y$ of a
prefix $p$ of any partial HDA is $\lst{Y}(p)=\lst{Y}(\lab(p))$, which is the open of the
prefixes $q$ of $Y$ with $\lab(q)$ in the open $\{w : \lab(p)\lle w\}$. A \emph{step
word} has a length and an ipomset at every time below it; $\SW_\Lambda$ is their set. A step
word $v$ \emph{extends} $u$ when $u$ is not longer than $v$ and $v$ has the ipomsets of $u$ below the length of
$u$. In $\JSW$ a set $S$ covers $u$ when $u\in S$ or when for some $n$ every extension of $u$
longer than $n$ is covered by $S$. The step word $\fgt(w)$ of a label word $w$
forgets its shapes and acceptance bits, and the step word of a prefix $p$ is
$\fgt(\lab(p))$. For a set $U$ of step words $\lsts{X}(U)$ holds the
prefixes whose step word extends a member of $U$.
\end{definition}

Sheaves on $(\LW_\Lambda,\JLW)$ form the topos $\BL$ of \emph{label streams}, the label topos, and sheaves on
$(\SW_\Lambda,\JSW)$ the topos $\BST$ of \emph{step streams}. The \emph{label map} $\ell_X$ acts on opens by $\lst{X}$,
which on an open $U$ of label words is $\{p\in\Pref(X) : \lab(p)\in U\}$.

\begin{definition}[Direct and inverse image]\label{def:label-images}
\begin{enumerate}
\item A \emph{presheaf on label words} has a set at each label word and restrictions along $\lle$ that respect
  identities and composites; a morphism of such presheaves is a family of maps commuting with
  restriction. For a set $S$ of label words a \emph{matching family} at $w$ has an element at each
  member of $S$ whose restriction to each later member of $S$ is the element there. An
  \emph{amalgamation} of it is an element at $w$ whose restriction to each member $v$ of $S$ with $w\lle v$ is the
  element at $v$. For a set $K$ of sieves on label words a \emph{data $K$-sheaf} is a presheaf
  with a function choosing for each family matching on a sieve of $K$ its only amalgamation. It is \emph{$K$-separated} when two elements at $w$ with equal restrictions to the
  members of a sieve of $K$ on $w$ are equal. The relation that holds of a label word $w$ and a set $S$ of label words when $S$
  covers $w$ in the sense of Definition~\ref{def:labels} is also written $\JLW$. A \emph{data sheaf} on label words is a data $\JLW$-sheaf.
\item On step words presheaves restrict along extension and their morphisms commute with
  restriction. Data $K$-sheaves on step words are as in (1). The relation that holds of a step word $u$ and a set $S$ of step words when $S$ covers $u$ in the sense of
  Definition~\ref{def:labels} is also written $\JSW$. A data sheaf on step
  words is a data $\JSW$-sheaf.
\item For a presheaf $\mathscr{Q}$ on label words the presheaf $\mathscr{Q}\circ\lab$ on $X$ has at $p$ the set of
  $\mathscr{Q}$ at $\lab(p)$, and a morphism $\alpha$ gives $\alpha\circ\lab$ with the components
  of $\alpha$. The \emph{inverse image} is $\fin\mathscr{Q}=(\mathscr{Q}\circ\lab)^{++}$,
  and $\fin\alpha=(\alpha\circ\lab)^{++}$. For a presheaf
  $\mathscr{Q}$ on step words $\mathscr{Q}\circ\fgt\circ\lab$ has at $p$ the set of $\mathscr{Q}$ at the step word of
  $p$, and $\alpha\circ\fgt\circ\lab$ has the components of $\alpha$.
\item A \emph{direct image} of a presheaf $\mathscr{F}$ on $X$ is a presheaf $\mathscr{E}$ on label words with
  bijections $\mathrm{Hom}(\mathscr{Q}\circ\lab,\mathscr{F})\cong\mathrm{Hom}(\mathscr{Q},\mathscr{E})$ natural in
  presheaves $\mathscr{Q}$ on label words. A \emph{direct image along $\lab$} assigns to
  each presheaf on $X$ a direct image and to each morphism of presheaves a morphism of their direct images so that the
  bijections are natural in $\mathscr{F}$ too. With presheaves on step words and
  $\mathscr{Q}\circ\fgt\circ\lab$ in place of $\mathscr{Q}\circ\lab$ the same two sentences define a direct image of
  $\mathscr{F}$ over step words and a direct image along
  $\fgt\circ\lab$.
\item For a presheaf $\mathscr{F}$ on $X$ the presheaf $\fsh\mathscr{F}$ on label words has at a label word $w$ the
  families of elements of $\mathscr{F}$ at the prefixes $p$ with $w\lle\lab(p)$ that are compatible under extension;
  its restriction along $w\lle v$ keeps the members at the prefixes over $v$. For a morphism $\theta$
  the map $\fsh\theta$ applies $\theta$ memberwise. The presheaf
  $\fshS\mathscr{F}$ has at a step word $u$ the families of elements of $\mathscr{F}$ at the prefixes whose step word
  extends $u$ that are compatible under extension, and $\fshS\theta$ applies $\theta$
  memberwise.
\item The \emph{transpose} of a morphism $\varphi\colon\mathscr{Q}\circ\lab\to\mathscr{F}$ is the morphism
  $\varphi^\sharp\colon\mathscr{Q}\to\fsh\mathscr{F}$ sending $y$ at $w$ to the family of the
  $\varphi_p(y|_{\lab(p)})$ at the prefixes $p$ with $w\lle\lab(p)$. The transpose
  $\varphi^\sharp\colon\mathscr{Q}\to\fshS\mathscr{F}$ of $\varphi\colon\mathscr{Q}\circ\fgt\circ\lab\to\mathscr{F}$
  sends $y$ at $u$ to the family of the $\varphi_p(y|_{\fgt(\lab(p))})$ at the prefixes $p$ whose step word extends
  $u$.
\end{enumerate}
\end{definition}

The inverse image of Definition~\ref{def:label-images}(3) sheafifies the precomposition with $\lab$, because a
presheaf on label words whose precomposition is a sheaf has exactly one section at the label word of each dead
prefix, while the direct image of $\Omega$ of a machine carries the two distinct
sections $\top$ and $\bot$ at every label word extended by the label word of one of its prefixes that
is not dead; on the recogniser $Z$ of $\LMp{\mathsf{T},\FT}$ of
Theorem~\ref{thm:muller-answer}(i) the two meet, so that $\fsh\Omega_Z\circ\lab$ has two sections at a dead prefix where
a sheaf has one.

\begin{definition}[Equivalence over the label topos]\label{def:labels-frame}
Let $X$ and $Y$ be partial HDAs over $\Lambda$. They are \emph{equivalent over the label topos}, written $X\eqB Y$, when some
equivalence $(h,h')$ of $\Op(X)$ with $\Op(Y)$ has $h(\lst{X}(U))\feq{Y}\lst{Y}(U)$ for every open $U$ of label
words. An equivalence $(h,h')$ \emph{witnesses} $X\eqB Y$ when
$h(\lst{X}(p))\feq{Y}\lst{Y}(p)$ for every prefix $p$ of $X$ and of $Y$.
\end{definition}

Since $\ell_X$ acts by $\lst{X}$, $X\eqB Y$ is an isomorphism over $\BL$ of the locales of $X$ and $Y$.

\begin{definition}[Equivalences of sheaves and of theories]\label{def:streams}\label{def:setup-desc}
\begin{enumerate}
\item An \emph{equivalence of data sheaves} of partial HDAs $X$ and $Y$ over $\Lambda$ is a pair of functors $\Phi,\Psi$ between presheaves on $X$ and on $Y$
  that preserve data sheaves, with natural maps $\mathscr{F}\to\Psi\Phi\mathscr{F}$ and
  $\mathscr{E}\to\Phi\Psi\mathscr{E}$ invertible at data sheaves. Its \emph{frame map}
  sends an open $A$ to the open of the prefixes at which $\Phi(\sbt{A})$ has an element.
  It is \emph{labelled} when its frame map $h$ has $h(\lst{X}(p))\feq{Y}\lst{Y}(p)$ for every prefix $p$ of $X$ and
  of $Y$. $X\eqLs Y$ when some equivalence of data sheaves
  has a frame map $h$ with $h(\lst{X}(U))\feq{Y}\lst{Y}(U)$ for every open $U$ of label words.
\item A data sheaf is \emph{describable} when each predicate on one of its section sets with exactly one solution
  is given its solution. A \emph{germ} of $\mathscr{F}$ at
  $r$ is an element at a prefix whose extensions cover $r$, up to agreement on a cover of $r$, and the \emph{push} of
  $\mathscr{F}$ along $(h,h')$ has at $q$ the matching families of germs on the prefixes covered by
  $h'(\upc{q})$. The \emph{label subterminal}
  of a prefix $R$ on a machine $X$ is inhabited exactly at the prefixes covered by $\lst{X}(R)$, with one element
  there.
\item Let $X\eqM Y$ when interpretations $I$ of $\FX{X}$ in $\FX{Y}$ and $I'$ back carry axioms to derivable sequents
  and make derivable
  \[
    I'I(\fat{p})\dashv\vdash\fat{p},\quad II'(\fat{q})\dashv\vdash\fat{q},\quad
    \fat{p}\Der\textstyle\bigvee_{\lab(p)\lle\lab(q)}I'(\fat{q}),\quad
    \fat{q}\Der\textstyle\bigvee_{\lab(q)\lle\lab(p)}I(\fat{p})
  \]
  for all prefixes $p$ of $X$ and $q$ of $Y$. We call $\eqM$ \emph{Morita equivalence over
  the label topos}.
\item Let $X\eqD Y$ when along some $(h,h')$ that witnesses $X\eqB Y$ both pushes preserve describable sheaves with
  invertible units, the push of $\sbt{A}$ has support $\feq{Y}$-equivalent to $h(A)$ for every open $A$, and the
  push of the label subterminal of every prefix of $X$ or of $Y$ on $X$ is isomorphic to its label subterminal on
  $Y$.
\end{enumerate}
\end{definition}

\subsection{The topos of a machine}\label{ssec:topos-a}

Acceptance cuts the topos of $X$ out of the topos of presheaves on its prefixes as a quotient theory, and the run
theory presents the same topos with the cells of $X$ named; together with the label map of item~(c) these
identifications make $\eqB$ a comparison of toposes over $\BL$.

\begin{theorem}[The topos of a machine]\label{thm:topos-a}\label{def:labels-morita}\label{thm:setup-quotient}\label{prop:setup-fbquot}\label{thm:cor638}\label{prop:comorphism}\label{thm:ladder}
Let $X$ and $Y$ be partial HDAs over any $\Lambda$.
\begin{enumerate}
\item[(a)] $\BB(p,n)$ is the sieve on $p$ of the acceptance axiom of index $n+1$, and a
  Grothendieck topology contains $\JX{X}$ exactly when it contains the sieves of all acceptance and totality
  axioms. $\FX{X}$ and $\FX{X}^{\mathrm{acc}}$ derive each other's
  axioms, and for every prefix $p$ the valuation making $\fat{q}$ true exactly when
  $q\ext p$ models the axioms of $\FX{X}$ other than its covering axioms and not $\FX{X}^{\mathrm{acc}}$.
\item[(b)] $\IX{X}$ carries each axiom of $\FX{X}$ to a sequent derivable in
  $\RX{X}$; a sequent over $\FX{X}$ whose image is derivable in $\RX{X}$ is
  derivable in $\FX{X}$; every sentence over $\RX{X}$ is equivalent in $\RX{X}$ to the
  image of a sentence over $\FX{X}$; and for every conclist $U$ the first two
  sequents of the diagram of $X$ at each cell of shape $U$ and its third sequent at $U$ are derivable in
  $\RX{X}$.
\item[(c)] \begin{enumerate}[label=(\roman*),ref=(\roman*)]
  \item For every prefix $p$ of $X$ and every $\JLW$-covering sieve $S$ on $\lab(p)$ the extensions $r$ of $p$
  with $\lab(r)\in S$ form a $\JX{X}$-covering sieve on $p$.
  \item The map $\lst{X}$ is
  monotone for covering, keeps binary intersections
  and the top open on the nose, and sends the union of a family to a least upper bound
  of its images.
  \item For presheaves $\mathscr{Q}$ on label words and $\mathscr{F}$ on $X$
  the map $\varphi\mapsto\varphi^\sharp$ of Definition~\ref{def:label-images} is a bijection from the morphisms
  $\mathscr{Q}\circ\lab\to\mathscr{F}$ to the morphisms $\mathscr{Q}\to\fsh\mathscr{F}$.
  It has $(\varphi\circ(\alpha\circ\lab))^\sharp=\varphi^\sharp\circ\alpha$ for every morphism
  $\alpha\colon\mathscr{Q}'\to\mathscr{Q}$ and
  $(\theta\circ\varphi)^\sharp=\fsh\theta\circ\varphi^\sharp$ for every morphism
  $\theta\colon\mathscr{F}\to\mathscr{F}'$. For presheaves $\mathscr{Q}$ on step words
  and $\mathscr{F}$ on $X$ the map $\varphi\mapsto\varphi^\sharp$ is a bijection from the morphisms
  $\mathscr{Q}\circ\fgt\circ\lab\to\mathscr{F}$ to the morphisms
  $\mathscr{Q}\to\fshS\mathscr{F}$, with
  $(\varphi\circ(\alpha\circ\fgt\circ\lab))^\sharp=\varphi^\sharp\circ\alpha$ and
  $(\theta\circ\varphi)^\sharp=\fshS\theta\circ\varphi^\sharp$.
  \item By (iii) $\fsh$ with
  $\varphi\mapsto\varphi^\sharp$ is a direct image along $\lab$, and $\fshS$ one along
  $\fgt\circ\lab$. Two direct images of one presheaf on $X$ are
  isomorphic, and so are two direct images of it over step
  words.
  \item A structure of data
  sheaf on $\mathscr{F}$ gives $\fsh\mathscr{F}$ a structure of data sheaf on label words.
  \item For every data sheaf $\mathscr{F}$ on $X$ and every presheaf $\mathscr{Q}$ on label words the map
  $\chi\mapsto(\chi\circ\eta)^\sharp$ with $\eta\colon\mathscr{Q}\circ\lab\to\fin\mathscr{Q}$ the unit is a bijection
  from the morphisms $\fin\mathscr{Q}\to\mathscr{F}$ to the morphisms
  $\mathscr{Q}\to\fsh\mathscr{F}$. It sends $\chi\circ\fin\alpha$ to
  $(\chi\circ\eta)^\sharp\circ\alpha$ and $\theta\circ\chi$ to
  $\fsh\theta\circ(\chi\circ\eta)^\sharp$.
  \item $\fin$ keeps the terminal
  presheaf, binary products and equalisers up to isomorphism of presheaves.
  \end{enumerate}
\item[(d)] $X\eqB Y$ exactly when some equivalence of $\Op(X)$ with $\Op(Y)$ witnesses
  $X\eqB Y$. $X\eqM Y$ (Definition~\ref{def:streams}(3)) exactly
  when some equivalence of $\Op(X)$ with $\Op(Y)$ witnesses $X\eqB Y$. Along
  every equivalence $(h,h')$ of $\Op(X)$ with $\Op(Y)$ the push of a describable sheaf is describable with invertible
  unit, and the push of $\sbt{A}$ has support $\feq{Y}$-equivalent to $h(A)$ for every
  open $A$. If $h(\lst{X}(R))\feq{Y}\lst{Y}(R)$ for a prefix $R$ of a partial HDA over
  $\Lambda$ then the push of the label subterminal of $R$ on $X$ is isomorphic to its label subterminal on
  $Y$. So every witness of $X\eqB Y$ meets the conditions of $X\eqD Y$
  (Definition~\ref{def:streams}(4)), and some equivalence of $\Op(X)$ with $\Op(Y)$ witnesses $X\eqB Y$ exactly when
  $X\eqD Y$. $X\eqLs Y$ exactly when some labelled equivalence of data
  sheaves exists, and such an equivalence gives
  an equivalence of $\Op(X)$ with $\Op(Y)$ that witnesses $X\eqB Y$.
\end{enumerate}
\end{theorem}

\begin{proof}
(a) The acceptance axiom of index $n+1$ holds in $M_r$ when $r$ accepts after time $n$, so the axiom sieves
contain the basic covers and inductions on covers give the rest;
no $q\ext p$ lies in $\BB(p,\dimn{p})$.
(b) Prefixes holding together are comparable. The second and
third claims of (b) translate $\mathsf{at}_n(x)\wedge\nm{c}(x)$ into
$\bigvee_{p_n=c}\fat{p}$,
and the sequents of its last claim are axioms of $\RX{X}$ by clause (1) of
Definition~\ref{def:machine-theories}.
(c) The extensions of $p$ longer than $n$ contain $\BB(p,n)$, so basic covers
lift; preimages are computed
pointwise, and matching families of $\fsh\mathscr{F}$ glue on
lifted sieves. Restriction inverts
transposition;
$\mathscr{F}\to\mathscr{F}^{++}$ is universal against data sheaves (Proposition~\ref{thm:setup-built}(4)), and
precomposition and the plus construction keep finite
limits.
(d) Given $I$ let $h(A)$ be the open where some $I(\fat{p})$ with $p\in A$ is forced, and conversely
$I(\fat{p})=\bigvee_{q\in h(\upc{p})}\fat{q}$.
Descriptions pass to germs and to the push, where they invert the
unit;
label opens are inverse images with union $\lst{X}(U)$ over
$U$.
\end{proof}

By (a), Caramello's duality of quotients and subtoposes
\cite{Caramello09,CL25a}
cuts the topos of $X$ out of the presheaf topos by the acceptance and totality axioms,
properly when $X$ has a prefix.

In the form used here, Corollary~6.38 of \cite{Caramello19} states that an interpretation $I$
of a geometric theory $\Tt$ in a geometric theory $\Tt'$ induces a Morita equivalence exactly when three
conditions hold. In the first a sequent over $\Tt$ whose image is provable in $\Tt'$ is provable in $\Tt$. In
the second every formula over the signature of $\Tt'$ in a context $I(\vec x)$ is provably equivalent to
$I(\phi)$ for a formula $\phi$ over the signature of $\Tt$ in the context $\vec x$; the source writes
``$\Tt$-provably'' for a formula that only $\Tt'$ can prove, and its own example proves this equivalence in
$\Tt'$. In the third the object $\{y^B.\top\}$ is covered in the
syntactic topology of $\Tt'$ by objects in the image of $I$ for every sort $B$ of $\Tt'$. For $\IX{X}$
the first clause of (b) makes it an interpretation and its second clause is the first condition; since $\FX{X}$
has no sort its only context is empty, so that the second condition concerns sentences and is the third clause
of (b). For each cell $c$ of shape $U$ the formula $\nm{c}(y)$ is an arrow from $\{.\top\}=\IX{X}(\{.\top\})$ to
$\{y^U.\top\}$ in the syntactic category of $\RX{X}$, since the first two sequents of the diagram make it functional.
By its third sequent these arrows cover $\{y^U.\top\}$, so that the last clause of (b) is
the third condition and the topos of $X$ classifies its run theory in every dimension. The interpretation $I$ of
$X\eqM Y$ meets the first two conditions, and the third is empty since $\FX{Y}$ has no
sort; so $X\eqB Y$ gives a Morita equivalence of $\FX{X}$ and $\FX{Y}$. Under
$\UC$ the label maps are geometric morphisms (Theorem~\ref{thm:prices-uc}), and the two toposes and the label topos are
localic. Since a geometric morphism between localic toposes is determined up to isomorphism by its frame
map \cite[Section~C1.4]{Johnstone02} and the frame maps of $I$ and $I'$ fix the label opens by the proof of (d), this Morita
equivalence commutes with $\ell_X$ and $\ell_Y$ up to isomorphism. Under $\UC$ the
pair $(\fin,\fsh)$ of (c) is a geometric morphism (Theorem~\ref{thm:prices-uc}), the one Caramello induces from that
comorphism \cite{Caramello19}.

\begin{definition}[Image]\label{def:image}\label{def:image-equiv}\label{def:image-st}
Let $X$ and $Y$ be partial HDAs over $\Lambda$.
\begin{enumerate}
\item A set $K$ of sieves on label words \emph{carries} $\fsh$ when $\fsh\mathscr{F}$ has a structure of data $K$-sheaf
  for every data sheaf $\mathscr{F}$ on $X$. The \emph{image topology} $\Jim{X}$ is the
  largest set of sieves on label words that carries $\fsh$. $X$ and $Y$ have \emph{one image},
  $\Ima X=\Ima Y$, when for every label word $w$ and every open $S$ of label words the prefixes
  of $X$ with label word in $S$ cover in $\JX{X}$ every prefix of $X$ whose label word extends $w$ exactly when this
  holds in $Y$.
\item A set $K$ of sieves on step words \emph{carries} $\fshS$ when $\fshS\mathscr{F}$ has a structure of data
  $K$-sheaf for every data sheaf $\mathscr{F}$ on $X$. The \emph{image topology over the
  step words} $\JimS{X}$ is the largest set of sieves on step words that carries $\fshS$. $X$
  and $Y$ have \emph{one image over the step words}, $\ImaST X=\ImaST Y$, when for every step word $u$ and every set
  $S$ of step words closed under extension the prefixes of $X$ whose step word lies in $S$ cover in $\JX{X}$ every
  prefix of $X$ whose step word extends $u$ exactly when this holds in $Y$.
\end{enumerate}
\end{definition}

The image of $\ell_X$ is the smallest subtopos of $\BL$ through which its direct image factors, and computing its
topology from covers of prefixes makes it an invariant of $\eqB$ written on label words alone
(Theorem~\ref{tp:img}).

\begin{theorem}[The image topology]\label{thm:image-up}
Let $X$ and $Y$ be partial HDAs over any alphabet $\Lambda$.
\begin{enumerate}
\item A set $K$ of sieves on label words carries $\fsh$ exactly when every sieve $S$ of $K$ on a word $w$ has the
  prefixes $r$ with $\lab(r)\in S$ covering in $\JX{X}$ every prefix $p$ of $X$ with
  $w\lle\lab(p)$. The relation in which $S$ covers $w$ under this condition is the image
  topology $\Jim{X}$, and a relation is a largest set of sieves carrying $\fsh$ exactly when it
  lets the same sieves cover. $\Jim{X}$ is a Grothendieck topology on label
  words, and every sieve covering in $\JLW$ covers in
  $\Jim{X}$. A set $K$ of sieves on step words carries $\fshS$ exactly when every sieve
  $S$ of $K$ on a step word $u$ has the prefixes whose step word lies in $S$ covering in $\JX{X}$ every prefix of $X$
  whose step word extends $u$. The relation in which $S$ covers $u$ under this condition
  is the image topology over the step words $\JimS{X}$, and a relation is a largest set
  of sieves carrying $\fshS$ exactly when it lets the same sieves cover. $\JimS{X}$ is
  a Grothendieck topology on step words, and every sieve covering in $\JSW$ covers
  in $\JimS{X}$. A set $K$ of sieves on label words has $\fsh\mathscr{F}$ $K$-separated
  for every sheaf $\mathscr{F}$ on $X$ exactly when every sieve of $K$ on a word $w$ covers $w$ in
  $\Jim{X}$.
\item For opens $U,V$ of label words every word of $U$ is covered by $V$ in $\Jim{X}$ exactly when
  $\lst{X}(U)\fle{X}\lst{X}(V)$. $\Ima X=\Ima Y$ holds exactly when some equivalence
  between the preorder on sets of label words given by $\lst{X}(U)\fle{X}\lst{X}(V)$ and the one given by
  $\lst{Y}(U)\fle{Y}\lst{Y}(V)$ sends each singleton to a set equivalent to it.
  Such an equivalence exists exactly when $\lst{X}(U)\fle{X}\lst{X}(V)$ and $\lst{Y}(U)\fle{Y}\lst{Y}(V)$ are
  equivalent for all sets $U,V$ of label words.
\end{enumerate}
\end{theorem}

\begin{proof}
(1) The covers of $\Jim{X}$ lift by definition.
Amalgamations on lifted covers glue the families of $\fsh\mathscr{F}$; conversely $\top$
and the characteristic map of the closed sieve generated by the inverse image of $S\in K$ agree on $S$, hence at its
word in $\fsh\Omega_X$. A sheaf is
separated on each lifted sieve, and $\Omega_X$ is a
sheaf. Covers of
$\Jim{X}$ inherit stability and locality from $\JX{X}$, and covers of $\JLW$ lift by
Theorem~\ref{thm:topos-a}(c); these
arguments carry over to the step words.
(2) A sieve $S$ covers $w$ in $\Jim{X}$ exactly when
$\lst{X}(w)\fle{X}\lst{X}(S)$.
\end{proof}

By Theorem~\ref{thm:image-up}(2) a sieve $S$ covers $w$ in $\Jim{X}$ exactly when every prefix whose label word
extends $w$ is covered by the prefixes whose label words lie in $S$, so that $\Jim{X}$ is the
topology that Caramello coinduces from $\JX{X}$ along $\lab$, whose sheaves form the image of the geometric morphism
that $\lab$ induces into presheaves on label words \cite{Caramello19}; every
$\JLW$-covering sieve covers in $\Jim{X}$, so this image lies in $\BL$, and under unique choice, where $\ell_X$ is a geometric morphism
(Theorem~\ref{thm:prices-uc}), it is the image of $\ell_X$, the smallest subtopos of $\BL$ through which the direct image
of $\ell_X$ factors.

\subsection{Transport}\label{ssec:transport}

For an equivalence $(h,h')$ of $\Op(X)$ with $\Op(Y)$ and a point $\xi$ of $X$ let
$h_\bullet\xi=\{\,q\in\Pref(Y) : h'(\upc{q})\text{ contains a member of }\xi\,\}$, and $h'_\bullet$ with $h$ in place of
$h'$.

\begin{definition}[Carrying labels]\label{def:carries}
Let $X$ and $Y$ be partial HDAs over $\Lambda$. A point $\eta$ of $Y$ \emph{carries the labels} of a point $\xi$ of $X$
when each $p\in\xi$ has some $q\in\eta$ with $\lab(p)\lle\lab(q)$. A \emph{homeomorphism over the labels} is a pair of mutually inverse
\emph{continuous} maps $\varphi$ and $\psi$ between the points of $X$ and of $Y$ with $\varphi(\xi)$ carrying the
labels of $\xi$ and $\psi(\eta)$ those of $\eta$, where $\varphi$ is continuous when for $q\in\varphi(\xi)$ some
$p\in\xi$ has $q\in\varphi(\xi')$ for every point $\xi'$ containing $p$ and $\psi$ is continuous with $X$ and $Y$
exchanged.
\end{definition}

Transport makes derived sets, ranks, counts of points by rank and compactness of the frame invariant under every
equivalence of $\Op(X)$ with $\Op(Y)$ over any alphabet and with no finiteness, with labels carried along when the
equivalence witnesses $\eqB$, so that one difference in any of them proves two machines inequivalent.

\begin{theorem}[Transport]\label{thm:transport}
Let $X$ and $Y$ be partial HDAs over any alphabet $\Lambda$, with no finiteness, and $(h,h')$ an equivalence of
$\Op(X)$ with $\Op(Y)$.
\begin{enumerate}[label=(\arabic*),ref=\thetheorem(\arabic*)]
\item\label{tp:pts} $h_\bullet$ and $h'_\bullet$ are mutually inverse continuous maps between the points of $X$ and
  of $Y$.
\item\label{tp:der} For every point $\xi$ of $X$ and every $n$ we have $\xi\in\dset{n}{X}$ exactly when $h_\bullet\xi\in\dset{n}{Y}$, and $\xi$ has
  rank $n$ exactly when $h_\bullet\xi$ has.
\item\label{tp:derop} $h$ and $h'$ send $n$-th derived opens to $n$-th derived opens.
\item\label{tp:count} For all $n$ and $k$ the machine $X$ has $k$ pairwise different points of rank $n$ exactly when
  $Y$ has, and $k$ pairwise different points exactly when $Y$ has.
\item\label{tp:lat} $h$ and $h'$ send a least upper bound of a family to a least upper bound of its
  images. $h(A\cap B)\feq{Y}h(A)\cap h(B)$,
  $h(A\cup B)\feq{Y}h(A)\cup h(B)$ and
  $h(\Pref(X))\feq{Y}\Pref(Y)$, while $h(\emptyset)\fle{Y}\emptyset$ and
  $h'(\emptyset)\fle{X}\emptyset$.
\item\label{tp:cpt} $h$ and $h'$ send complemented opens to complemented opens and compact opens to compact
  opens, and $\Op(X)$ is compact exactly when $\Op(Y)$
  is.
\item\label{tp:lab} If $(h,h')$ witnesses $X\eqB Y$ then $(h',h)$ witnesses $Y\eqB X$, and
  $h(\lst{X}(p))\feq{Y}\lst{Y}(p)$ and $h'(\lst{Y}(p))\feq{X}\lst{X}(p)$ for every prefix $p$ of every partial HDA
  over $\Lambda$. If $(h,h')$ witnesses $X\eqB Y$ then for every set $W$ of prefixes
  of one partial HDA over $\Lambda$ the open $h(\bigcup_{p\in W}\lst{X}(p))$ is $\feq{Y}$-equivalent to
  $\bigcup_{p\in W}\lst{Y}(p)$.
\item\label{tp:homeo} If $(h,h')$ witnesses $X\eqB Y$ then for every point $\xi$ of $X$ the point $h_\bullet\xi$
  carries the labels of $\xi$ and $\xi$ carries the labels of
  $h_\bullet\xi$, and $h'_\bullet\eta$ carries the labels of $\eta$ for every point
  $\eta$ of $Y$; so $(h_\bullet,h'_\bullet)$ is a homeomorphism over the
  labels. For such $(h,h')$ a label word is extended by the label word of
  a member of $\xi$ exactly when it is extended by that of a member of $h_\bullet\xi$.
\item\label{tp:img} If $(h,h')$ witnesses $X\eqB Y$ then $\lst{X}(U)\fle{X}\lst{X}(V)$ and $\lst{Y}(U)\fle{Y}\lst{Y}(V)$
  are equivalent for all sets $U,V$ of label words, so that
  $\Ima X=\Ima Y$ by Theorem~\ref{thm:image-up}(2).
\item\label{tp:sh} The frame maps of an equivalence of data sheaves and of its reverse form an equivalence of $\Op(X)$
  with $\Op(Y)$, to which (1) to (6) apply.
\end{enumerate}
\end{theorem}

\begin{proof}
(1) By Theorem~\ref{thm:pts-x}(2) a point is a completely prime filter of $\Op(X)$; its preimage under $h'$ is one,
and a member of $\xi$ in $h'(\upc{q})$ witnesses
continuity.
(2) and (3) Induction on $n$, since $h$ keeps isolating
opens
with least upper bounds and the opens in which exactly one point lies. (4) follows from (1) and
(2). (5) and (6) Meets, least
upper bounds, complements and finite subcovers are defined by $\fle{X}$ and so kept by $h$ and $h'$ up to
$\feq{}$.
(7) Label opens of prefixes and their unions are inverse images of opens of label words, which a
witness fixes, and an open fixed by $h$ is fixed by
$h'$.
(8) A completely prime filter containing the label open of a member of $\xi$ contains $\upc{q}$ for a prefix $q$
carrying its labels, and by (7) the filter of $h_\bullet\xi$ contains the label opens of the members of
$\xi$. (9) Fixing each label open fixes
the order of their unions, which is $\Ima X$ by Theorem~\ref{thm:image-up}(2).
(10) The frame maps are monotone.
\end{proof}

\subsection{Safety, liveness and trimming}\label{ssec:safety}

\begin{definition}[Closed and dense parts]\label{def:setup-parts}
Let $U$ be a set of prefixes closed under extension. The \emph{closed topology} of $U$ is the least Grothendieck
topology in which the empty sieve covers each member of
$U$. A relation ``$S$ covers $p$'' is a \emph{closed
topology} when for some such $U$ it lets a sieve $S$ on $p$ cover $p$ exactly when $p\in S\cup
U$. It lies \emph{in} $\JX{X}$ when every sieve it lets cover $p$ covers $p$ in
$\JX{X}$. The \emph{safety part} $C_X$ of $X$ is the greatest closed topology in
$\JX{X}$. A topology is \emph{dense} when the empty sieve covers nothing. $D_X$ is the
least Grothendieck topology covering the sieves $\mathrm{D}(p,n)=\BB(p,n)\cup\{\,r : p\ext r,\ r\text{ dead}\,\}$,
and $J\vee J'$ the least above both.
\end{definition}

\begin{theorem}[Safety and liveness]\label{thm:setup-safety}
Let $X$ be a partial HDA over any alphabet.
\begin{enumerate}
\item Let $U$ be a set of prefixes closed under extension. A relation is the closed topology of $U$ exactly when
  the sieves $S$ on $p$ it lets cover $p$ are those with $p\in S\cup U$. This topology lies in $\JX{X}$
  exactly when $U$ has only dead members. The dead prefixes are closed under
  extension, the topology of which lies in $\JX{X}$ and
  contains each such topology that lies in $\JX{X}$. So $C_X$ is the relation in which
  $S$ covers $p$ when $p\in S$ or $p$ is dead, and a relation
  is a greatest closed topology in $\JX{X}$ exactly when it lets the same sieves cover.
\item Let $B$ be a family of sieves. Some relation is the least Grothendieck topology covering every member of $B$,
  and the relations that are such a least topology are those covering the same sieves as
  it. $D_X$ exists. For any relations $J$
  and $J'$ the join $J\vee J'$ exists in the same sense. $\JX{X}$ is
  $C_X\vee D_X$, and $D_X$ is dense.
\item $p$ is dead exactly when $\emptyset^{+}$ has an element at $p$.
\item The points of $\JX{X}$ are those of both $C_X$ and $D_X$.
\item Over $\Lambda$ with decidable equality an $\omega$-track $\beta$ that avoids the cells $\ast_U$ and starts
  has $X_\beta\models\Tacc$ exactly when no initial prefix is dead and each $n$ has an accepting $\beta_m$ with $m>n$
  or a dead initial prefix.
\end{enumerate}
\end{theorem}

\begin{proof}
(1) The relation covering $p$ by $S$ when $p\in S\cup U$ is the closed topology of $U$,
unique up to the sieves it covers. A closed topology contained in $\JX{X}$ covers with the empty
sieve only dead prefixes, so
$C_X$ is the closed topology of the dead prefixes.
(2) Closing $B$ or the covers of $J$ and $J'$ under pullback and local character gives the
topologies, unique up to the sieves they cover.
$\JX{X}$ covers the same sieves as the generated join of $C_X$ and $D_X$, since members of $\mathrm{D}(q,n)$ lie in $\BB(q,n)$ or are
dead; an empty
$D_X$-cover makes the member $q$ both dead and not dead. (3) and (4) unfold the
definitions, and Theorem~\ref{thm:pts-x}(4) gives
(5).
\end{proof}

We read $C_X$ as safety and $D_X$ as liveness, which makes (5) the decomposition of Alpern and Schneider \cite{AS85}.
In (5) the first conjunct says that the chain of $\beta$ is a point of $C_X$ and the second, with its disjunct on dead
initial prefixes, that it is a point of $D_X$; the
first conjunct makes that disjunct idle, and it stays so that each conjunct is the condition of one part alone. Since joins of
topologies are meets of subtoposes \cite{CL25b}\cite[Section~A4.5]{Johnstone02},
the topos of $X$ is a dense subtopos met with the closed complement of the open subtopos of the dead prefixes
\cite{CL25b},
which by (1) is its closure \cite{Caramello09}; neither part is
unique at the torus $\TB{\emptyset}$.

\begin{definition}[Live prefixes and trimming]\label{def:trim}
A prefix is \emph{live} when it lies on the chain of an $\omega$-track $\beta$ with
$X_\beta\models\Tacc$, and $X$ is \emph{trimmed} when every prefix is
live. Let $\Lambda$ have decidable equality and $X$ be finite. A cell is \emph{live} when it
reaches through proper steps an accepting cell on a cycle of at least one proper
step. The machine $\trim{X}$ has the cells of $X$ and keeps a face
$\delta_{A,B}(y)=x$ of $X$ exactly when $x$ and $y$ are both live or both not; its acceptance and
finiteness structure are those of $X$ and its start cells the live start cells of
$X$.
\end{definition}

Over $\Lambda$ with decidable equality live prefixes are not dead, and
conversely under $\EM$ and $\DC$.

\begin{lemma}\label{lem:no-point-dead}
Over an alphabet with decidable equality a prefix of a finite partial HDA contained in no point is dead.
\end{lemma}

\begin{proof}
Liveness of the last cell is decided; a live last cell puts the prefix on an accepted track, whose chain is a
point by Theorem~\ref{thm:pts-x}(4), and a dead one makes the prefix
dead.
\end{proof}

\begin{proposition}[Trimming]\label{prop:trim}
Let $\Lambda$ have decidable equality and $X$ be finite. Then $\trim{X}$ is finite and trimmed, $X\eqLs\trim{X}$, and
the live prefixes of $X$ and $\trim{X}$ correspond preserving extension and label words. If $X$ is rigid or has
closed acceptance with constant $\Delta$, so has $\trim{X}$.
\end{proposition}

\begin{proof}
Accepted tracks visit live cells only
and other prefixes are dead, so the tree correspondence carries
sheaves. $\trim{X}$ has the
cells of $X$ and fewer steps, and Theorem~\ref{thm:topos-a}(d) gives
$X\eqB\trim{X}$.
\end{proof}
\section{Logical strength}\label{sec:prices}

\begin{definition}[Enough points]\label{def:setup-bar}
A partial HDA $X$ has \emph{enough points}, written $\BarS{X}$, when every sieve on a prefix $p$ that meets each
point containing $p$ is $\JX{X}$-covering.
\end{definition}

\begin{definition}[Comparison lemma]\label{def:glue}
Let $\mathscr{F}$ be a presheaf on a partial HDA $X$. A \emph{section} of $\Ext{\mathscr{F}}$ over an open $U$ is a
family of $x_r\in\mathscr{F}(r)$, one at each prefix $r$ covered by $U$, with $x_r|_{r'}=x_{r'}$ for
$r\ext r'$. Sections restrict along $V\fle{X}U$ to the prefixes covered by $V$. A \emph{cover} of $U$ is a family of
opens $V_i\fle{X}U$ with $U\fle{X}\bigcup_iV_i$. The \emph{comparison lemma holds on $X$} when $\Ext{\mathscr{F}}$ is a
sheaf on $\Op(X)$ for these covers, with amalgamations given, for every data sheaf $\mathscr{F}$ on $X$.
\end{definition}

$\BarS{X}$ is spatiality of the frame of $X$ stated on sieves, and the comparison lemma on $X$ is that of \cite{Caramello19} and
\cite[Theorem~C2.2.3]{Johnstone02} applied to the basis $p\mapsto\upc{p}$ of
$\Op(X)$, whose equivalence is the right Kan extension $\mathscr{F}\mapsto\Ext{\mathscr{F}}$.

The logical strength of a construction on machines is the principle under which it holds; each item names it for
sheaves to carry their amalgamations as operations, which makes the label map geometric, for the sieves meeting every
point to cover, for points to be tracks and for the comparison lemma to hold, and each recovers its principle from the
construction on some class of machines.

\begin{theorem}[Logical strength]\label{thm:prices}
\begin{enumerate}[label=(\arabic*),ref=\thetheorem(\arabic*)]
\item\label{thm:prices-uc} \emph{Sheaves and data sheaves.} Let $X$ be a partial HDA over any alphabet. Under
  $\UC$ every sheaf on $X$ carries a structure of data sheaf, and so does
  $\mathscr{F}^{++}$ for every presheaf $\mathscr{F}$ on $X$. With no principle
  the canonical comparison maps of $\fin$ for the terminal presheaf, binary products and equalisers of presheaves on
  label words are isomorphisms, and so are those of
  $\mathscr{Q}\mapsto(\mathscr{Q}\circ\fgt\circ\lab)^{++}$ on presheaves on step
  words. Under $\UC$ the functor $\fin$ lands in sheaves and
  $(\fin,\fsh)$ is a geometric morphism from the topos of $X$ to $\BL$, with its adjunction natural in both
  variables and $\fin$ left exact by these comparison maps. Under $\UC$ the
  map of prefixes to their step words gives the same with $\fshS$ to $\BST$. If $X$ has a dead prefix
  and one function gives every sheaf on $X$ a structure of data sheaf then
  $\UC$ holds.
\item\label{thm:setup-bar} \emph{Covers and bars.} Let $X$ be a partial HDA over any alphabet. Under $\BarS{X}$ a
  sieve on $p$ covers $p$ exactly when it meets each point containing $p$. The
  generic model of $\FX{X}$ satisfies every sequent $\fat{p}\Der\bigvee_{r\in S}\fat{r}$, for $S$ a sieve on $p$,
  that holds in every model of $\FX{X}$ in sets exactly when $\BarS{X}$ holds.
  Over $\ab$ each of these two properties holds of every finite partial HDA exactly when $\BIb$
  holds.
\item\label{thm:pt-track} \emph{Points and tracks.} Under $\CUC$ every point of every partial HDA over every
  alphabet is the chain of an $\omega$-track. Under $\BC$ every point of a
  finite rigid partial HDA over an alphabet with decidable equality is the chain of an
  $\omega$-track, and $\BC$ holds exactly when this is so on all those
  with closed acceptance.
\item\label{thm:setup-price} \emph{The frame and the sheaves.} $\CD$ holds exactly when the comparison lemma holds
  on every finite rigid partial HDA over every alphabet with decidable equality, and exactly when it holds on those
  over $\ab$. $\DCS$ holds exactly when the comparison lemma holds on every
  partial HDA over every alphabet, and exactly when it holds on those over $\ab$. $\UC$ is equivalent to $\DCS\wedge\CS$, to $\CS$ with $\DCS$ for
  one-element index sets, and to $\CS$ with the comparison lemma on every partial HDA over every
  alphabet.
\end{enumerate}
\end{theorem}

\begin{proof}
(1) A matching family on a covering sieve of a sheaf has exactly one amalgamation, and $\UC$ chooses
it. Gluing in $\mathscr{F}^+$ asks for unique representatives over sets of prefixes that
are not enumerable, which $\UC$
supplies.
Precomposition with $\lab$ keeps products and equalisers. The comparison map of sheafification for products
composes two comparisons of plus constructions, each with an explicit inverse;
so does the comparison map for equalisers, and the only map into the terminal presheaf is an
isomorphism.
$\fin\mathscr{Q}$ is such a sheafification, and Theorem~\ref{thm:topos-a}(c) at the data sheaves that $\UC$ provides
gives the rest. Conversely let $R$ hold of exactly one element of a set $A$; the constant presheaf at $\{a\in A : R(a)\}$ is a
sheaf, whose amalgamation of the empty family at a dead prefix is that element.
(2) Covering sieves on $p$ meet every point through $p$, and $\BarS{X}$ is the
converse. A sequent $\fat{p}\Der\bigvee_{r\in S}\fat{r}$
holds in every model in sets exactly when $S$ meets every point through $p$, and in the generic model exactly when it
is derivable (Theorem~\ref{thm:generic}(3)), that is when $S$ covers
$p$;
so the second sentence restates $\BarS{X}$. Under $\BIb$ let a sieve $S$ on a
prefix $p$ of a finite $X$ meet every point through $p$. A prefix with a dead last cell is dead; otherwise the
extensions of $p$ through basic covers between live cells form a spread \cite{TvD88} whose branches generate points, and $\BIb$
turns the bar $S$ into a
cover.
Conversely, on an $a$-loop and an accepting $b$-loop at one starting vertex the points through the root are the
branches of blocks $a^kb$, so $\BarS{X}$ there gives $\BIb$.
(3) Members of a point agree where both are defined and exist at every
length, so at each time the long members carry exactly one
cell and one next cell with the step between them; $\CUC$ assembles these into a
track. On a finite rigid machine $\BC$ once per listed cell decides which cell
is at time $n$, and rigidity fixes the step.
For the reversal take $\wGt$ of Example~\ref{exa:dec-witnesses}, finite and rigid with closed acceptance of
constant $2$; for decided propositions
$(P_n)$ the prefixes that are at $e_1$ at time $2i+1$ exactly when $P_i$ form a
point, and a track
through it is at $e_1$ at time $2n+1$ exactly when $P_n$.
(4) The sections of a data sheaf at a prefix form a complete set, and the comparison lemma holds
on $X$ exactly when every data sheaf on $X$ is describable, which $\DCS$
gives. Conversely a family of complete sets is a data sheaf on the graph over $\ab$
with one starting and accepting vertex and one $a$-loop at it for each
index. On a finite
rigid machine the gluing problems form a sequence of complete sets, which $\CD$ solves.
Conversely let $(A_n)$ be a sequence of complete sets and $\beta_n$ the track of $\wGt$ at $v$ at even times, at $e_0$
at the odd times $2j+1$ with $j<n$ and at $e_1$ at the other odd times. The sheaf on
$\wGt$ whose sections at $q$ give an element of $A_n$ for each $n$ with $q$ on $\beta_n$ is a data sheaf, since each
$A_n$ is complete; only $\beta_n$ passes its prefix of length
$2n+2$, so a description of this sheaf there is a description operator on $A_n$, and the
comparison lemma on $\wGt$ gives $\CD$
back. $\UC$ gives
the operators and the values of constant families, and
conversely $\DCS$ and $\CS$ describe the pair $(i,a)$ in $\coprod_iA_i$, which gives
$\UC$.
The second sentence gives the form with the comparison lemma, and $\CD$ is $\DCS$ for $I=\Nn$.
\end{proof}

\begin{proposition}[The comparison lemma over the label topos]\label{prop:prices-glue}
Let $X,Y$ be partial HDAs over $\Lambda$.
(1) Under the comparison lemma on $X$ and $Y$, $X\eqB Y$ gives a labelled equivalence of data
sheaves.
(2) Under $\UC$ the comparison lemma holds on every partial HDA, $X\eqB Y$
exactly when some labelled equivalence of data sheaves exists, and $X\eqM Y$ exactly when
$X\eqLs Y$.
(3) If $\Pref(X)$ is a point of $(\Pref(X),\JX{X})$ then $\UC$ holds exactly when the
comparison lemma holds on $X$ and every constant presheaf on $X$ carries a structure of data
sheaf.
(4) An equivalence of data sheaves makes the comparison lemmas on $X$ and $Y$
equivalent.
\end{proposition}

\begin{proof}
(1) Push data sheaves along the frame equivalence through their extensions
$\Ext{\mathscr{F}}$. (2) Use
Theorems~\ref{thm:setup-price} and~\ref{thm:topos-a} with (1).
(3) Constant presheaves are sheaves, since covering sieves are inhabited and directed, and $\UC$
makes them data sheaves; conversely descriptions on the
constant presheaf at one prefix give
$\UC$. (4) A section at
$p$ is a morphism out of the subterminal of $\upc{p}$, and an equivalence of data sheaves carries these with
their descriptions.
\end{proof}

The hypothesis of Proposition~\ref{prop:prices-glue}(3) holds on an $a$-loop at a
starting accepting vertex and on a square from a starting vertex to itself with
undefined edges, either of them accepting.

\begin{corollary}[Languages over the label topos]\label{thm:mb-lang}
Let $X$ and $Y$ be partial HDAs over any alphabet with $X\eqB Y$. Then $\LangPt{X}=\LangPt{Y}$.
If $\Lambda$ has decidable equality then $\LangA{X}=\LangA{Y}$ under $\CUC$, and
under $\BC$ if $X$ and $Y$ are finite and rigid.
\end{corollary}

\begin{proof}
By Theorems~\ref{tp:pts} and~\ref{tp:homeo} the bijection of points carries labels both ways, and a point carrying
the labels of a point reads its sequence. Chains of
accepted tracks are points, and Theorem~\ref{thm:pt-track} turns points into accepted tracks; so
$\LangA{X}=\LangPt{X}$.
\end{proof}

Under $\EIO$ accepted and PAF-accepted tracks of finite
machines agree (Convention~\ref{conv:buchi}), so for finite $X$ and $Y$ the corollary holds for PAF-accepted languages.

\section{Acceptance}\label{sec:muller}

\subsection{Languages and Muller acceptance}\label{ssec:muller-defs}

An \emph{$\omega$-ipomset} over $\Lambda$ has the events $\Nn$ labelled in $\Lambda$ with decidable relations $<$ of
precedence and $\dashrightarrow$ of event order, and decidable sets of source events and of \emph{marked} events
$\Xi$; its isomorphisms are the label-preserving bijections of $\Nn$ that keep sources and
marks, which preserve precedence and the event order of incomparable events. For
an $\omega$-track $\beta$ with steps $s_0,s_1,\dots$ and a decidable $\Xi\subseteq\Nn$ the \emph{event
$\omega$-ipomset} $\evo{\Xi}{\beta}=\ip(s_0)*\ip(s_1)*\cdots$ glues the ipomsets of the steps and marks $\Xi$
\cite{PAF25}.

\begin{definition}[$\omega$-languages]\label{def:muller-lang}
Let $\Lambda$ have decidable equality and $X$ be a partial HDA over $\Lambda$. A \emph{marking} $\mu$ assigns to
each $\omega$-track $\beta$ of $X$ a decidable $\mu(\beta)\subseteq\Nn$; since $\beta$ has unboundedly many
events, $\evo{\mu(\beta)}{\beta}$ is defined. For a property $\mathcal{A}$ of $\omega$-tracks of $X$ the set
$L_\mu(X,\mathcal{A})$ holds every $Q$ with $\evo{\mu(\beta)}{\beta}\cong Q$ for some $\omega$-track $\beta$ with
$\mathcal{A}$. For PAF-acceptance $\mathcal{P}$ (Convention~\ref{conv:buchi}) the set
$L_\mu(X,\mathcal{P})$ is the \emph{B\"uchi language} of $X$ at $\mu$.
\end{definition}

The \emph{$\omega$-language} of $\mathcal{A}$ is the union $\Lang{X}{\mathcal{A}}=\bigcup_{\mu}L_\mu(X,\mathcal{A})$, and the \emph{B\"uchi language} $\LB{X}$ is $\Lang{X}{\mathcal{P}}$. Passemard, Amrane and Fahrenberg mark exactly the $<$-maximal events
\cite{PAF25}, and their language $\Lmax{X,\mathcal{A}}$ holds the
members of $\Lang{X}{\mathcal{A}}$ so marked; our language equalities give
theirs. Steps are proper, and no language is closed under
subsumption.

\begin{definition}[Muller acceptance]\label{def:muller}
A \emph{Muller partial HDA} $(X,F_X)$ is a partial HDA with a finite list $F_X=(F_1,\dots,F_k)$ of decidable sets
of cells. Let $\Inf(\beta)=\{c : \forall n\,\exists m>n\ \beta_m=c\}$. An $\omega$-track
$\beta$ from a start cell avoiding the cells $\ast_U$ is \emph{Muller-accepted} when
$\exists i\ \Inf(\beta)=F_i$, and \emph{positively Muller-accepted} when it eventually stays in some $F_i\subseteq\Inf(\beta)$.
\end{definition}

$\LM{X,F_X}$ and $\LMp{X,F_X}$ are the $\omega$-languages of the two
properties; accepting cells play no part, and
$\Inf(\beta)=F_i$ is Muller's condition in the form $\Inf(\alpha)\in F_X$ of
\cite{PAF25}.

The torus $\mathsf{T}$ has Muller form $(\mathsf{T},\FT)$ with $\FT=(\{q_0,q_{ab}\})$, and
$\mathsf{T}^{A}$ its B\"uchi form.

\subsection{Event-geometric languages}\label{ssec:muller-events}

The \emph{event signature} over $\Lambda$ has one sort and relations $<,\dashrightarrow,\lambda_a\,(a\in\Lambda),\mathsf{src},\Xi$
of $\omega$-ipomsets on $\Nn$, and an $\omega$-ipomset is
\emph{valid} when it is an ipomset with finite pasts and finite $<$-antichains that marks exactly its
$<$-maximal events \cite{PAF25}.

\begin{definition}[Event-geometric languages]\label{def:muller-eg}
A set $L$ of $\omega$-ipomsets over $\Lambda$ is \emph{event-geometric} when a geometric theory $\Tt$ in the
event signature has $Q\models\Tt\iff Q\in L$ for all valid $Q$.
\end{definition}

\begin{definition}[Positive limits]\label{def:muller-poslim}
A \emph{piece} $f\colon Q|_D\to R$ on a finite set $D$ of events is a map $D\to\Nn$ preserving every relation of the
event signature. A set $L$ is \emph{closed under positive limits} when every valid $Q$ satisfies
\[
  \bigl(\forall n\;\exists R\in L\;\exists b\colon Q|_{[n]}\to R\;\;\forall M\;\exists w\colon R|_{[M]\cup b[n]}\to Q\;\;w\circ b=\mathrm{id}\bigr)\Rightarrow Q\in L
\]
with $R$ valid and $[n]=\{0,\dots,n-1\}$.
\end{definition}

Closure under positive limits is the test by which a set of $\omega$-ipomsets is shown not to be event-geometric,
and under $\EM$ it is the only obstruction to being so.

\begin{theorem}[Event-geometric criterion]\label{thm:muller-eg}\label{prop:muller-events}
Every event-geometric set of $\omega$-ipomsets over any $\Lambda$ is closed under positive limits, with no
principle. Under $\EM$ a set of $\omega$-ipomsets over any $\Lambda$ is
event-geometric exactly when it is closed under positive limits.
\end{theorem}

\begin{proof}
Truth of a geometric formula in a structure on $\Nn$ is supported on finitely many
events, so a left side true in $Q$ on $[n]$ is carried by $b$ into $R$ and the right side
returns by $w$.
Conversely under $\EM$ the sequents between finite positive diagrams true in $L$ cut out $L$; a failure of the
hypothesis at $n$ gives one false in $Q$, from the diagram of $Q$ on $[n]$ to the configurations with no piece back
into $Q$.
\end{proof}

Over $\ab$ the unmarked $\Theta_N=(ab)^N(a\parallel b)^\omega$ has events $2k$ labelled $a$ and $2k+1$ labelled $b$ with
$2k\dashrightarrow 2k+1$; $x<y$ holds when $x<y$ in $\Nn$ unless $\{x,y\}=\{2k,2k+1\}$ with $k\geq N$, while
$(ab)^\omega$ has the same labels and the same event order $2k\dashrightarrow 2k+1$ with $x<y$ whenever $x<y$ in
$\Nn$.

\begin{corollary}[Languages of the torus]\label{cor:muller-events}
(1) A partial HDA $Z$ over $\ab$ with $\Theta_N\in\LB{Z}$ for all $N$ and $(ab)^\omega\notin\LB{Z}$ has $\LB{Z}$
not event-geometric.
(2) $\LB{\mathsf{T}^{\{q_{ab}\}}}$ is not event-geometric.
(3) $\LMp{\mathsf{T},\FT}$ is not event-geometric.
\end{corollary}

\begin{proof}
(1) A geometric theory holds in a structure $S$ on $\Nn$ when for every $n$ some model of it has every atom of $S$
on $[n]$ and no atom that $S$ lacks, since by finite support the left side of an axiom true in $S$ holds in such a model
and its right side returns to $S$ along the identity of $\Nn$, which is then a
homomorphism; for $S=(ab)^\omega$ and $n=2N$ the valid $\Theta_N$,
whose precedence is contained in that of $(ab)^\omega$ and agrees with it on $[2N]$, is such a
model of any theory whose valid models form
$\LB{Z}$.
(2), (3) The track reading $ab$ $N$ times and then $a\parallel b$ for ever reads $\Theta_N$ and recurs in $q_0$ and $q_{ab}$, staying
in them from time $4N$ on. An
accepted track visits the square, while all events of $(ab)^\omega$ are comparable, so (1) applies to $\mathsf{T}^{\{q_{ab}\}}$
and to a partial HDA with B\"uchi language $\LMp{\mathsf{T},\FT}$.
\end{proof}

By item~(2) the event signature does not suffice for B\"uchi languages, and the signature of $\TR$ has a sort of
cells for each conclist.

\subsection{Observability}\label{ssec:muller-obs}

With $\Tacc$ as in Definition~\ref{def:theory} and the structure $X_\beta$ defined after it, the \emph{track axioms} are those of $\Tacc$ other than the
axioms of partial HDAs and of acceptance, and $\Tt^{\mathrm{acc}}_{\ab}$ is countable.

\begin{theorem}[Observability]\label{thm:muller-obs}
Let $\Lambda$ have decidable equality.
(i)\label{thm:muller-obs-acc} If $\beta$ is PAF-accepted by a partial HDA $X$ then $X_\beta\models\Tacc$. The converse holds for finite $X$
under $\EIO$. If it holds for all finite $X$ over $\ab$ then every $u\colon\Nn\to\{0,1\}$ takes the value $1$ or
the value $0$ infinitely
often.
(ii)\label{thm:muller-obs-torus} Assume $\EM$ and $\DC$. Let $\mathbb{S}$ be a countable theory over the signature of $\Tacc$ and
$\mathbb{S}^+$ the theory whose axioms are those of $\mathbb{S}$, those of HDAs with every face defined and the track
axioms. Then it is not the case that for every track $\beta$ of $\mathsf{T}$ the structure $\mathsf{T}_\beta$ models $\mathbb{S}^+$
exactly when $(\mathsf{T},\FT)$ Muller-accepts $\beta$.
\end{theorem}

\begin{proof}
(i) A recurring accepting cell is accepting after every time, and conversely
$\EIO$ applied to the listed accepting cells finds one visited infinitely often. On the torus with only its vertex not
accepting, the track at $e_a$ or $q_{ab}$ at time $2n+1$ according to $u_n$ models $\Tt^{\mathrm{acc}}_{\ab}$; its
recurring accepting cell says which value
recurs. (ii) A geometric
formula true in $\mathsf{T}_\beta$ is true in $\mathsf{T}_\gamma$ whenever $\gamma$ agrees with $\beta$ long
enough, so the tracks modelling a countable theory are those $\gamma$ with
$\gamma\in O_j\Rightarrow\gamma\in O'_j$ for every $j$, for open sets $O_j,O'_j$ of
tracks. Suppose the Muller-accepted tracks of $\mathsf{T}$ were so cut out.
A list $(a_1,\dots,a_m)$ of naturals names the track that for each $i$ in turn runs the square loop $a_i$ times and
then passes $e_a$ once, and runs the square loop for ever after; it is Muller-accepted.
As each $O'_j$ is open, $\EM$ gives a count $a$ such that every track agreeing with that of the list for $a$ more
rounds of the square lies in $O'_j$ for each $j\leq m$ with the track of the list in $O_j$,
and $\DC$ gives lists $\langle\rangle,(a_1),(a_1,a_2),\dots$ each extending the last by such a
count. Their limit passes $e_a$ infinitely often and is not Muller-accepted, while each $O_j$
that contains it contains the track of a long enough list, whose count puts the limit in
$O'_j$.
\end{proof}

The converse in (i), asked of every finite $X$ over $\ab$, thus lies between $\EIO$, which gives it, and the principle
that every $u\colon\Nn\to\{0,1\}$ takes the value $1$ or the value $0$ infinitely often, which it gives; neither
reverse implication is proved.

\subsection{Muller and B\"uchi}\label{ssec:muller-answer}

\begin{definition}[The statements]\label{def:mb-statements}
$\MinB$ (Muller in B\"uchi) states that for every alphabet $\Lambda$ with decidable equality every finite Muller partial HDA
$(X,F_X)$ over $\Lambda$ has a finite partial HDA $Z$ over $\Lambda$ with $\LB{Z}=\LM{X,F_X}$. $\BinM$ (B\"uchi in Muller) states that
every finite $Z$ has a finite $(X,F_X)$ with the same equation. $\MinBt$ asks $X$ and $Z$ to have all faces
defined. The superscript $\mathrm{dec}$ restricts the inputs to machines carrying a point of $\TR$ with decided
faces (Theorem~\ref{thm:points}); $\MinBtd$ also asks the point's relational form and $Z$ to be total.
\end{definition}

The classical equivalence of Muller and B\"uchi acceptance on finite partial HDAs is measured by logical principles,
$\SDNE$ for Muller in B\"uchi and $\EIO$ for both directions together, and its form for machines with every face
defined is refuted.

\begin{theorem}[Muller and B\"uchi]\label{thm:muller-answer}
Let $\Lambda$ have decidable equality and all machines be finite.
(i)\label{thm:muller-pos} Every Muller partial HDA $(X,F_X)$ over $\Lambda$ has a partial HDA
$Z$ over $\Lambda$ with $\LB{Z}=\LMp{X,F_X}$.
(ii) $\MinB\wedge\BinM$ holds exactly when $\EIO$ does.
(iii)\label{thm:muller-split} $\MinB$ holds exactly when $\SDNE$ does; under $\SDNE$, $\BinM$
holds exactly when $\PLEM$ does.
(iv) $\MinBt$ is false. No partial HDA $Z$ over $\ab$ with all
faces defined has $\LB{Z}=\LM{\mathsf{T},\FT}$, or $\Lmax{Z,\mathcal{P}}=\Lmax{(\mathsf{T},\FT),\mathcal{M}}$ for Muller
acceptance $\mathcal{M}$; under $\EIO$
some partial HDA $Z$ satisfies each.
(v)\label{thm:muller-points} $\MinBd$ holds exactly when $\MinB$ does; under $\SDNE$, $\BinMd$ holds
exactly when $\BinM$ does. Hence $\MinBd\wedge\BinMd$ is equivalent to $\EIO$, and $\MinBtd$ is
false.
\end{theorem}

The separation in item~(iv) is the non-inclusion half of
Theorem~37 of \cite{PAF25}.

\begin{proof}
(i) Fix $i$ and write $F=F_i$, a decidable set of defined cells. A cell $c\in F$ is \emph{bypassed} by a proper
step $p\to q$ between cells of $F$ when two proper steps $p\to c\to q$ in one direction glue to its
ipomset. An $\omega$-track \emph{skips through} $F$ when it starts at a start cell and eventually stays in $F$, and when
each cell of $F$ is visited or bypassed by it at arbitrarily late times. Rerouting a bypass through $c$ composes no
faces, so skipping through $F$ and positive acceptance at $F$ give one
$\omega$-language. A track is
\emph{alternating} when its up-steps and down-steps alternate. A run of up-steps raises the dimension at every
step, so it is finite since the shapes of $X$ are finitely many; so is a run of down-steps. Composing each maximal
run into one step keeps the event $\omega$-ipomset, since the lax law applies at the defined lower end of the run; a cell in the middle of
a run is bypassed by the composite, so the composed track is an alternating track that still skips through
$F$.
It remains to recognise the alternating tracks that skip through $F$, and for $F$ empty the empty partial HDA does.

Otherwise let $e_0$ be a cell of $F$ of least dimension and $e_1,\dots,e_k$ list every cell of $F$, $e_0$ among
them, and write $v^+$ for $v+1$
modulo $k+1$; since $e_0$ has least dimension, no cell of $F$ is a proper face of $e_0$.
For a defined cell $x$ of shape $V$ and a proper lower coface $d^0_A$ into $V$ a cell $c$ lies \emph{strictly
between} $\delta^0_A(x)$ and $x$ when its dimension lies strictly between theirs, $c=\delta^0_{A_2}(x)$ and
$\delta^0_{A_1}(c)=\delta^0_A(x)$ for lower cofaces with $d^0_{A_2}\circ d^0_{A_1}=d^0_A$. Strictly between $x$ and
$\delta^1_B(x)$ is defined with upper cofaces. The last
clause asks that the face $\delta^0_{A_1}(c)$ be defined, which the lax law does not supply. A
cell of $F$ strictly between two cells of $F$ joined by a step is bypassed by that step, and every bypass passes
this test, because the coface of a bypass is the composite of the two steps through the bypassed
cell.
A move from $v$ is \emph{justified} at $x$ entered along $d^0_A$ and left along $d^1_B$ when $e_v$ is
$\delta^0_A(x)$ or $x$, or lies strictly between $\delta^0_A(x)$ and $x$ or between $x$ and
$\delta^1_B(x)$.

The \emph{gadget} $Z_i$ has as cells of shape $V$ the tuples $(x,\pi,I,J,a,b)$ with $x$ a defined cell of $X$ of shape
$V$ and $\pi$ either \emph{valley} or \emph{peak}, where $I$ and $J$ are sets of inhabited sets of coordinates of
$V$ and $a,b\in\{\bot,0,\dots,k\}$ are \emph{counters}. The members $A$
of $I$ name the proper lower cofaces $d^0_A$ by which the peak may be entered, and those $B$ of $J$ the proper upper
cofaces $d^1_B$ by which it may be left. A lower coface is determined by the set $A$ where the top corner of its face
is $0$, and an upper one by the set $B$ where the bottom corner of its face is $1$; so each entry and each exit names
exactly one coface. The tuples kept are the following.
\begin{enumerate}[label=(G\arabic*),nosep,leftmargin=*]
\item A valley has $I=J=\emptyset$ and $b=\bot$, and $x\in F$ when $a\neq\bot$.
\item A peak has $a=b=\bot$, or $a=\bot$ and $x\in F$, or $a=v$ and $b=t$ with $x\in F$ and two conditions. For all
  $A\in I$ and $B\in J$ with $\delta^0_A(x)$ and $\delta^1_B(x)$ defined, $t=v$ or else $t=v^+$ with the move
  justified at $x$ entered along $d^0_A$ and left along $d^1_B$. For every $A\in I$ with $\delta^0_A(x)=e_v$ we have
  $t=v^+$.
\end{enumerate}
The face of a cell along $d_{\emptyset,\emptyset}$ is the cell. The face of a peak along $d^0_A$ with $A\in I$ is
the valley $(\delta^0_A(x),\emptyset,\emptyset,a,\bot)$ and along $d^1_B$ with $B\in J$ the valley
$(\delta^1_B(x),\emptyset,\emptyset,b,\bot)$, each defined when it is a kept tuple, and every other face is
undefined.
A valley has no proper face, so of two defined faces that compose one is along $d_{\emptyset,\emptyset}$, and the
lax law $\delta_{C,D}\delta_{A,B}\subseteq\delta_{A\cup C,B\cup D}$ holds.
The start cells of $Z_i$ lie over start cells of $X$ with $a=\bot$, and its one accepting cell is the valley
$(e_0,\emptyset,\emptyset,0,\bot)$. A cell of $Z_i$ is a cell of
$X$ of a listed shape with a bit, two sets of sets of coordinates and two counters; these are finitely many,
equality of tuples is decided componentwise and the conditions (G1) and (G2) are decided, so $Z_i$ is
finite.

By the faces just listed an up-step of $Z_i$ goes from a valley to a peak and keeps the counter, and a down-step goes
from a peak to a valley whose counter is the peak's $b$,
so the tracks of $Z_i$ alternate. Forgetting all but $x$ maps a track of $Z_i$
to a track of $X$ through defined cells with the same steps.
By (G1) and (G2) a counter once set stays set and its cell lies in $F$; at a peak it
stays at $v$ or moves to $v^+$ by a justified move, and it must move when $e_v$ is the cell entered
from.

\emph{Soundness.} Let a track of $Z_i$ visit the accepting valley at arbitrarily late times. From its first visit the
counter is set, so the projected track $\gamma$ eventually stays in $F$. The step after each visit enters a peak from
$e_0$ with counter $0$, so the counter moves to $0^+$ there; a counter that moves by one step round
$\{0,\dots,k\}$ and leaves $0$ after every time leaves every value $v$ after every time.
Each move from $v$ at the peak $\gamma_{n+1}$ is justified, so $e_v$ is $\gamma_n$ or $\gamma_{n+1}$ or is bypassed
by the step at $n$ or at $n+1$. The cell $e_0$ is visited at each accepting time, and every cell of $F$ is some $e_v$, so
$\gamma$ skips through $F$.

\emph{Completeness.} Let $\gamma$ be an alternating track through defined cells that skips through $F$ and stays in
$F$ from time $N$. Its \emph{lift} has at time $t$ the valley $(\gamma_t,\emptyset,\emptyset,c_t,\bot)$ when $s_t$ is
an up-step and the peak $(\gamma_t,I_t,J_t,c_t,c_{t+1})$ when $s_t$ is a down-step, where $I_t=\{A\}$ when $s_{t-1}$
is an up-step along $d^0_A$ and $I_0=\emptyset$, and $J_t=\{B\}$ when $s_t$ is along
$d^1_B$. The counter starts at $c_0=\bot$ and
is kept along up-steps; at a peak $t$ it switches on to $0$ when $c_t=\bot$ and $t\geq N$, and from $c_t=v$ it moves
to $v^+$ exactly when the move is justified at $\gamma_t$ entered along $s_{t-1}$ and left along
$s_t$. Counters are set only after $N$, where $\gamma$ is in $F$. Each
entry set and exit set holds one set of coordinates, so (G2) speaks only of the steps actually taken, and a move
made exactly when justified meets it. So the lift is a track of $Z_i$ with the steps of $\gamma$ from a start
cell. Its counter is set from
time $N+2$ on. After any time some value $v$ is held, the skip condition gives a later visit or bypass of $e_v$, and
by alternation the peak at that time or the next one reads it as a justification. So the counter moves after every
time, and it leaves $0$ after every time. At a peak
$\gamma_t$ where it leaves $0$ the cell $e_0$ is not $\gamma_t$, of which $\gamma_{t-1}\in F$ is a proper face, and
lies strictly between no two cells of $F$.
So $\gamma_{t-1}=e_0$, and the valley at $t-1$ has counter $c_{t-1}=c_t=0$ and is the accepting cell; the lift is
therefore accepted with no pigeonhole.
The sum over $i$ of the gadgets recognises $\LMp{X,F_X}$.

(ii), (iii) Under $\SDNE$ a cell outside $F_i$ that does not recur is eventually unvisited, so (i) gives
$\MinB$.
Under $\PLEM$ the Muller form of $Z$ whose accepting sets are those meeting an accepting cell
\cite{PAF25} realises $\Inf(\beta)$ as one of them. The reversals run on $\mathsf{E}$. For
$u\colon\Nn\to\{0,1\}$ the word $w_u=a\,x_0\,a\,x_1\cdots$ has sequential events with $x_n=b$ when
$u_n=1$ and $x_n=a$ when $u_n=0$, so that one track of $\mathsf{E}$ reads it. For $\MinB\Rightarrow\SDNE$ take $F_{\mathsf{E}}=(\{v,e_a\})$, whose Muller
language holds $w_u$ when $u$ is not infinitely often $1$ and no word with $b$-events beyond every bound. A
B\"uchi recogniser $Z$ starts no $b$-event after a visit of its recurring accepting cell, since pumping that loop would
give such a word; so $w_u\in\LB{Z}$ makes $u$ eventually
$0$. For $\BinM\Rightarrow\PLEM$ under $\SDNE$ let
$v$ alone accept, so that every $w_u$ lies in $\LB{\mathsf{E}_v}$. $\SDNE$ gives the \emph{limited principle of omniscience} $\LPO$ that every Boolean
sequence is $1$ somewhere or $0$ everywhere. A finite
Muller recogniser accepts $w_u$ at some $F_i$, and whether $F_i$ holds a cell with $b$ in its shape is decided. If
it does then $u$ is not eventually $0$, and $\LPO$ on the tails of $u$ makes $u$ infinitely often $1$; otherwise
$\SDNE$ makes such cells unvisited from some time on, which refutes $I(u)$. Then
(ii) follows from $\EIO\iff\SDNE\wedge\PLEM$.

(iv) A total recogniser of $\LM{\mathsf{T},\FT}$ accepts the word of the square loop. Rerouting round the boundaries of
their squares the passages of its track that miss its recurring accepting cell and every second one that visits it
gives an accepted track with $a$-events running alone beyond every bound, which on $\mathsf{T}$ run on $e_a$ and violate
$\FT$; under $\EIO$
item~(ii) applies to $(\mathsf{T},\FT)$. (v) Point
hypotheses only restrict the inputs, and $\mathsf{E}$, $\mathsf{E}_v$ carry points with decided faces by
Theorem~\ref{thm:points};
the torus with every cell accepting has a finite total point through its square loop, where (iv) refutes
$\MinBtd$.
\end{proof}

Dimension two enters item~(i) through the square loop of $\mathsf{T}$. Let the edge $e_a$ of the torus lack its lower face and
let its starting vertex and its square form the one accepting set; the recogniser of item~(i) then accepts the word of
the Muller-accepted track alternating between them.

Excluded middle gives $\EIO$, and through it the form of the conjecture recalled in
Remark~\ref{rem:muller-conjecture} that compares the languages $L^{\max}$ holds for finite partial HDAs over an
alphabet with decidable equality.

\begin{corollary}[Muller and B\"uchi under $\EM$]\label{cor:muller-classical}
Assume $\EM$ and let $\Lambda$ have decidable equality. Every finite Muller partial HDA $(X,F_X)$ over $\Lambda$ has
a finite partial HDA $Z$ over $\Lambda$ with $\LB{Z}=\LM{X,F_X}$, and every finite partial HDA $Z$ over $\Lambda$ has
a finite Muller partial HDA $(X,F_X)$ over $\Lambda$ with
$\LM{X,F_X}=\LB{Z}$.
Each of these equations gives $\Lmax{Z,\mathcal{P}}=\Lmax{(X,F_X),\mathcal{M}}$.
\end{corollary}

\begin{proof}
$\EM$ decides $\exists n\,p(n)$ for every sequence of decided propositions $p(n)$, and
this gives $\EIO$, from which Theorem~\ref{thm:muller-answer}(ii) gives
$\MinB\wedge\BinM$. A language $L^{\max}$ holds the correctly marked members of
the corresponding $\omega$-language (Section~\ref{ssec:muller-defs}), so equal $\omega$-languages have equal languages
$L^{\max}$.
\end{proof}

\begin{remark}[The conjecture]\label{rem:muller-conjecture}
Bazille, Dubut, Fahrenberg and Ziemia\'nski write that \cite{PAF25} ``notices that the usual equivalence between
Muller and B\"uchi acceptance does not hold for HDAs but conjectures that this problem disappears when passing to
pHDAs'' \cite{BDFZ26}. The languages on infinite tracks of
\cite{PAF25} are those of HDAs whose faces all exist, and \cite{PAF25} names partial HDAs as a possible route to a Kleene theorem. For
HDAs \cite{PAF25} collects the event $\omega$-ipomsets of accepted tracks with exactly the $<$-maximal events marked
and closes no language under subsumption. Carried to partial HDAs this is the language
$\Lmax{X,\mathcal{A}}$ of Section~\ref{ssec:muller-defs}, and in that reading the conjecture holds under $\EM$ by
Corollary~\ref{cor:muller-classical}; its form for the unions over markings is equivalent to $\EIO$ with no
principle (Theorem~\ref{thm:muller-answer}(ii)).
\end{remark}

Under $\EM$ and $\DC$ no countable theory, taken with the axioms of HDAs and the track axioms, cuts out Muller acceptance on the torus
(Theorem~\ref{thm:muller-obs}(ii)), and under $\EIO$ a change of machine makes the Muller language of every finite Muller partial HDA the language of the
theory $\Tacc$.

\begin{corollary}\label{cor:muller-geometric}
Let $\Lambda$ have decidable equality. Under $\EIO$ every finite Muller partial HDA $(X,F_X)$ over $\Lambda$ has a
finite partial HDA $Z$ over $\Lambda$ with $\LM{X,F_X}=\Lang{Z}{\beta\mapsto
Z_\beta\models\Tacc}$.
\end{corollary}

\begin{proof}
Theorem~\ref{thm:muller-answer}(ii) gives a finite partial HDA $Z$ whose B\"uchi language is $\LM{X,F_X}$, and
Theorem~\ref{thm:muller-obs}(i) identifies B\"uchi acceptance of an $\omega$-track $\beta$ of $Z$ with
$Z_\beta\models\Tacc$; on machines
with every face defined this is the earlier form.
\end{proof}

\section{Image and fibres}\label{sec:image}

\subsection{Fibres}\label{ssec:image-fibres}

\begin{definition}[Fibres]\label{def:fibres}\label{def:fibres-carry}
Let $X$ and $Y$ be partial HDAs over $\Lambda$. A point $\eta$ of $Y$ and a point $\xi$ of $X$ \emph{lie over one
label word} when each carries the labels of the other (Definition~\ref{def:carries}). The
\emph{fibres} of $X$ are \emph{bounded by $k$} when no $k+1$ points of $X$ that are pairwise \emph{different},
that is unequal as sets of prefixes, lie pairwise over one label word; $X$ is
\emph{finitely ambiguous} when some $k$ bounds them. $X$ is \emph{ambiguous} when
two different points of $X$ lie over one label word, and \emph{label-injective} when each
point of $X$ carrying the labels of a point $\xi$ of $X$ is $\xi$.
\end{definition}

$X$ is \emph{unambiguous} when its fibres are bounded by $1$, which holds exactly when $X$ is not
ambiguous. Two points lie over one label word when they read one infinite
sequence of cell shapes, acceptance bits and step ipomsets. L\"oding and Pirogov count the number
$\mathrm{da}(X,w)$ of accepting runs over each input word $w$ of a B\"uchi automaton, its degree of ambiguity
\cite{LP18}. On a finite rigid machine whose points are chains of
tracks a bound by $k$ on the fibres is the bound $\mathrm{da}(X,w)\leq k$ for the automaton of Section~\ref{ssec:image-bounded}, whose letters record the shapes and
bits of the cells that a step joins; it is no bound on the runs over words of $\Lambda$ alone. Amrane, Bazille,
Fahrenberg and Ziemia\'nski call an HDA \emph{$k$-ambiguous} when each member of its language of finite ipomsets is
the event ipomset of at most $k$ sparse accepting paths
\cite{ABFZ23}. Their count runs over every path with one event
ipomset, while a fibre holds only points that also agree in shapes and acceptance bits at every time. A regular
language has \emph{bounded ambiguity} when some $k$-ambiguous HDA recognises it,
and whether this is decidable is open. Theorems~\ref{thm:unamb}, \ref{thm:ida} and~\ref{thm:cyc}
decide bounds on the fibres of one given machine, which is a question about the machine and not about its language.

\subsection{Embedding}\label{ssec:image-trivial}

\begin{definition}[Embedding]\label{def:embed}\label{def:image-is}\label{def:generate}
A partial HDA $X$ over $\Lambda$ \emph{embeds} over the label topos when every open $V$ of $X$ has an open $U$ of label
words with $V\feq{X}\lst{X}(U)$. The \emph{label opens generate} the frame of $X$ when every
open $V$ has a set $W$ of prefixes with $V\feq{X}\bigcup_{p\in W}\lst{X}(p)$.
\end{definition}

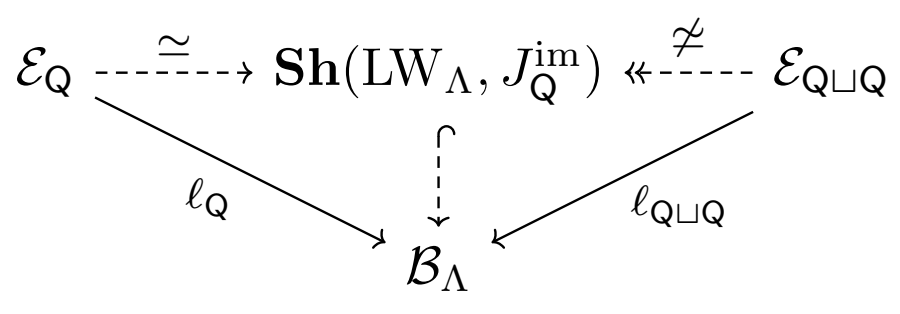
\begin{figure}[htb]
\centering
\begin{tikzpicture}[v4,x=1cm,y=1cm]
  \node[v4obj] (eq) at (-2.6,0) {$\Ee_{\wQ}$};
  \node[v4obj] (ed) at (2.6,0) {$\Ee_{\wQ\dsum \wQ}$};
  \node[v4obj] (im) at (0,0) {$\Sh(\LW_\Lambda,\Jim{\wQ})$};
  \node[v4obj] (bl) at (0,-1.3) {$\BL$};
  \draw[v4read] (eq) -- node[v4lab,above=1.5pt] {$\simeq$} (im);
  \draw[v4epiread] (ed) -- node[v4lab,above=1.5pt] {$\not\simeq$} (im);
  \draw[v4inclread] (im) -- (bl);
  \draw[v4map] (eq) -- node[v4lab,below left] {$\ell_{\wQ}$} (bl);
  \draw[v4map] (ed) -- node[v4lab,below right] {$\ell_{\wQ\dsum \wQ}$} (bl);
\end{tikzpicture}
\caption{The label maps of $\wQ$ and $\wQ\dsum \wQ$ factor through one image since $\Ima\,\wQ=\Ima(\wQ\dsum \wQ)$
(Theorem~\ref{thm:separation}(2)). Every open of $\wQ$ is equivalent to $\lst{\wQ}(U)$ for an open $U$ of label words, so $\wQ$ embeds and its factor is an
equivalence, while $\wQ\dsum \wQ$ has two points that read one infinite sequence of labels and does not embed (Definition~\ref{def:embed} and
Theorem~\ref{thm:unamb}). Solid arrows are geometric morphisms under $\UC$ (Theorem~\ref{thm:prices-uc}), and under $\UC$ the dashed arrows factor them through the image, the subtopos of sheaves for the topology coinduced along $\lab$
\cite{Caramello19} (Theorem~\ref{thm:image-up}).}
\label{fig:embed}
\end{figure}

An \emph{inclusion of locales} is a map of locales whose frame homomorphism is onto. By
Theorem~\ref{thm:topos-a}(c)(ii) the map $\lst{X}$ is a frame homomorphism from the opens of label words to $\Op(X)$
taken up to $\feq{X}$, and Definition~\ref{def:embed} says that it is onto; so $X$ embeds exactly when $\ell_X$ is an
inclusion of locales (Figure~\ref{fig:embed}).

\begin{definition}[Square search]\label{def:square}
Let $\Lambda$ have decidable equality and $X$ be finite. The defined cells of $X$ are then
listed and equality of its cells is decided, and so is whether a proper step
joins two cells.
For cells $c,c'$ joined by a proper step let $s(c,c')$ be the proper step from $c$ to $c'$ that this
decision returns. If $c$ and $c'$ are defined and $X$ is rigid then $s(c,c')$ is the only
one. The \emph{synchronised square} $\Sq{X}$ is the graph on
pairs of cells with an edge $(c,d)\to(c',d')$ when $s(c,c')$ exists and some proper step $d\to d'$ has its
ipomset. A pair is \emph{admissible} when its cells are defined and have one shape and
one acceptance bit. The \emph{square search} succeeds when a walk of $\Sq{X}$ through
admissible pairs leads from a pair of start cells through a pair of different cells to an accepting pair
on a cycle of positive length through such pairs.
\end{definition}

Embedding says that the frame of $X$ is recovered from its label words, and on finite rigid machines the theorem
turns this property of the topos into unambiguity, which the square search decides.

\begin{theorem}[Embedding]\label{thm:unamb}\label{thm:is-image}
(1) Let $X$ be a partial HDA over any alphabet. $X$ embeds exactly when its label opens generate its
  frame. If they do then every point of $X$
  carrying the labels of a point $\xi$ of $X$ contains $\xi$, and $X$ is not
  ambiguous.
(2) Let $\Lambda$ have decidable equality and $X$ be a finite rigid partial HDA over $\Lambda$ of any dimension,
  with neither trimming nor closed acceptance asked. Then the following are
  equivalent:
  \begin{enumerate}[label=(\alph*),nosep,leftmargin=*]
  \item $X$ embeds;
  \item $X$ is label-injective;
  \item $X$ is unambiguous;
  \item the square search on $X$ fails.
  \end{enumerate}
  If the search succeeds then $X$ is ambiguous. Each such $X$ is label-injective or
  ambiguous.
\end{theorem}

\begin{proof}
(1) An open $\lst{X}(U)$ is the union of the label opens of its members. If $X$
embeds and $p\in\xi$, a cover of $\upc{p}$ by label opens meets every point carrying the labels of $\xi$, which
holds $p$.
(2) Let $\nu$ exceed the number of listed pairs. If the search fails the \emph{labels settle the cells}, in that
$q_i=r_i$ for $i\leq n$ whenever $r$ has $\nu$ accepting times after $n$ and $q$ carries its label word, since otherwise
the pairs $(r_t,q_t)$ walk in $\Sq{X}$ by rigidity and two accepting times with one pair close a
cycle. A point holds the
extensions of a prefix by $\nu$ accepting times, which cover it, so (d) gives (a) and
(b).
A successful search run round its cycle gives two different points over one label
word, which (1) excludes; the search is
decided.
Under (d) the label-injective $X$ has its fibres bounded by $1$, which is (c).
Conversely an unambiguous $X$ is not ambiguous, hence label-injective by the
dichotomy, so (a) to (d) are equivalent.
\end{proof}

\subsection{The image}\label{ssec:image-image}

\begin{definition}[Surjection and live words]\label{def:fibres-live}\label{def:image-lp}
Let $X$ and $Y$ be partial HDAs over $\Lambda$. The label map $\ell_X$ is a \emph{surjection} when
$\lst{X}(U)\fle{X}\lst{X}(V)$ implies that $V$ covers every word of $U$ in $\JLW$, for all opens $U,V$ of label
words. A label word is \emph{dead in $X$} when the empty sieve covers it
in $\Jim{X}$. With live prefixes as in Definition~\ref{def:trim}, $Y$ \emph{has the live label
words} of $X$ when each live prefix $p$ of $X$ has a live prefix $q$ of $Y$ with $\lab(q)=\lab(p)$, where $\lab(q)=\lab(p)$ says that $q$ and $p$ have one length and at every time one shape and one acceptance bit
of their cells and one ipomset of their steps. For $\Lambda$ with
decidable equality the \emph{language of live prefixes} $\LangLp{X}$ is the set of sequences $w$ of ipomsets such
that for every $n$ some live prefix of $X$ of length $n$ has $\ip(s_i)=w_i$ at each step
$s_i$.
\end{definition}

On finite rigid machines with closed acceptance the comparison of images, and on label-injective ones the relation $\eqB$ itself, reduce to
a comparison of live label words, which the subset construction of Section~\ref{ssec:decide-inj} decides.

\begin{theorem}[Image]\label{thm:image-live}
(1) Let $X$ and $Y$ be partial HDAs over any alphabet. $\ell_X$ is a surjection exactly when every cover in $\Jim{X}$ is a cover in
$\JLW$, and no finite $X$ over an inhabited alphabet is a
surjection. The map $\fgt$ is a comorphism of sites. A sieve $S$ covers $u$ in $\JSW$ (in
$\JimS{X}$) exactly when the label words whose step word lies in $S$ cover in $\JLW$ (in $\Jim{X}$) each label word
whose step word extends $u$; so $\Ima X=\Ima Y$ implies
$\ImaST X=\ImaST Y$. No open of step words has the inverse image along $\fgt$
equivalent in $\JLW$ to the open of the words extending the word of length zero at an
accepting vertex.
(2) In items (2) to (4) $\Lambda$ has decidable equality and $X$, $Y$ are finite rigid partial HDAs over $\Lambda$
of any dimension. $\Ima X=\Ima Y$ implies that $Y$ has the live label words of $X$.
If $X$ and $Y$ have closed acceptance then $\Ima X=\Ima Y$ holds exactly when each has the live label words of
the other, and $\ImaST X=\ImaST Y$ exactly when
$\LangLp{X}=\LangLp{Y}$. If $X$ has closed acceptance then a sieve $S$ covers
$w$ in $\Jim{X}$ exactly when $S$ together with the words dead in $X$ covers $w$ in
$\JLW$.
(3) $X\eqB Y$ implies $\LangLp{X}=\LangLp{Y}$.
(4) If $X$ and $Y$ are label-injective then $X\eqB Y$ holds exactly when $\Ima X=\Ima
Y$, and exactly when each has the live label words of the
other.
\end{theorem}

On finite rigid machines of any acceptance one image already gives each machine the live label words of the other, and whether the converse holds in that generality stays open.

\begin{proof}
(1) A sieve $S$ on $w$ covers it in $\Jim{X}$ exactly when $\lst{X}(\{v : w\lle v\})\fle{X}\lst{X}(S)$, and one
on a step word $u$ covers it in $\JimS{X}$ exactly when $\lsts{X}(u)\fle{X}\lsts{X}(S)$; with
Theorem~\ref{thm:image-up}(2) this gives the criterion for a surjection, and
a finite $X$ carries no word of length zero at a conclist longer than its
shapes.
Covers lift both ways along $\fgt$; the
words of length zero at an accepting and at a rejecting vertex have one step word and no common
extension.
(2) On a finite rigid machine the label open of a prefix is dead or holds a live prefix with its label
word; under closed
acceptance covers are bars of live extensions of bounded length, which matched live words transfer between $X$ and
$Y$.
For the last sentence a cover of $w$ in $\Jim{X}$ gives the finitely many prefixes carrying $w$ a common bar of
live prefixes, past which a word is carried through the bar or only by dead prefixes; the converse lifts covers
of $\JLW$.
That sentence fails on a finite rigid graph over $\ab$ without closed acceptance, through a live walk that never accepts
after time $0$.
(3) On a finite rigid $Y$ the label open in $Y$ of a prefix $p$ of $X$ lies below $\emptyset$ or holds a live prefix
with the label word of $p$; by Theorems~\ref{tp:lat} and~\ref{tp:lab} the first case
makes $p$ dead.
(4) On label-injective machines the
labels settle the cells after $\nu$ accepting times (proof of Theorem~\ref{thm:unamb}), so extending members by $\nu$
accepting times gives an equivalence over the label topos, and conversely (3) matches live label
words.
By (2) one image matches live label words, and by Theorem~\ref{tp:img} $X\eqB Y$ gives one
image.
Unambiguous machines are label-injective by Theorem~\ref{thm:unamb}(2),
so (4) covers them.
\end{proof}

Applied to (2), the description of closed subtoposes by Caramello and Lafforgue
\cite{CL25b} identifies the image of a finite
rigid machine with closed acceptance with the closed subtopos complementary to the open of its dead words;
at the machine $\wQ$ of Example~\ref{exa:dec-witnesses} that open is neither empty nor everything, an instance of
the failure of surjectivity that (1) proves for every finite machine over an inhabited alphabet. Caramello's theory of
comorphisms \cite{Caramello19} turns the comorphism $\fgt$ of (1) into a
geometric morphism $\BL\to\BST$, which by (1) is a surjection and no inclusion, and by (1) the image topology of $X$
over the step words is computed from that over the label words through $\fgt$.

\subsection{Bounded fibres}\label{ssec:image-bounded}

L\"oding and Pirogov prove that a trim nondeterministic B\"uchi automaton is finitely ambiguous exactly when it
has no \emph{IDA pattern}, a pair of different states $p,q$ with paths $p\to p$, $p\to q$ and $q\to q$ over one
word; they decide it in cubic time
\cite{LP18};
whether bounded ambiguity of a regular HDA language is decidable is open \cite{ABFZ23},
and the fibres of Section~\ref{ssec:image-fibres} measure the ambiguity of one machine by its points that read one
infinite sequence of labels.
With its labels a finite rigid partial HDA is a B\"uchi automaton whose states are its defined cells and whose
transitions are its proper steps, read as letters through their ipomsets and the labels of their ends. On this
automaton the square search is the classical search of the self-product for two accepting runs over one word, and the
IDA search of Definition~\ref{def:ida} is the pattern test of Weber and Seidl \cite{WS91} in the form of \cite{LP18}. Both are reachability
questions on $\Sq{X}$ and $\Cu{X}$, graphs on $n^2$ and $n^3$ vertices for $n$ listed cells, and each is a Boolean
computed by an explicit search of its graph from the finiteness structure. Theorems~\ref{thm:unamb}, \ref{thm:ida}
and~\ref{thm:cyc} state these tests as properties of the topos. Their fibres hold points of the locale of $X$, which on finite rigid
machines with closed acceptance are all chains of $\omega$-tracks exactly when $\BC$ holds
(Theorem~\ref{thm:pt-track}). Theorem~\ref{thm:unamb} identifies unambiguity with the property that $\ell_X$
embeds, and all three hold with no principle. On graphs the criterion of \cite{LP18} is more general. With $X$ as in Definition~\ref{def:square} a cell is \emph{reachable} when a walk of proper steps through defined cells reaches it from a
start cell.

\begin{definition}[IDA search and cycle class]\label{def:ida}\label{def:generate-cycle}
With $s(c,c')$ as in Definition~\ref{def:square} the \emph{synchronised cube} $\Cu{X}$ is the graph on
triples of cells with an edge $(c,d,e)\to(c',d',e')$ when $s(c,c')$ exists and some proper steps $d\to d'$
and $e\to e'$ have its ipomset. A triple is \emph{admissible} when its cells are defined
and have one shape and one acceptance bit. The \emph{IDA search} succeeds when for a reachable cell $p$
and a cell $q\neq p$ some walk of $\Cu{X}$ of positive length through admissible triples leads from $(p,p,q)$ to
$(p,q,q)$. $X$ is in the \emph{cycle class} when for every reachable cell
$c$ and proper step $c\to d$ to a defined cell $d$ some walk of proper steps through defined cells leads from $d$ to
$c$.
\end{definition}

On a finite rigid machine with closed acceptance, success of the IDA search is a finite certificate, found on the
cell graph, that for every $k$ some label word has more than $k$
points of the topos over it.

\begin{theorem}\label{thm:ida}
Let $\Lambda$ have decidable equality and $X$ be a finite rigid partial HDA over $\Lambda$ of any dimension with
closed acceptance. If the IDA search on $X$ succeeds then for every $k$ some $k+1$ pairwise different points lie
over one label word, so the fibres of $X$ are bounded by
no $k$, and $X$ is not finitely
ambiguous. If $X$ is finitely ambiguous the IDA search
fails.
\end{theorem}

\begin{proof}
Under closed acceptance the chain of every $\omega$-track from a start cell is a point
(Theorem~\ref{thm:pts-x}(4)). Let $u$ be a walk from a start cell to $p$ of length
$m_0$ and $\pi$ a walk of $\Cu{X}$ of length $L>0$ from $(p,p,q)$ to $(p,q,q)$; the
track $\beta^i$ follows $u$, the first coordinate of $\pi$ for $i$ rounds, its second once and its third for ever.
Admissibility and rigidity give the $\beta^i$ one label word, and for
$i<j$ at time $m_0+(i+1)L$ the track $\beta^i$ is at $q$ and $\beta^j$ at
$p$.
\end{proof}

On the cycle class the failure of the IDA search bounds the fibres, so that under closed acceptance finite
ambiguity, a property of the points of the topos, is decided there.

\begin{theorem}\label{thm:cyc}
Let $\Lambda$ have decidable equality and $X$ be a finite rigid partial HDA over $\Lambda$ of any dimension in the
cycle class, with $n$ listed cells.
(1) If the IDA search on $X$ fails then the fibres of $X$ are bounded by $n^2$, with neither trimming nor closed
  acceptance asked.
(2) With closed acceptance $X$ is finitely ambiguous exactly when the IDA search
  fails. Over $\ab$ some function $\mathsf{fa}$ from finite partial HDAs to
  $\{0,1\}$ has $\mathsf{fa}(X)=1$ exactly when $X$ is finitely ambiguous, for every such $X$ with closed
  acceptance.
\end{theorem}

\begin{proof}
(1) Let $\xi_0,\dots,\xi_{n^2}$ be pairwise different over one label word. Since the goal is absurd, double negation
commutes with the finitely many pairs and gives one length $T$ at which the members of every pair
differ. By pigeonhole two
members $a,b$ of length $T$ share first and last cells, and the
cycle class gives a walk through defined cells from their last cell back to their first. Let
$s\leq T$ be a time at which the cells $a_s$ and $b_s$ differ. The rotation of Weber and
Seidl \cite{WS91}, whose classes \cite{LP18} restates, runs through $(a_t,a_t,b_t)$ for $t$ from $s$ to $T$,
then through the return walk in all three coordinates, then through $(a_t,b_t,b_t)$ for $t$ from $0$ to $s$; as $a$
and $b$ lie over one label word and the return walk is shared, its triples are admissible and its steps have one
ipomset, so it is a walk of $\Cu{X}$ of positive length from $(a_s,a_s,b_s)$ to $(a_s,b_s,b_s)$ with $a_s$ reachable and
$b_s\neq a_s$. The IDA search fails, so $a$ and $b$ agree at every time up to $T$, against the choice of
$T$.
(2) is Theorem~\ref{thm:ida} with (1), and $\mathsf{fa}(X)=1$ exactly when the IDA search on $X$
fails.
\end{proof}

\subsection{Sheets and the position of a fibre}\label{ssec:image-sheets}

\begin{definition}[Sheets and position]\label{def:image-sheets}\label{def:isol}\label{def:image-split}
Let $X$ be a partial HDA over $\Lambda$ and $(p_n)$ a sequence of prefixes of a partial HDA
over $\Lambda$.
(1) The label opens \emph{generate on} an open $U$ of $X$ when every open $V\fle{X}U$ has a set $W$ of prefixes
with $V\feq{X}U\cap\bigcup_{p\in W}\lst{X}(p)$. $X$ has \emph{$k$ sheets over} an open $V$ when
$V\fle{X}U_1\cup\dots\cup U_k$ for opens $U_i$ on which the label opens
generate.
(2) $X$ has an \emph{isolated point over $(p_n)$} when
some complemented $U\not\fle{X}\emptyset$ has $U\fle{X}\lst{X}(p_n)$ for all $n$.
$X$ is \emph{split over $(p_n)$} when some complemented $U$ with complement $V$ has $U\cap\lst{X}(p_n)\not\fle{X}\emptyset$ and $V\cap\lst{X}(p_n)\not\fle{X}\emptyset$ for
all $n$.
\end{definition}

One sheet over $\Pref(X)$ is generation by the label opens, and $k$ sheets
bound the fibres by $k$, since $k+1$ different points over one label word each meet
one of the $k$ opens, two of them meet the same one by pigeonhole, and on that open the label opens generate, so that
points over one label word meeting it coincide.
The least number of sheets is thus an upper bound on the fibres, the counterpart over an open of the degree of
ambiguity of \cite{LP18}.

\begin{definition}[Shape sequent]\label{def:image-shape}
A prefix $p$ of $X$ \emph{passes} a conclist $U$ over $\Lambda$ when some cell of $p$ has shape $U$. The sequent $\mathrm{pass}_U\vdash\bot$ \emph{holds} in $X$ when every prefix of
$X$ passing $U$ is covered by the empty sieve, and $X$ is \emph{free
of $U$} when no member of a point of $X$ passes $U$.
\end{definition}

The prefixes passing $U$ form a sieve, and $\mathrm{pass}_U\vdash\bot$ is the
geometric sequent of the site which says that the subterminal generated by this sieve is null.

\begin{lemma}\label{prop:image-shape}
Let $\Lambda$ have decidable equality, $U$ be a conclist over $\Lambda$ and $X$ be a finite partial HDA over
$\Lambda$. Then $X$ is free of $U$ exactly when $\mathrm{pass}_U\vdash\bot$ holds in $X$.
\end{lemma}

\begin{proof}
Points meet covering sieves and miss $\emptyset$; conversely
Lemma~\ref{lem:no-point-dead} applies.
\end{proof}

\begin{corollary}[Invariants of the fibres]\label{thm:image-profile}\label{def:image-profile}
Let $X$ and $Y$ be partial HDAs over any alphabet with $X\eqB Y$, with no finiteness or rigidity asked.
(1) The label opens generate the frame of $X$ exactly when they generate that of
$Y$.
(2)\label{thm:fib-inv} For every $k$ the fibres of $X$ are bounded by $k$ exactly when those of $Y$
are, and $X$ is finitely ambiguous exactly when $Y$
is.
(3) $X$ and $Y$ have the same \emph{ambiguity profile}, which is to say that for every $k$ and every prefix $p$ of
$X$ or of $Y$ the machine $X$ has $k$ sheets over $\Pref(X)$ or over $\lst{X}(p)$ exactly when $Y$ has $k$ sheets
over $\Pref(Y)$ or over $\lst{Y}(p)$ respectively.
(4) Let $(p_n)$ be a sequence of prefixes of any partial HDA over $\Lambda$. Then $X$ has an isolated point over
$(p_n)$ exactly when $Y$ has one, and
$X$ is split over $(p_n)$ exactly when $Y$ is.
(5) $X$ is free of a conclist $U$ exactly when $Y$
is.
\end{corollary}

\begin{proof}
Let $(h,h')$ witness $X\eqB Y$. By Theorems~\ref{tp:lat}, \ref{tp:cpt} and~\ref{tp:lab} $h$ and $h'$ keep
meets, unions and complemented opens and fix unions of label opens, so they carry
generation for (1), which with
Theorem~\ref{thm:unamb}(1) makes $X$ embed exactly when $Y$ does; generation on an open and sheets over
it for
(3). For (4) they carry isolated points and
splits in both directions, since the inverse of $(h,h')$ witnesses
$Y\eqB X$. By Theorems~\ref{tp:pts} and~\ref{tp:homeo} the bijection of points carries labels both ways, which gives (2)
since carrying labels is
transitive.
(5) Each member $p$ of a point $\xi$ has its label word below that of a member of $h_\bullet\xi$, which passes every
conclist that $p$ passes; $h'_\bullet$ acts in reverse.
For finite $X$ and $Y$ over $\Lambda$ with decidable equality Lemma~\ref{prop:image-shape} turns (5) into
invariance of the sequent $\mathrm{pass}_U\vdash\bot$.
\end{proof}

\subsection{Doubling and suspension}\label{ssec:image-double}

\begin{definition}[Bisimilarity]\label{def:image-bisim}\label{def:image-stepbisim}
Let $X$ and $Y$ be partial HDAs over $\Lambda$ and $\rho$ a relation between live prefixes of $X$ and of $Y$ with
$p\mathrel{\rho}q\Rightarrow\lab(p)=\lab(q)$. $X$ and $Y$ are \emph{run-tree bisimilar} when some such $\rho$ is total
on both sides and matches each live extension of $p$ or of $q$ by a live extension of the other related to
it. They are \emph{step bisimilar} when some such $\rho$ relates each live prefix of length $0$
of either to one of the other and matches each live extension of $p$ or of $q$ by one step with an extension of the
other related to it.
\end{definition}

Van Glabbeek extends history-preserving and ST-bisimulation to HDAs \cite{vG06},
and Zouari, van Glabbeek and Ziemia\'nski give the bisimulations of HDAs their modal logics \cite{ZvGZ26}.

$X$ is \emph{label-deterministic} when every prefix carrying the label word of a prefix $p$ extends
$p$.

\begin{proposition}[Doubling]\label{thm:image-double}\label{prop:image-double}
Let $X$ be a partial HDA over any alphabet.
(1) If $X$ is label-deterministic its label opens generate its frame. If $X$
  has a point then $X\dsum X$ is ambiguous. If both hold then $X\eqB X\dsum X$
  fails.
(2) If $\Lambda$ has decidable equality and $X$ and $X\dsum X$ are finite and rigid then they are run-tree
  bisimilar, and each live prefix of either has a live prefix of the other with
  its label word and the label words of its live extensions.
\end{proposition}

\begin{proof}
(1) An open of a label-deterministic machine is the union of the label opens of its
members, and the copies of a point are different points of $X\dsum X$
over one label word; Theorem~\ref{thm:unamb}(1) and
Corollary~\ref{thm:image-profile}(1) separate
them. (2) Relate
a prefix to its copies.
\end{proof}

A \emph{prefix isomorphism} over a map $\zeta$ of conclists from $X$ to $Y$ is a pair of maps
$e\colon\Pref(X)\to\Pref(Y)$ and $e'$ back, monotone for $\ext$ and keeping lengths and acceptance bits; they
replace the shape $U$ of each cell by $\zeta(U)$ and are mutually inverse up to the equivalence $p\ext p'\ext p$
of prefixes. $X$ has \emph{vertex steps} when a proper step between defined cells of shapes $U$ and
$V$ has the ipomset $\starter{V}{V}$ if $U$ is empty and $\terminator{U}{U}$ otherwise, as when its defined cells
have dimension $0$ or a common $n$. The
\emph{suspension} $\Sigma X$ of a graph $X$ over $\ab$ with $n$ vertices and $n$ edges of shape $a$ and possibly
undefined ends replaces each edge by a square of shape $ab$ with its bits and with the ends of the edge as its only
faces, along $d^0_{ab}$ and $d^1_{ab}$. The construction lists the vertices and the edges by one index set $\{0,\dots,n-1\}$, which is why their counts are
equal; padding with cells that neither start nor are faces adds no prefix, so the equal counts lose
no generality.

\begin{proposition}[Suspension]\label{prop:suspension}
(1) Let prefix isomorphisms over one $\zeta$ go from $X_i$ to $X_i'$ for $i=1,2$, partial HDAs over any alphabet.
  Then $\Op(X_i)\eqFrF\Op(X_i')$. If $X_1$ and $X_1'$ have vertex steps then each of
  embedding and a bound $k$ on the fibres holds of $X_1$ exactly when it holds of $X_1'$;
  if $X_2$ and $X_2'$ have them too then each of $\eqB$ and one image holds between $X_1$ and $X_2$ exactly when it
  holds between $X_1'$ and $X_2'$.
(2) Let $\zeta$ exchange the conclists $a$ and $ab$ and fix the others. Every graph $X$ as above and $\Sigma X$
  have vertex steps and a prefix isomorphism over $\zeta$, so that (1) applies with
  $X_i'=\Sigma X_i$.
(3) Prefix isomorphisms over this $\zeta$ take $\wG$ and $\wGt$ with their doubles to $\wQ$ and $\wQ^\circ$ with
  their doubles and $\wG^{(4)}$ to $\wQ^{(4)}$.
  $\wG$ and $\wQ$ with their doubles and $\wG^{(4)}$ and $\wQ^{(4)}$ have vertex steps. The defined cells of
  $\wGt$ and its double have dimension $0$ or $1$, and
  those of $\wQ^\circ$ and its double have dimension $0$ or $2$.
\end{proposition}

\begin{proof}
(1) Covers carry along $e'$, so $e$ and $e'$ carry opens.
With vertex steps the label order depends only on shapes and bits, so label opens carry with points and
image covers.
(2) A prefix passes no undefined face and $d_{a,b}$, $d_{b,a}$ are neither lower nor upper, so $\Sigma X$ has
the steps of $X$.
(3) The cell maps are computed; those into $\wQ$ exchange $s_0$ and $s_1$.
Defined cells have dimension $0$ or $n$ if all shapes listed by the finiteness data do, and $\wGt$,
$\wQ^\circ$ and their doubles have vertex steps.
\end{proof}
\section{Deciding over the label topos}\label{sec:decide}

A relation is \emph{decided} on a class of pairs of finite machines when a function to $\{0,1\}$, built with no
principle from the two machines with their finiteness structures, takes the value $1$ exactly on the related pairs of
the class, and the bits $\dgB$, $\matchB$, $\moB$ and $\pfB$ below are such functions. Each is defined by recursion
over the lists and decided equalities of the two finiteness structures, with every iteration given in advance a number
of rounds computed from those lists, so that a positive decision is an explicit finite search and thus an
algorithm. No bound on their complexity is
proved.

\subsection{Stone duality, plain and labelled}\label{ssec:decide-stone}

\begin{definition}[Clopens; Boolean algebras with decided order]\label{def:dec-clopen}\label{def:dec-bool-alg}\label{def:dec-bool}
An open is \emph{clopen} when it is compact and complemented (Definition~\ref{def:frame}). A \emph{countable Boolean algebra with decided order} is a
set $\alpha$ with a preorder $\leq$ decided on every pair, elements $\bot,\top$ and operations $\wedge,\vee,\neg$
satisfying the Boolean laws up to the equivalence of $\leq$, and a map $\Nn\to\alpha$ meeting every class. A
\emph{typing} of it in a commutative monoid $\mathfrak M$ is a map $\tau\colon\alpha\to\mathfrak M$ constant on classes, with
$\tau(a\vee b)=\tau(a)+\tau(b)$ when $a\wedge b\leq\bot$ and $\tau(a)=0$ exactly when $a\leq\bot$, such that whenever
$\tau(a)=s+s'$ for values $s,s'$ of $\tau$ some $b,c$ with $b\wedge c\leq\bot$ and $b\vee c$ equivalent to $a$ have
$\tau(b)=s$ and $\tau(c)=s'$.
\end{definition}

\begin{lemma}[Back-and-forth]\label{thm:dec-backforth}
Let two countable Boolean algebras with decided order have typings in one commutative monoid with decidable
equality. If the typings take the same values and agree on $\top$, the two preorders are equivalent, with no
principle.
\end{lemma}

\begin{proof}
The relation of equal types meets Vaught's back-and-forth criterion for isomorphism of countable Boolean algebras
\cite{Vaught54};
to build the isomorphism, refine matched partitions by the $n$-th element of each side in turn, splitting partner blocks into pieces of equal
types found by search, and send an element to the join of its blocks' partners; primitive Boolean spaces are
classified by partitions labelled by finite PO systems
\cite{Apps23}.
\end{proof}

Stone duality turns an equivalence of frames into an equivalence of countable Boolean algebras of clopens presented
by finite lists, and $\eqB$ into one that fixes the label opens, so that the back-and-forth of
Lemma~\ref{thm:dec-backforth} applies to machines.

\begin{theorem}[Stone duality over $1$ and over $\BL$]\label{prop:dec-clopen}\label{thm:dec-stone}
Let $\Lambda$ have decidable equality and $X$, $Y$ be finite rigid partial HDAs over $\Lambda$ with closed
acceptance, of any dimension.
\begin{enumerate}[nosep]
\item Up to $\feq{X}$ the clopens of $X$ are the finite unions of cylinders $\upc{p}$, and every open is the least upper
  bound of the clopens below it.
\item Let $X$ be trimmed, and let a length $n$ with a list of sequences of $n+1$ cells present the open of the prefixes
  of length at least $n$ whose first $n+1$ cells form a listed sequence. The order between presented opens is decided
  on the lists; under it these pairs form a countable Boolean algebra with decided
  order, whose presented opens are compact with every cylinder $\feq{X}$-equivalent to one of
  them.
\item $\Op(X)\eqFrF\Op(Y)$ exactly when the preorders of clopens of $X$ and $Y$ are equivalent.
\item $\lst{X}(p)$ is clopen for every prefix $p$ of every partial HDA over $\Lambda$.
\item $X\eqB Y$ exactly when some equivalence of their preorders of clopens sends every clopen
  $\feq{X}$-equivalent to $\lst{X}(p)$, for $p$ a prefix of $X$ or of $Y$, to one $\feq{Y}$-equivalent to
  $\lst{Y}(p)$.
\end{enumerate}
\end{theorem}

\begin{figure}[htb]
\centering
\begin{tikzpicture}[v4,x=1cm,y=1cm]
  \node[v4obj] (lw) at (0,0.9) {$\Op(\LW_\Lambda)$};
  \node[v4obj] (ox) at (-1.9,0) {$\Op(X)$};
  \node[v4obj] (oy) at (1.9,0) {$\Op(Y)$};
  \node[v4obj] (cx) at (-1.9,-1.0) {$\mathrm{Clop}(X)$};
  \node[v4obj] (cy) at (1.9,-1.0) {$\mathrm{Clop}(Y)$};
  \draw[v4map] (lw) -- node[v4lab,above left] {$\lst{X}$} (ox);
  \draw[v4map] (lw) -- node[v4lab,above right] {$\lst{Y}$} (oy);
  \draw[v4map] (ox) -- node[v4lab,above] {$(h,h')$} node[v4lab,below] {$\simeq$} (oy);
  \draw[v4incl] (cx) -- (ox);
  \draw[v4incl] (cy) -- (oy);
  \draw[v4map] (cx) -- node[v4lab,above] {$\simeq$} node[v4lab,below=1pt] {$\lst{X}(p)\mapsto\lst{Y}(p)$} (cy);
\end{tikzpicture}
\caption{On finite rigid machines with closed acceptance over an alphabet with decidable equality, where the frame $\Op(\LW_\Lambda)$ of opens of label words maps to each machine by its label map, $X\eqB Y$ holds exactly when an equivalence of the preorders
$\mathrm{Clop}(X)$ and $\mathrm{Clop}(Y)$ of clopens fixes the label opens (Theorem~\ref{thm:dec-stone}). The bottom row is decided unlabelled (Theorems~\ref{thm:dec-stone}(3) and~\ref{thm:dec-decide}), and labelled when one machine is unambiguous (Theorem~\ref{thm:dec-morita-inj}).}
\label{fig:stone}
\end{figure}
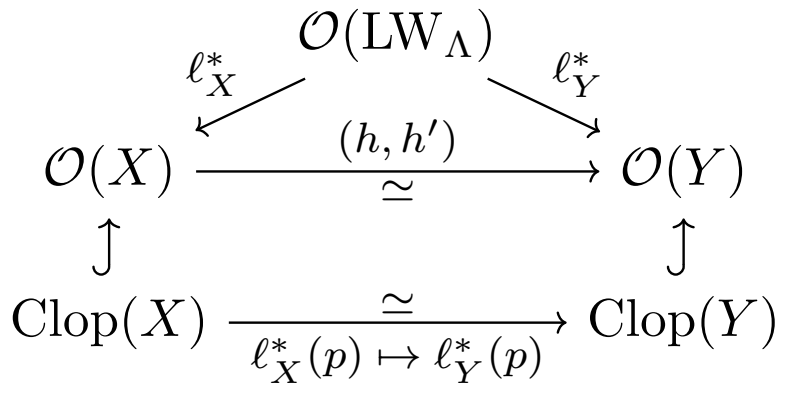

\begin{proof}
Closed acceptance with constant $\Delta$ puts every extension of $q$ of length $\dimn{q}+n+\Delta$ in $\BB(q,n)$, so cylinders
are compact; listed heads compare at the larger
length. In (3) and (5) an equivalence keeps compact opens by Theorem~\ref{tp:cpt}, and in (5)
it fixes the label opens by Theorem~\ref{tp:lab} (Figure~\ref{fig:stone}); conversely
equivalences extend from clopens by unions. $\lst{X}(p)$ is the union of the
cylinders of the finitely many prefixes of length $\dimn{p}$ carrying
$\lab(p)$.
\end{proof}

\subsection{Over $1$}\label{ssec:decide-total}

\begin{definition}[Label-free terms and diagrams]\label{def:dec-terms}\label{def:dec-diagram}
The \emph{label-free terms}
$t::=\top\mid\neg t\mid t\wedge t\mid t\vee t\mid\tcl\,t\mid\tder\,t$ are read at points, $\neg$, $\wedge$ and $\vee$
pointwise; $\tcl\,t$ holds at $\xi$ when every member of $\xi$ lies on a point at which $t$ holds, and $\tder\,t$ when $t$
holds at $\xi$ and $\xi$ is not isolated in the set of points at which $t$ holds (Definition~\ref{def:derived}). The term $t$
\emph{has $k$ points} on $X$ when $X$ has at least $k$ pairwise different points at which $t$ holds,
and it \emph{entails} $t'$ on $X$ when $t'$ holds at every point of $X$ at which $t$ holds. $X$
and $Y$ have \emph{isomorphic diagrams} $X\eqDg Y$ when for all terms $t,t'$ and every $k$ each of `$t$ has $k$ points' and
`$t$ entails $t'$' holds on $X$ exactly when it holds on $Y$.
\end{definition}

Let $X$ and $Y$ be finite over $\Lambda$ with decidable equality and their cells listed without repetition. A
\emph{code} on $X$ is a finite list of sets of cells, and the code of a label-free term follows Head's construction
on the cell graph \cite{Head86}. The code of $\top$ holds the set of all cells and that of
$\neg t$ the set of the cells in no set of the code of $t$, while meets and joins act on the sets of the codes. The code of
$\tcl\,t$ holds the set $\KK$ of the cells reaching a cell on a cycle that lies in a set of the code of
$t$. The code of $\tder\,t$ meets each set of the code of $t$ with the set $\DD$ of the cells
reaching a cell with two distinct successors in $\KK$. The \emph{normal form} of a term on
$X$ records which vectors of values on the listed cells the sets of its code take; normal forms have one length, so
there are finitely many. The \emph{saturation} of the pair starts from the
list $[\top]$ and adds every operation applied to its members until a round adds no new pair of normal forms on $X$
and on $Y$. The \emph{count} of a code is $\omega$ when a cell
on a cycle reached from a start cell lies in $\DD$ of the code; otherwise it is the number of initial segments of
length $n$ of live walks whose last cell lies in $\KK$ of the code, for $n$ the number of listed
cells. For terms $t,t'$ the bit $\mathsf{sub}_X(t,t')$ is $1$
when every cell of $X$ on a cycle reached from a start cell that lies in a set of the code of $t$ lies in a set of the
code of $t'$. The bit $\dgB(X,Y)$ is $1$ exactly when every term of the saturation has one count on $X$ and on $Y$, and every
two terms $t,t'$ of it have $\mathsf{sub}_X(t,t')=\mathsf{sub}_Y(t,t')$.

Deciding equivalence of frames on finite machines makes the frame of the topos over $1$ an invariant computed from
finitely many counts on the cell graph.

\begin{theorem}\label{thm:dec-decide}
Let $\Lambda$ have decidable equality and $X$, $Y$ be finite rigid partial HDAs over $\Lambda$ with closed acceptance,
of any dimension.
(1) If $X\eqDg Y$ then $\Op(X)\eqFrF\Op(Y)$; conversely $\Op(X)\eqFrF\Op(Y)$ implies
$X\eqDg Y$ for all partial HDAs over any alphabet, with no finiteness. (2) One
procedure uniform in the alphabet computes the bit $\dgB(X,Y)$ from two finite partial HDAs, and $\dgB(X,Y)=1$ exactly
when $\Op(X)\eqFrF\Op(Y)$ and exactly when $X\eqDg Y$.
\end{theorem}

\begin{proof}
\emph{Codes compute terms.} A term holds at $\xi$ exactly when $\xi$ eventually stays, up to double negation, in a
set of its code \cite{Head86}, so that the points at which $t$ holds are read off the cell
graph. This rests on the invariant that every set of a code is a union of strongly connected components of the cell
graph, since the set of $\top$ and the sets that closure and derivative produce are such unions and the Boolean
operations keep them. A point eventually stays, up to double
negation, in such a set or in its complement, because its cells end in one component and that component lies wholly
on one side; the set of the cells in no set of the code of $t$ is therefore the code of
$\neg t$. Hence $t$ has $k$ points exactly when the count of its code is at least $k$, and the
double negation of $t$ entails that of $t'$ on $X$ exactly when $\mathsf{sub}_X(t,t')=1$.
Terms with one normal form have the same points.

\emph{The saturation ends and represents every term.} The normal form of $\neg t$, $\tcl\,t$, $\tder\,t$, $t\wedge t'$
or $t\vee t'$ depends only on the normal forms of $t$ and $t'$. A round of the saturation that is not the
last attains a new pair of normal forms among finitely many, so the saturation ends.
Once a round adds no new pair, induction on a label-free term finds a member of the saturation with the pair of normal
forms of the term, and hence with its points on each machine; the saturation then represents every label-free term
on both machines.

(1) By Theorem~\ref{tp:pts},~\ref{tp:der} a frame equivalence carries points injectively with their isolation, and
the evaluation of terms at points is carried, so that
$k$ different points at which $t$ holds go to $k$ such points on the other side and an entailment holds on one side
exactly when it holds on the other.

Conversely let $X\eqDg Y$. Trimming keeps the data sheaves and the diagrams, so by Proposition~\ref{prop:trim} we may
take $X$ and $Y$ trimmed. Let $\mathfrak S$ be the saturation of the pair, so that every label-free term has on each machine the points of some member of
$\mathfrak S$. The frames are compared through the Boolean algebras of
clopens of Theorem~\ref{thm:dec-stone}(2), to which Lemma~\ref{thm:dec-backforth} applies once they carry
typings that take the same values and agree on $\top$.

\emph{The typing.} Type each clopen $x$ by the vector $\tau(x)$ of the counts inside $x$ of the members of $\mathfrak S$, with
entries in $\Nn\cup\{\omega\}$; the vectors add pointwise with $\omega$ absorbing. The count of a
member of $\mathfrak S$ inside $x$ depends only on the points of $x$, and the points of a join are
those of either side, so $\tau$ is constant on classes and adds over disjoint joins; since
$\top$ lies in $\mathfrak S$, $\tau(x)=0$ exactly when $x\leq\bot$. At
the top it is the vector of the counts on the whole machine, so the counts of $X\eqDg Y$ make the two typings agree
on $\top$.

\emph{Germ cells.} A \emph{germ cell} of $x$ is a cell on a cycle reached from the last cell of a listed head of $x$
along cells on no cycle, and its \emph{profile} records for each $t\in \mathfrak S$ whether it lies in
$\KK$ and whether it lies in $\DD$ of the code of $t$. A germ cell gives the count of $t$ the
value $\omega$ in $\DD$, the value $1$ in $\KK$ outside $\DD$ and $0$ outside $\KK$, a vector that depends only on its
profile, and $\tau(x)$ is the sum of these vectors over the germ cells of $x$; so two clopens, on $X$ or on $Y$, whose germ cells carry the
same profiles with the same multiplicities have one type. The \emph{atom} of a profile $P$ is the meet over $t\in \mathfrak S$
of $\tcl\,t$ or $\neg\tcl\,t$ and of $\tder\,\tcl\,t$ or $\neg\tder\,\tcl\,t$ as $P$
prescribes; it is a label-free term, so it has on each machine the points of a member $t_P$ of
$\mathfrak S$.

\emph{Each value of the typing of $X$ is a value of that of $Y$.} Let $x$ be a clopen of $X$. The entry of $\tau(x)$
at $t_P$ counts the points of the atom of $P$ inside $x$, and in the sum of germ vectors each germ cell of profile $P$
gives that entry at least $1$. Hence the atom of $P$ has at least as many points inside $x$ as $P$ occurs among the
germ cells of $x$, and so at least as many points on $X$; the counts of $X\eqDg Y$ give it as
many points on $Y$. Through them, one profile at a time, one finds pairwise disjoint cylinders of prefixes of $Y$
ending at cycle cells, with the profiles of the germ cells of $x$ and the same multiplicities;
pairwise disjoint germ cylinders of one length have as type the sum of the vectors of their last cells, so their union
is a clopen $y$ of $Y$ with $\tau(y)=\tau(x)$. Symmetrically
every value of the typing of $Y$ is one of that of $X$. On the trimmed machines only the
counts of $X\eqDg Y$ enter.

\emph{Splitting.} Let $\tau(a)=s+s'$ for a clopen $a$ of $X$ and values $s=\tau(x)$, $s'=\tau(y)$. The same carving
inside $a$, with the counts read off $\tau(a)$ as they were off $\tau(x)$, gives one germ cylinder inside $a$ for each germ
cell of $x$ and of $y$, pairwise disjoint, those for $x$ of type $s$ and those for $y$ of type
$s'$. Cut what is left of $a$ at a length past one
window of cells, so that each of its cylinders passes a cycle cell $e$ inside it. As
$\tau(a)$ is both $s+s'$ and $s+s'$ plus the type of the rest, and the rest has a point of the atom of $e$, the entry of
$s+s'$ at that atom is $\omega$; so $s$ or $s'$ has $\omega$ there and absorbs the vector of the
cylinder. The absorption holds entry by entry. The vector of the
cylinder is that of $e$, whose entry at a term $t$ is nonzero only where $e$ lies in $\KK$ of the code of
$t$; the entry $\omega$ of the summand at the atom of $e$ comes
from a cell of the summand lying in $\DD$ of the code of that atom,
and that cell lies in $\DD$ of the code of every $t$ whose $\KK$ contains $e$, so that
the summand has $\omega$ wherever the vector of $e$ is nonzero. Joining each such cylinder to the part of a summand that
absorbs it splits $a$ into two disjoint clopens of types $s$ and $s'$, which is the splitting clause, so that $\tau$ is a
typing in the sense of Definition~\ref{def:dec-bool}.

Lemma~\ref{thm:dec-backforth} now makes the preorders of clopens equivalent and Theorem~\ref{thm:dec-stone}(3) gives
the frames.

(2) The bit $\dgB(X,Y)$ compares finitely many counts and bits computed on the two cell graphs for the finitely many
members of the saturation. Since the saturation represents every label-free term, $\dgB(X,Y)=1$ exactly when the
two diagrams have the same counts and the same inclusions of double negations. These
clauses follow from $X\eqDg Y$; conversely they give the frames, since the proof of (1) used only the counts, and the
frames give $X\eqDg Y$ by the first half of (1). The three conditions are therefore
equivalent.
\end{proof}

For $X$ finite over $\Lambda$ with decidable equality and its $n$ cells $c_i$
listed without repetition, $\grph{X}$ is the graph over $\ab$ with vertices and edges $0,\dots,n^2+n$; the edge $ni+j$
has shape $a$ and goes from $i$ to $j$ when a proper step leads from $c_i$ to $c_j$, and every other edge is a loop at
the vertex $n$. Its start vertices are the start cells, and every cell accepts.

\begin{proposition}\label{prop:dec-graph}
Let $X$ be a finite rigid partial HDA over $\ab$ with closed acceptance, of any dimension.
Then $\Op(X)\eqFrF\Op(\grph{X})$.
\end{proposition}

\begin{proof}
A prefix $p$ of $X$ of length $L$ is \emph{read} in $\grph{X}$ as the prefix of length $2L$ that is at time $2t$ at
the vertex $i$ and at time $2t+1$ on the edge $ni+j$, where $c_i$ and $c_j$ are the cells of $p$ at times $t$ and
$t+1$. It starts at a start vertex, and its consecutive cells are joined by
proper steps of $\grph{X}$, since a proper step from $c_i$ to $c_j$ makes $ni+j$ an edge from $i$ to
$j$. Conversely a prefix $z$ of $\grph{X}$ starts at the vertex
of a start cell, and a vertex $i<n$ is left only along an edge $ni+j$ of a proper step, which ends at the vertex $j$;
so $z$ is at a vertex below $n$ at every even time, and its cells at even times, read as cells of $X$, form a prefix of
$X$ of length $\lfloor\dimn{z}/2\rfloor$, its \emph{reading back}.

Reading preserves extension, and on the rigid $X$ it reflects extension, since there the cells of a prefix determine
its steps; reading back and then reading gives a prefix that
$z$ extends. Relate $p$ to the prefixes of length $\dimn{p}$ that its reading extends,
and $z$ to the prefixes $p$ of length $\lfloor\dimn{z}/2\rfloor$ whose reading $z$ extends. Both relations are
monotone, a round trip from a prefix ends at a prefix that it extends, related prefixes are no longer than the
prefix they come from, and every prefix is covered by extensions whose round trips are as long as wanted; a pair of
relations with these four properties is \emph{cover data}. Every cell of $\grph{X}$ accepts,
so $\grph{X}$ has closed acceptance, and cover data between two machines with closed
acceptance give an equivalence of their frames, each half sending an open to the extensions of the prefixes related to
its members.
\end{proof}

Over $\ab$ the decision of Theorem~\ref{thm:dec-decide} therefore reduces to graphs; finite rigid partial HDAs $X$
and $Y$ with closed acceptance have equivalent frames exactly when $\grph{X}$ and $\grph{Y}$ have, and the dimension of
their cells plays no part in it.

\begin{remark}\label{rem:dec-head}
Theorem~\ref{thm:dec-decide} is a point-free and choice-free counterpart of Head's Theorem~3, which decides
homeomorphism of adherences of regular languages by structural diagrams counting cardinals, and
Proposition~\ref{prop:dec-graph} is the counterpart of his Theorem~4, which realises every zero-dimensional compact metrizable space
of finite type as an adherence, by a reduction of frames of every dimension to frames of graphs over $\ab$
\cite{Head86};
we take run spaces as adherences of live finite runs and prove no comparison of points with an adherence.
\end{remark}

The graph $\wB$ over $\ab$ with edges of shape $a$ has two edges from the vertex $0$ to $1$, two loops at $1$ and two
edges from $1$ to $0$; only $0$ starts and accepts. In $\wBh$ every cell accepts.

\begin{theorem}[Invariants over $1$]\label{thm:dec-cb}\label{prop:dec-compact}
(1) If $\Lambda$ has decidable equality then $\Op(X)$ is compact for every finite rigid partial HDA $X$ over $\Lambda$
of any dimension with closed acceptance.
(2) The graphs $\wB$ and $\wBh$ have isomorphic diagrams, $\Op(\wBh)$ is compact and $\Op(\wB)$ is not, and no
equivalence of data sheaves from $\wB$ to $\wBh$ exists.
\end{theorem}

\begin{proof}
(1) Induction on covers with finite branching. (2) Both point spaces are
perfect, and the prefixes of $\wB$ accepting between times $1$ and $k$ cover with
no finite subcover; an equivalence of data sheaves keeps compactness by
Theorem~\ref{tp:cpt} and~\ref{tp:sh}.
\end{proof}

\begin{remark}[Cell graphs]\label{prop:dec-maps}\label{def:dec-celliso}\label{def:dec-maps}
On finite rigid machines over $\Lambda$ with decidable equality and closed acceptance a bijection between sets of
defined cells that contain the start cells and are closed under proper steps, keeping start bits and proper steps both
ways, gives an equivalence of data sheaves whatever cells accept.
Closed acceptance makes each machine equivalent to the one in which every cell accepts, and
between those two the bijection keeps acceptance and carries prefixes to prefixes step by
step.
\end{remark}

\subsection{On embeddings}\label{ssec:decide-inj}

For finite $X$ and $Y$ over $\Lambda$ with decidable equality the subset construction $\matchB(X,Y)$ follows each proper
step of $X$ from a cell $c$ with the list of cells of $Y$ reached along the same labels; it returns $1$ when no pair
reached has $c$ live and no listed cell live, exactly when $Y$ has the live label words of
$X$.

Ambiguity is invariant under $\eqB$, and on pairs with an unambiguous member it is all that $\eqB$ sees beyond the
live label words, which makes $\eqB$ decidable on such pairs.

\begin{theorem}\label{thm:dec-morita-inj}
Let $\Lambda$ have decidable equality and $X$, $Y$ be finite rigid partial HDAs over $\Lambda$ of any dimension and any
acceptance.
(1) If $X$ is unambiguous and $Y$ ambiguous, neither $X\eqB Y$ nor $Y\eqB X$.
(2) If one of them is unambiguous, $X\eqB Y$ exactly when $\moB(X,Y)=1$, where $\moB(X,Y)=1$ when the square search
(Definition~\ref{def:square}) fails on both and
$\matchB(X,Y)=\matchB(Y,X)=1$.
\end{theorem}

\begin{proof}
(1) Label opens generate the frame of $X$ and fail to generate that of $Y$ (Theorem~\ref{thm:unamb}), and
Corollary~\ref{thm:image-profile}(1) carries generation. (2) The search
separates the two cases, and on unambiguous machines $X\eqB Y$ is equality
of live label words (Theorem~\ref{thm:image-live}(4)), while (1) refutes it otherwise. Theorem~\ref{thm:topos-a}
adds Morita equivalence over the label topos.
\end{proof}

Acceptance enters Theorem~\ref{thm:dec-morita-inj} only through the label words, which record at each time
whether the cell accepts, so that equality of live label words compares acceptance together with shapes. When one machine
is unambiguous, $\eqB$ therefore adds to equality of live label words only the requirement that the other is
unambiguous too.

\subsection{The general case}\label{ssec:decide-game}

\begin{definition}[Decision]\label{def:decision}
$\Cls$ is the class of pairs of finite partial HDAs over $\ab$ with their finiteness structures, both rigid and trimmed
with closed acceptance. A \emph{decision} of $\eqB$ on $\Cls$ is a function $d$ from pairs of finite partial HDAs over
$\ab$ with their finiteness structures to $\{0,1\}$ with $X\eqB Y\iff d(X,Y)=1$ on $\Cls$; a \emph{decision on the
ambiguous pairs} asks this only where both members are ambiguous.
\end{definition}

The \emph{labels of the pair} $(X,Y)$ are the opens $\lst{X}(p)$ and $\lst{Y}(p)$ for $p$ a prefix of $X$ or of $Y$.

\begin{definition}[The game]\label{def:game}\label{def:game-strategy}
A \emph{position} is a finite list of blocks $(U,V)$ with $U\in\Op(X)$ and $V\in\Op(Y)$; its \emph{split}
$\splitP(\pi;x,x';y,y')$ replaces each block by $(U\cap x,V\cap y)$ and $(U\cap x',V\cap y')$. A
\emph{strategy} for Duplicator from $\pi_0$ is a set $\mathcal W$ of positions containing $\pi_0$ in which every block has
$U\fle{X}\emptyset$ exactly when $V\fle{Y}\emptyset$. It comes with \emph{answer functions} sending a position and two
opens of one machine to two opens of the other. Every $\pi\in \mathcal W$ has (i)
$\splitP(\pi;\lst{X}(p),x';\lst{Y}(p),y')\in \mathcal W$ for each prefix $p$ of $X$ or $Y$ and all complements $x'$ and $y'$ of the two
label opens; (ii) if $x'$ is a complement of $x$ then the answer $(y,y')$ to $(\pi,x,x')$ has $y'$ a complement of $y$ and
$\splitP(\pi;x,x';y,y')\in \mathcal W$, and symmetrically. Duplicator \emph{wins} when a strategy from $[(\Pref(X),\Pref(Y))]$ exists.
\end{definition}

\begin{theorem}\label{thm:game}
(1) For partial HDAs over every alphabet with no finiteness, $X\eqB Y$ implies that Duplicator
wins. (2) Let $\Lambda$ have decidable equality and $X$, $Y$ be finite rigid trimmed
partial HDAs over $\Lambda$ with closed acceptance, of any dimension. Then $X\eqB Y$ exactly when Duplicator
wins.
\end{theorem}

\begin{proof}
(1) Answer by the equivalence; it keeps meets and joins together with complements by Theorem~\ref{tp:lat} and fixes the
label opens by Theorem~\ref{tp:lab}. (2) Enumerate
the clopens of Theorem~\ref{thm:dec-stone}(2) of both machines with their complements, and let Spoiler split by the
$n$-th clopen of $X$ at stage $2n+1$ and by the $n$-th clopen of $Y$ at stage $2n+2$, and let the strategy answer each
split. After the stage of a clopen each block lies inside it or meets it
in $\emptyset$, and a block of a later stage lies inside a block of an earlier one or meets it in $\emptyset$, on both
sides at once. Let $h(U)$ be the union of the partners of the blocks below $U$ at all
stages, and $h'$ the same from $Y$; both are monotone, so they respect $\feq{X}$ and
$\feq{Y}$. $U$ is the union of the clopens below it, each settled at its
stage, so $U\fle{X}h'h(U)$; conversely a block whose partner lies below $h(U)$ has that partner compact and covered
by the partners of finitely many blocks below $U$, and at a stage after all of them the dichotomy and the matching at
$\emptyset$ put below $U$ each block of that stage inside the first. The forced split by a label open and a
complement keeps a block disjoint from that complement exactly when its partner is, so $h$ fixes the label opens, which are
clopen. Theorem~\ref{thm:topos-a} adds Morita equivalence over the label topos.
\end{proof}

\begin{definition}[Relative type]\label{def:type}
The \emph{frame below} an open $U$ of $X$ is $\Op(X)$ preordered by $A\fle{U}B$ when $A\cap U\fle{X}B$. For families
$l=(l_i)$ and $l'=(l'_i)$ of opens of $X$ and of $Y$ on one index set, $U$ and an open $V$ of $Y$ have \emph{one type}
$U\tyl{l,l'}V$ when an equivalence $(f,g)$ of the frames below $U$ and $V$ has each $f(l_i)$ $\fle{V}$-equivalent to
$l'_i$. We write $\tyl{l}$ for $\tyl{l,l}$, and $U\tyeq V$ when $l$ and $l'$ are the labels of the pair.
\end{definition}

\begin{theorem}\label{thm:types}
(1) For partial HDAs over every alphabet with no finiteness, $X\eqB Y$ exactly when
$\Pref(X)\tyeq\Pref(Y)$. (2) If $U_1\cap U_2\fle{X}\emptyset$ and
$V_1\cap V_2\fle{Y}\emptyset$, with $U_1\tyl{l,l'}V_1$ and $U_2\tyl{l,l'}V_2$, then $U_1\cup U_2\tyl{l,l'}V_1\cup
V_2$. (3) Under the hypotheses of Theorem~\ref{thm:game}(2), clopens $U$ of $X$ and $V$ of
$Y$ have $U\tyeq V$ exactly when Duplicator wins the game played on the frames below $U$ and
$V$.
\end{theorem}

\begin{proof}
(1) Below the top the order is that of the frame. (2) Glue $f(A)=(f_1(A)\cap
V_1)\cup(f_2(A)\cap V_2)$ and $g$ symmetrically. (3) is Theorem~\ref{thm:game} below two
clopens.
\end{proof}

\begin{lemma}[Germ lemma]\label{lem:germ}\label{prop:dec-germ}
Let $X$ and $Y$ be rigid partial HDAs over any alphabet, of any dimension, with no finiteness. For prefixes $p,q$ of $X$
an extension $r'$ of $q$ is the \emph{shift} of an extension $r$ of $p$ when $\dimn{r'}-\dimn{q}=\dimn{r}-\dimn{p}$ and
$r'_{i-\dimn{p}+\dimn{q}}=r_i$ for $\dimn{p}\leq i\leq\dimn{r}$.
(1) If $p$ and $q$ end at one cell, the shift gives an equivalence of the frames below $\upc{p}$ and below $\upc{q}$,
its image and preimage read along the shift.
(2) For prefixes $p_X,q_X$ of $X$ and $p_Y,q_Y$ of $Y$ with $\dimn{p_X}=\dimn{p_Y}$, $\dimn{q_X}=\dimn{q_Y}$ and
$\lab(q_X)\lle\lab(q_Y)$, the shifts $r'$ of $r$ from $p_X$ to $q_X$ and $s'$ of $s$ from $p_Y$ to $q_Y$ have
$\lab(r')\lle\lab(s')$ when $\lab(r)\lle\lab(s)$.
(3) If $p_1,q_1$ are prefixes of $X$ ending at one cell and $p_2,q_2$ prefixes of $Y$ ending at one cell, with
$\lab(p_1)=\lab(p_2)$ and $\lab(q_1)=\lab(q_2)$, then $\upc{p_1}\tyeq\upc{p_2}$ implies
$\upc{q_1}\tyeq\upc{q_2}$.
(4) If $\Pref(X)\fle{X}l_i$ for every member of a family $l$ of opens of $X$, prefixes $p,q$ of $X$ ending at one cell
have $\upc{p}\tyl{l}\upc{q}$.
\end{lemma}

\begin{proof}
On a rigid machine steps are read off cells, so the shift is a correspondence of extensions keeping covers and the
label order. Conjugating by it gives
(3), and (4) holds as frame equivalences keep the opens above the top.
\end{proof}

On $\Cls$ one type of cylinders over one label word is thus a relation
between last cells, defined by a quantifier over prefixes. A partial HDA is
\emph{fibre-perfect} when through each member of each point $\xi$ passes a point over one label word with $\xi$
(Definition~\ref{def:fibres-carry}) having a member outside $\xi$; it \emph{carries one
label word} when live prefixes $p,q$ with $\dimn{p}\leq\dimn{q}$ have $\lab(p)\lle\lab(q)$. For
finite $X$ and $Y$ over $\Lambda$ with decidable equality the bit $\pfB(X,Y)$ is $1$ when both or neither have a live
start cell and live cells reached together from live start cells have one label, with one ipomset on their steps to
live successors.

Brouwer's characterisation of Cantor space compares frames with no count once branching is bounded, and over one
label word it makes $\eqB$ a check of labels and steps on cells reached together.

\begin{theorem}[Brouwer over $1$ and over one label word]\label{thm:dec-brouwer}\label{cor:dec-grow}
Let $\Lambda$ have decidable equality and $X$, $Y$ be finite rigid partial HDAs over $\Lambda$ of any dimension,
presented on lists of their cells without repetition. Say that $N$ \emph{bounds branching} in $X$ when for every $i$
each live prefix of length at least $i+N$ has a cell with two distinct live successors at one of the positions
$i,\dots,i+N-1$.
(1) If $X$ and $Y$ are trimmed with closed acceptance and
some $N_1$ and $N_2$ bound branching in $X$ and in $Y$, and $X$ has a prefix exactly when $Y$ does, then
$\Op(X)\eqFrF\Op(Y)$.
(2) If $X$ is fibre-perfect, one more than the number of its listed
cells bounds branching in $X$.
(3) If $X$ and $Y$ are fibre-perfect of any acceptance
and $X$ carries one label word, then $X\eqB Y$ exactly when $\pfB(X,Y)=1$, and so does Morita equivalence over the
label topos.
(4) Let $\wQtt=\Wq{2}{3}{\{s_0,s_1,s_2\}}$ (Example~\ref{exa:dec-witnesses}). Then
$\wQt\eqB\wQtt$.
\end{theorem}

\begin{proof}
(1) This is a frame form of Brouwer's characterisation of Cantor space \cite{Brouwer10}. Bounded branching makes inhabitation a typing of the clopens of Theorem~\ref{thm:dec-stone}(2) in
$\{0,\omega\}$ with $\omega+\omega=\omega$, and Lemma~\ref{thm:dec-backforth} applies. (2) A
branch-free window of that length repeats a cell, and forced continuations then contradict fibre-perfection. (3)
By Theorem~\ref{tp:pts} and~\ref{tp:homeo} points carry labels both
ways. Conversely code a live
prefix by the choices it makes among the live start cells and then among the live successors of each of its cells,
the $j$-th of $k$ choices written $1^j0$ for $j<k-1$ and the last written $1^{k-1}$, so that two choices from one list without repetition whose codes are comparable coincide, and every word of bits
of length at least $k-1$ begins with the code of some choice, and relate a live prefix $p$ of $X$ to the longest live prefix of $Y$ of
length at most $\dimn{p}$ whose code is an initial segment of that of $p$, and symmetrically. On
rigid machines these relations are monotone, a round trip from $p$ ends at a prefix that $p$ extends, and $\pfB(X,Y)=1$ makes
them carry labels;
by the bound of (2) on branching every prefix is covered by prefixes whose round trips are as long as
wanted. Interpreting $\fat{q}$ as the join of the $\fat{p}$ with
$p$ related to an extension of $q$, and symmetrically, gives $X\eqM Y$, hence $X\eqB Y$ by
Theorem~\ref{thm:topos-a}.
(4) The converse half of (3) uses fibre-perfection only through the bound of (2), and so holds for rigid machines
on lists of cells without repetition whenever some bound on branching holds in each and $\pfB(X,Y)=1$, the bit
supplying the labels and the points. For $\wQt$ and $\wQtt$ these four hypotheses are
checked by a finite computation on their listed shapes and cells. Between any two listed shapes there is at most one
lower and at most one upper coface, which makes the machines rigid; the lists of cells
have no repetition; every walk of live successors from a live cell meets a cell with two
distinct live successors among its first $4$ cells, so $4$ bounds branching; and
$\pfB(\wQt,\wQtt)=1$. They give Morita equivalence over the
label topos, hence $\wQt\eqB\wQtt$.
\end{proof}

\begin{definition}[Absorption, copies and cycles]\label{def:absorb}\label{def:absorb-cycle}
Let $l$ be a family of opens of $X$. An open $U$ \emph{absorbs} $V$ when $U\cap V\fle{X}\emptyset$ and $U\cup V\tyl{l}U$.
\emph{Copies} of $V$ in $U$ are opens $c_0,c_1,\dots\fle{X}U$ with $c_i\cap c_j\fle{X}\emptyset$ for $i<j$ and $c_j\tyl{l}V$.
For prefixes $p\ext p'$ the \emph{rest} $\rest(p,p')$ is the open of the extensions $q$ of $p$ with $q_i\neq p'_i$ for some
$i\leq\min(\dimn{q},\dimn{p'})$. A \emph{cycle} inside $U$ with \emph{base} $p$ and \emph{piece} $b$ is a pair $p\ext p'$ with
$\dimn{p}<\dimn{p'}$, $p_{\dimn{p}}=p'_{\dimn{p'}}$ and $\upc{p'}\fle{X}U$, with a complemented open
$b\fle{X}\rest(p,p')$.
\end{definition}

\begin{theorem}\label{thm:absorb}
Let $\Lambda$ have decidable equality and $X$ be a finite rigid trimmed partial HDA over $\Lambda$ of any dimension with
closed acceptance, with clopens $U$ and $V$. (1) For every family $l$ of opens, copies of $V$ in
$U$ give finitely many cycles inside $U$ whose bases have pairwise disjoint cylinders and whose pieces have a union
$\tyl{l}V$. (2) If $\Pref(X)\fle{X}l_i$ for every $i$ and
$U\cap V\fle{X}\emptyset$, then $U$ absorbs $V$ exactly when there are copies of $V$ in $U$, and exactly when there are
cycles as in (1).
\end{theorem}

\begin{proof}
(1) The bases are the minimal prefixes above $U$ of bounded length with a longer extension ending at their last cell;
by pigeonhole they partition long prefixes above $U$. Pumping the return of a base $w$ gives
prefixes $\rho_w$ above $w$ ending at its last cell, as long as wanted, with $\upc{\rho_w}\fle{X}U$. Present one more
pairwise disjoint copy of $V$ than there are bases by prefixes of one length $M$. If each copy had a prefix agreeing with
some $\rho_w$ up to time $M$, two copies would share a $\rho_w$, and their two prefixes, which have the same cells, would
put one inhabited cylinder inside two disjoint copies. So some copy has no prefix on any $\rho_w$; each of its prefixes
extends a base $w$ and leaves $\rho_w$ by time $M$, so lies in $\rest(w,\rho_w)$, and the pieces, its prefixes above
each base, have as union this copy, of the type of $V$.
(2) Iterate $x\mapsto f(x)\cap U$ on $V$ for $(f,g)$
witnessing $U\cup V\tyl{l}U$. Conversely pumping a cycle $p\ext p'$ gives
prefixes repeating $p'$ from time $\dimn{p}$ on, and the rests between consecutive ones are pairwise disjoint clopens
inside $\upc{p'}$, carried onto each other by the shift and of one type by Lemma~\ref{lem:germ}(4), which converge to
the run of the pumped prefixes. Restricted to the pieces and joined over the cycles they
give copies of $V$ converging to finitely many such runs, and Hilbert's hotel, which
sends $V$ to the first copy and each copy to the next while it fixes the limits, is an equivalence over the labels of the
frames below $U\cup V$ and below $U$.
\end{proof}

In $\wC{1}$ and $\wCh{1}$, $\kappa_0\ext\kappa_2$ is a cycle inside $\upc{\kappa_2}$ with clopen piece
$\rest(\kappa_0,\kappa_2)$.

\begin{proposition}\label{prop:flip}
(1) Under the labels of $(\wC{1},\wC{2})$ read in $\wC{1}$ the open $\upc{\kappa_2}$ absorbs $\rest(\kappa_0,\kappa_2)$,
which does not lie below the empty open. (2) In $\wCh{1}$ the prefix $\kappa_2$ extends
$\kappa_0$, and under the labels of $(\wCh{1},\wCh{2})$ read in $\wCh{1}$ the open $\upc{\kappa_2}$ does not absorb
$\rest(\kappa_0,\kappa_2)$.
\end{proposition}

\begin{proof}
On $\wC{1}$ only vertices accept, so the labels are above the top and Theorem~\ref{thm:absorb}(2)
applies. On $\wCh{1}$ the label open of $\kappa'_2$ meets the rest and misses
$\upc{\kappa_2}$, while one type keeps which label opens meet an
open.
\end{proof}

\subsection{Separation}\label{ssec:decide-sep}\label{ssec:image-separation}

Let $(a\parallel b)^\omega$ be the $\omega$-ipomset $\Theta_0$ of Section~\ref{sec:muller}. The graphs
of items (5) to (8) are over $\ab$ with edges of shape $a$, and in items (5) to (7) every vertex accepts.
$\mathsf{L}_3=\Wq{1}{3}{\{v,e_2\}}$ and $\mathsf{L}_4=\Wq{1}{4}{\{v,e_2,e_3\}}$, with $\mathfrak p_n$ the prefix of length $n$ of
the run of $\mathsf{L}_3$ round $e_2$. $\KA$ starts at $p_0$ and $q_1$
with rejecting cycles $p_0\to p_1\to p_0$ and $q_1\to q_0\to q_1$ and accepting edges $p_0\to t$, $q_0\to t$ and $t\to t$.
$\KB{1}$ starts at $r$ and $s$ with a rejecting loop at each and accepting edges $r\to t$ and $t\to t$; $R_n$ is
the prefix of length $n$ of the loop at $s$. $\mathsf{H}$ has
vertices $0$ to $5$ with loops at $1$ and $2$ and the edges $0\to1$, $0\to2$, $3\to1$ and $4\to2$; its edges $0\to2$
and $4\to2$ accept. $\mathsf{H}_1$ starts it at $0$ and $\mathsf{H}_2$ at $3$ and
$4$. $\mathsf{L}_1=\Wq{1}{1}{\{v\}}$, and
$\mathsf{L}_2$ has two start vertices with an edge each way and no accepting edge.
$\mathsf O_1$ and $\mathsf O_2$ have loops at vertices $u$ and $w$ and an edge $u\to w$; $u$ starts, and every cell
accepts in $\mathsf O_1$ and $w$ alone in $\mathsf O_2$.

A label word records at each time the shape of a cell and whether it accepts; the label topos keeps this record,
and a B\"uchi language keeps the partially ordered events alone. In item (1) the frames of $\mathsf P$ and $\mathsf S$
already differ, since $\mathsf P$ has one run and $\mathsf S$ branches at every round, whereas $\wQ^\circ$ has the B\"uchi language and the frame of $\mathsf S$, which identify the two, while their step
words already differ and the label topos separates them; $\mathsf S$ passes a cell of shape $a$
and $\wQ^\circ$ has none, and the sequent $\mathrm{pass}_a\vdash\bot$ is this difference of shapes read in the topos (Figure~\ref{fig:separation}). In item (4) the two
machines differ only in which cells accept. Items (2), (3), (5) and (6) pair machines with the same live label
words, and in each of them $\eqB$ separates the two machines by their points that read one infinite sequence of labels. The pairs of dimension two in items (2) and
(3) receive their separations from pairs of graphs through the prefix isomorphisms of
Proposition~\ref{prop:suspension}(3).

\begin{theorem}[Separation]\label{thm:separation}\label{thm:image-sq}\label{thm:image-concurrency}\label{thm:image-incomparable}\label{prop:lang-not-image}\label{thm:pair}
\begin{enumerate}[nosep]
\item (\emph{concurrency}) $\mathsf P$ and $\mathsf S$ are finite, and both have a cell of
  shape $ab$ and a point. $\LB{\mathsf P}=\LB{\mathsf S}\ni(a\parallel b)^\omega$, and a valid $\omega$-ipomset
  (Section~\ref{ssec:muller-events}) lies
  in $\LB{\mathsf S}$ exactly when it is isomorphic to $(a\parallel
  b)^\omega$. No partial HDA over $\ab$ whose cells all have
  dimension at most one has $(a\parallel b)^\omega$ in its B\"uchi
  language. $\matchB(\mathsf S,\mathsf P)=0$, so $\mathsf S$ has a live label word
  that $\mathsf P$ lacks. $\mathrm{pass}_a\vdash\bot$ holds in $\mathsf P$ and fails in
  $\mathsf S$, and $\mathsf P\eqB\mathsf S$ fails; neither
  $\Op(\mathsf P)\eqFrF\Op(\mathsf S)$ nor $\Op(\wQ^\circ)\eqFrF\Op(\mathsf P)$ holds.
  $\wQ^\circ$ and $\mathsf S$ are finite, rigid and trimmed with closed acceptance of constants $2$ and $4$, and both have
  a cell of shape $ab$. $\LB{\wQ^\circ}=\LB{\mathsf S}$ and
  $\Op(\wQ^\circ)\eqFrF\Op(\mathsf S)$, while
  $\mathrm{pass}_a\vdash\bot$ holds in $\wQ^\circ$ and fails in $\mathsf S$ and $\wQ^\circ\eqB\mathsf S$
  fails. The ST-word whose first step starts $a$ alone is the ST-word of a PAF-accepted
  $\omega$-track of $\mathsf S$, while no live prefix of $\wQ^\circ$ reads it; so
  $\LangA{\wQ^\circ}\neq\LangA{\mathsf S}$, $\LangPAF{\wQ^\circ}\neq\LangPAF{\mathsf S}$ and
  $\LangLp{\wQ^\circ}\neq\LangLp{\mathsf S}$, and $\ImaST\wQ^\circ\neq\ImaST\mathsf S$ and
  $\Ima\wQ^\circ\neq\Ima\mathsf S$.
\item (\emph{ambiguity, suspended}) $\wQ$ is finite and so is $\wQ\dsum
  \wQ$. Both are rigid with closed acceptance of constant $2$, with one language of the points and
  one accepted and one PAF-accepted language. They are run-tree bisimilar with
  $\Op(\wQ)\eqFrF\Op(\wQ\dsum \wQ)$ and $\Ima
  \wQ=\Ima(\wQ\dsum \wQ)$. $\wQ$ is unambiguous and
  embeds, while $\wQ\dsum \wQ$ does not embed and
  has fibres bounded by $2$ and not by $1$;
  $\wQ\eqB \wQ\dsum \wQ$ fails.
\item (\emph{sheets, suspended}) $\wQ\dsum \wQ$ and $\wQ^{(4)}$ are ambiguous and run-tree bisimilar with the same
  live label words and $\Op(\wQ\dsum \wQ)\eqFrF\Op(\wQ^{(4)})$. $\wQ\dsum \wQ$ has $2$ sheets over its prefixes and not
  $1$, and $\wQ^{(4)}$ has $4$ and not $3$ with fibres bounded by $4$ and not by $3$;
  $\wQ\dsum \wQ\eqB \wQ^{(4)}$ fails.
\item (\emph{acceptance bit, dimension two}) $\wQt$ is finite and so is
  $\wQ$. Both are rigid with closed acceptance of constant $2$, with $\LangA{\wQt}=\LangA{\wQ}$ and $\ImaST\wQt=\ImaST \wQ$ while $\Ima\wQt=\Ima \wQ$ fails; so the converse of the
  last implication of Theorem~\ref{thm:image-live}(1) fails.
\item (\emph{profile, on graphs}) $\mathsf L_3$ is finite and so is
  $\mathsf L_4$. Both are ambiguous rigid trimmed graphs with closed acceptance of constant $2$ and have the same
  ambiguity profile, a degenerate one in which no $\mathsf L_i$ has $k$ sheets over $\Pref(\mathsf L_i)$ for any
  $k$, the same live label words and $\Op(\mathsf L_3)\eqFrF\Op(\mathsf L_4)$. $\mathsf L_4$ is split over
  $(\mathfrak p_n)$ and $\mathsf L_3$ is not; neither
  $\mathsf L_3\eqB\mathsf L_4$ nor $\mathsf L_4\eqB\mathsf L_3$ holds.
\item (\emph{isolated point, on graphs}) $\KA$ is finite and so is
  $\KB{1}$. Both are ambiguous rigid trimmed graphs with closed acceptance of constant $2$, fibres bounded by $2$ and the same live label words. Only $\KB{1}$ has an isolated point over
  $(R_n)$; neither $\KA\eqB\KB{1}$ nor $\KB{1}\eqB\KA$ holds.
\item (\emph{bisimilarity, on graphs}) $\mathsf H_1\eqB\mathsf H_2$ holds and $\mathsf H_1$ and $\mathsf H_2$ are
  not step bisimilar. $\mathsf L_1$ and $\mathsf L_2$ are step bisimilar and $\mathsf L_1\eqB\mathsf L_2$
  fails.
\item (\emph{Booleanness, on graphs}) $\LB{\mathsf O_1}=\LB{\mathsf O_2}$ is inhabited. $\mathsf O_2$ is discrete and
  $\mathsf O_1$ is not, and $\Omega_{\mathsf O_1}$ is not Boolean while $\Omega_{\mathsf O_2}$ is Boolean under $\EM$
  alone.
\end{enumerate}
\end{theorem}

\begin{proof}
(1) The track round $v\to s_0\to v$ is accepted in both, and each of the four ways a round of $\mathsf S$
enters and leaves $s_0$ reads $a\parallel b$;
a track of dimension at most one orders its events.
$\mathsf P$ has no cell of shape $a$ while an accepted track of $\mathsf S$ passes $e_a$, and
Lemma~\ref{prop:image-shape} with Corollary~\ref{thm:image-profile}(5)
concludes.
A frame equivalence carries isolated points, and $\mathsf P$ has one while $\wQ^\circ$ and $\mathsf S$ have none.
$\wQ^\circ$ and $\mathsf S$ are rigid and trimmed with closed acceptance and $2$ bounds branching in each, so that
Theorem~\ref{thm:dec-brouwer}(1) gives $\Op(\wQ^\circ)\eqFrF\Op(\mathsf S)$; the
round $v\to s_0\to v$ of $\wQ^\circ$ reads $a\parallel b$ as in $\mathsf S$, and since $\wQ^\circ$ has no cell of
shape $a$ the argument for $\mathsf P$ separates $\wQ^\circ$ from $\mathsf S$.
The target interface of a step is the shape of the cell it enters, so in a machine
with no cell of shape $a$ the empty sieve covers every step word with a step into shape $a$;
the first step of the staircase track of $\mathsf S$ enters $e_a$, and its step word is live in $\mathsf S$ and covered
by the empty sieve in the image of $\wQ^\circ$.
(2) The cell graphs give rigidity and closed acceptance with a trimmed label-deterministic $\wG$ that has a
point; a track of a double stays in one copy, and $\wG$ and $\wG\dsum\wG$ branch within a bounded window and match their
live label words.
Proposition~\ref{thm:image-double} and Theorems~\ref{cor:dec-grow}(1), \ref{thm:image-live}(2) and~\ref{thm:unamb}
make $\wQ$ and $\wQ\dsum \wQ$ run-tree bisimilar and separate $\wG$ from $\wG\dsum\wG$, and Proposition~\ref{prop:suspension}(3)
carries the separation.
(3) Copies of a point refute fewer sheets, since the labels settle the cells on each copy of
$\wQ$;
Corollary~\ref{thm:image-profile}(3) with Proposition~\ref{thm:image-double}(2) and Theorem~\ref{cor:dec-grow}(1)
concludes, as does
Proposition~\ref{prop:suspension}(3) from $\wG\dsum\wG$ and $\wG^{(4)}$.
(4) The machines differ only in their accepting cells, and under closed acceptance a track is accepted exactly when it
starts; so
Theorem~\ref{thm:image-live}(2) gives one image over the step
words;
over the label words a live prefix of $\wQt$ at the rejecting $v$ has no partner in
$\wQ$.
(5) Padding with $k$ rounds of $e_0$ and swapping one for $e_1$ gives $k+1$ points over one label word through
every prefix, so no inhabited open has $k$ sheets, and either machine has $k$ sheets over a label open exactly when
that open is empty, which by the matched live label words happens in both machines at once;
the first edge splits $\mathsf L_4$ over $(\mathfrak p_n)$ while in
$\mathsf L_3$ a prefix carrying the label word of $\mathfrak p_n$ has the cells of $\mathfrak p_n$, since $e_2$ is the only accepting
loop.
(6) The square search rooted at either start vertex fails, so two of any three points over one label word share a
root and are equal.
The prefixes of $\KB{1}$ rooted at $s$ form an inhabited complemented open inside every $\lst{\KB{1}}(R_n)$, while
in $\KA$ a member of such an open would extend to a live prefix ending at an accepting
edge.
(7) Past time $0$ the prefixes of $\mathsf H_1$ and $\mathsf H_2$ correspond with their
labels,
while the root $3$ of $\mathsf H_2$ matches no step of $\mathsf H_1$ along the accepting edge
$0\to2$. Relating live prefixes with one label word is a step bisimulation of
$\mathsf L_1$ and $\mathsf L_2$, and $\mathsf L_2$ has two points over the label word
of the single point of $\mathsf L_1$.
(8) Every track reads $a^\omega$, and the loop at $u$ is accepted in $\mathsf O_1$ only and is a limit of tracks
crossing to $w$.
Since $\mathsf O_1$ is not discrete, $\Omega_{\mathsf O_1}$ is not Boolean by
Proposition~\ref{thm:setup-truth}. Past $w$ the extensions of a prefix of $\mathsf O_2$ agree, so $\EM$ on membership in a closed sieve makes
$\Omega_{\mathsf O_2}$ Boolean directly, with no appeal to $\DC$.
\end{proof}

\begin{figure}[htb]
\centering
\begin{tikzpicture}[v4,x=1cm,y=1cm]
  \node[v4obj] (lw) at (0,1.0) {$\Op(\LW_\Lambda)$};
  \node[v4obj] (oq) at (-2.2,0) {$\Op(\wQ^\circ)$};
  \node[v4obj] (os) at (2.2,0) {$\Op(\mathsf S)$};
  \draw[v4map] (lw) -- node[v4lab,above left] {$\lst{\wQ^\circ}$} (oq);
  \draw[v4map] (lw) -- node[v4lab,above right] {$\lst{\mathsf S}$} (os);
  \draw[v4map] (oq) -- node[v4lab,below] {$\simeq$} (os);
  \foreach \x/\m/\r/\n in {-2.2/q/{\cong}/{\wQ^\circ},2.2/s/{\ncong}/{\mathsf S}} {
    \node[v4obj,anchor=east] at (\x-1.3,-1.0) {$\n$};
    \node[v4obj] (z\m) at (\x-0.85,-1.0) {$\sbt{\emptyset}$};
    \node[v4obj] (p\m) at (\x+0.55,-1.0) {$\sbt{\mathrm{pass}_a}$};
    \draw[v4mono] (z\m) -- node[v4lab,above] {$\r$} (p\m);
  }
\end{tikzpicture}
\caption{The bouquet $\wQ^\circ$ and the interleaved square $\mathsf S$ have one B\"uchi language, $\LB{\wQ^\circ}=\LB{\mathsf S}$, and the horizontal edge is an
equivalence of their frames, while no equivalence of the frames carries $\lst{\wQ^\circ}$ to $\lst{\mathsf S}$, so that
$\wQ^\circ\eqB\mathsf S$ fails
(Theorem~\ref{thm:separation}(1) and Definition~\ref{def:labels-frame}). The witness is the subterminal $\sbt{\mathrm{pass}_a}$ of the prefixes passing a
cell of shape $a$, least in $\wQ^\circ$ and not least in $\mathsf S$ (Theorem~\ref{thm:frame-sub}); the step words of the two
machines already differ at the first step of $\mathsf S$ that starts $a$ alone.}
\label{fig:separation}
\end{figure}

\subsection{The reduction and the acceptance bit}\label{ssec:decide-reduce}

With $\EM$ and $\UC$ every relation on $\Cls$ has a decision.

The reduction confines the question of a decision with no principle to the ambiguous pairs, and the game and the
relative types carry that question unchanged.

\begin{theorem}\label{thm:decide}
Each of (1) to (3) is equivalent to the existence of a decision of $\eqB$ on $\Cls$.
(1) A decision on the ambiguous pairs of $\Cls$ exists. (2) Some $d$ as in
Definition~\ref{def:decision} has $d(X,Y)=1$ exactly when Duplicator wins, on $\Cls$.
(3) Some $d$ as in Definition~\ref{def:decision} has $d(X,Y)=1$ exactly when $\Pref(X)\tyeq\Pref(Y)$, on
$\Cls$.
\end{theorem}

\begin{proof}
(1) A decision on $\Cls$ restricts; conversely apply the given decision where the square search finds both members
ambiguous and $\moB$ elsewhere (Theorem~\ref{thm:dec-morita-inj}(2)). (2) and
(3) Theorems~\ref{thm:game}(2) and~\ref{thm:types}(1) carry one function across pair by pair.
\end{proof}

\begin{openproblem}\label{oprob:decide}
Is there a decision of $\eqB$ on $\Cls$ with no principle?
\end{openproblem}

An open $U$ of $X$ is \emph{idempotent} when it is $\feq{X}$-equivalent to the union of two disjoint opens $U_1$ and
$U_2$ with $U_1\tyl{l}U$ and $U_2\tyl{l}U$, for $l$ the labels of the pair read in $X$.

Let $X$ be finite over $\Lambda$ with decidable equality. A \emph{state} is a pair reached by the subset construction
$\matchB(X,X)$, a cell $c$ with the list of cells reached along the label word of a walk to $c$.
Let $\mathcal R$ be a strongly connected component of the cell graph whose cells are live and lie on cycles reached from a start
cell. It is \emph{perfect} when one of its cells has two successors in $\mathcal R$, and
\emph{absorbed} when some state with its cell in $\mathcal R$ lists a live cell on a cycle that reaches $\mathcal R$ and is not reached
from $\mathcal R$. The \emph{label set} of $\mathcal R$ holds for each state $(c,L)$ with $c$ in $\mathcal R$ the shape
of $c$ and whether $c$ accepts, with the number of distinct live cells in $L$ cut off at $2$.
The \emph{entry count} of $\mathcal R$ adds the number of its start cells to the number of proper steps into it from live cells
outside it reached from a start cell. The \emph{type} of $\mathcal R$ is its label set with its
perfection and absorption, together with the type of each component entered from $\mathcal R$ and the number of walks that
leave $\mathcal R$ and enter that component through live cells on no cycle. Types add in the commutative
monoid in which $t+t=t$ and $s+t=t$ for every proper part $s$ of $t$ when $t$ is the type of a perfect
component. The \emph{top} of $X$ is the sum of the types of the components, each taken as many
times as it is entered from the start cells. A
\emph{matching} of a list of prefixes of $X$ with one of $Y$ in blocks is a list of pairs of blocks, a block of each
list in each pair, whose first members together are a reordering of the first list and whose second members are one of
the second, so that whether a matching exists does not depend on the order of either list.

\begin{proposition}\label{prop:presentations}
(1) (\emph{series}) Over any alphabet $X\eqB Y$ follows from a matching of the prefixes of one length in blocks of one
type, of equal sizes unless the block of $X$ has an idempotent cylinder and that of $Y$ is
nonempty. Although $\wC{1}\eqB\wC{2}$, no such matching exists for $\wC{1}$ and $\wC{2}$ at
any $n\geq1$ on the lists of the prefixes of length $n$ on the spine and off it.
(2) (\emph{fixpoints}) On $\wC{1}$ against itself the greatest relation over one label word matching one-step extensions one to one
after dropping absorbed ones relates $\kappa'_2$ to $\kappa_2$, whose cylinders differ in type;
dropping none it misses the roots of $\wC{1}$ and $\wC{2}$, which have one type over
each other's labels.
(3) (\emph{component poset}) The \emph{component poset} of a finite machine over an alphabet with decidable equality
orders by reachability the strongly connected components of live cells on cycles reached from a start
cell. Each node carries its label set, perfection, absorption and
multiplicity. Its multiplicity is $0$ when it is absorbed, the least of $1$ and its entry
count when it is perfect and not absorbed, and its entry count otherwise. The component poset
has one node for $\wQt$ and two for its double $\wQt\dsum\wQt$, which
is $\eqB$-equivalent to $\wQt$. It has one node for $\wGt$ and two for its double
$\wGt\dsum\wGt$, which is $\eqB$-equivalent to $\wGt$.
(4) (\emph{types of components}) The tops of $\wQt$ and $\wQt\dsum\wQt$ differ while
$\wQt\eqB\wQt\dsum\wQt$. Comparison of tops separates $\wGt$ from $\wGt\dsum\wGt$ while $\wGt\eqB\wGt\dsum\wGt$. It identifies
$\wQ\dsum\wQ$ with $\wQ^{(4)}$ while $\eqB$ fails between them; it identifies
$\wG\dsum\wG$ with $\wG^{(4)}$ while $\eqB$ fails between them.
\end{proposition}

\begin{proof}
(1) Blocks glue by Theorem~\ref{thm:types}(2). A cover correspondence gives $\Op(\wC{1})\eqFrF\Op(\wC{2})$, and on
these graphs every prefix carries the label word of every prefix of either machine no longer than
itself; each label open is then the top open, so the frame
equivalence fixes the label opens and $\wC{1}\eqB\wC{2}$. The cylinder of $\kappa_n$ splits and that of a prefix off the spine does not, so $\wC{2}$
has twice as many prefixes off the spine as $\wC{1}$, against (1).

(2) On $\wC{1}$ against itself relate, for every $n\geq1$, the prefix of length $n$ of the run that leaves the spine
at once to the spine prefix $\kappa_n$. This relation is a post-fixpoint of the operator that
drops absorbed extensions. The first prefix has one one-step extension, and $\kappa_n$ has the spine and, at even
times, an exit, which the spine absorbs at every round by Lemma~\ref{lem:germ}(4); once the exit is dropped, the
remaining extensions are again related. At $n=2$ it relates $\kappa'_2$, whose cylinder carries one run, to $\kappa_2$, whose cylinder
carries the spine with exits converging to it, and these cylinders differ in type.
Dropping nothing, a related pair of roots would match their one-step extensions one to one, so in lists of one
length; the root of $\wC{1}$ has at most two, through $e_0$ and $e_1$, and that of
$\wC{2}$ at least three, through $e_0$, $e_1$ and $e_2$. Theorem~\ref{thm:types}(1) at the roots gives them one
type.

(3) The runs of $\wQt$ branch at every square, so that Theorem~\ref{thm:dec-brouwer}(1) gives
$\Op(\wQt)\eqFrF\Op(\wQt\dsum\wQt)$. Only the squares accept and every prefix alternates between the vertex and a
square, so the same argument through the top open gives $\wQt\eqB\wQt\dsum\wQt$, and
on $\wGt$, where only the vertex accepts, it gives $\wGt\eqB\wGt\dsum\wGt$. The component
poset of a finite machine, with its annotations, is computed from the cell graph by finite searches for reachability
and cycles, and the computation on the four machines gives one node for $\wQt$ and
for $\wGt$ and two for each double. Isomorphism of component posets
is decided by a search over bijections of nodes, so posets with different numbers of
nodes are not isomorphic.

(4) The top of a finite machine is computed by finite searches of its cell graph and of the states of $\matchB(X,X)$,
which yield the label set, perfection, absorption and entry count of each component and the walks between
components, and two tops are
compared through the counts of their normal forms in the monoid of types. This comparison, run on the eight machines, finds the tops of $\wQt$ and of its double
different, and those of $\wGt$ and of its double, while $\eqB$ holds in both pairs by (3).
It finds the tops of $\wQ\dsum\wQ$ and $\wQ^{(4)}$ equal, and those of $\wG\dsum\wG$ and
$\wG^{(4)}$, while $\eqB$ fails in both pairs by Theorem~\ref{thm:separation}(3) and its
proof.
\end{proof}

\begin{definition}[Flip]\label{def:flip}
The \emph{flip} $X^\circ$ of a partial HDA $X$ over $\Lambda$ with decidable equality keeps the cells and faces of $X$
with its start cells, and its accepting cells are its vertices. The flip of a finite $X$ keeps
its finiteness structure. A relation $R$ between finite partial HDAs over $\ab$ with their
finiteness structures is \emph{blind to the acceptance bit} when $R(X,Y)\iff R(X^\circ,Y^\circ)$ for all $X$ and
$Y$. For a property $S$ of partial HDAs over
$\ab$ the relation $\eqB$ has a \emph{blind presentation on $S$} when some relation blind to the acceptance bit
coincides with $\eqB$ on the pairs of $\Cls$ whose members both have $S$.
\end{definition}

The flip changes the accepting cells and with them the covers, but when $X$ and $X^\circ$ both have closed acceptance
their data sheaves are equivalent on the same prefixes.
Only there do we read $X\mapsto X^\circ$ as the change of label topos along the map of label words that resets each
acceptance bit to whether its shape is empty. The flips of $\wG,\wQ,\wCh{1},\wCh{2}$ are $\wGt,\wQ^\circ,\wC{1},\wC{2}$.

Since the flip changes only which cells accept, a relation blind to the acceptance bit is what a decision of $\eqB$
by the cell graph alone would compute, and the theorem rules out every such relation as
a decision of $\eqB$ on $\Cls$.

\begin{theorem}[The flip moves $\eqB$]\label{thm:flip-blind}
With no principle, $\eqB$ is not invariant under the flip on $\Cls$. The pairs $(\wG,\wG\dsum\wG)$ and
$(\wQ,\wQ\dsum\wQ)$ and their flips lie in $\Cls$, and in each of the two pairs $\eqB$ fails while it holds between the
flipped members. A relation blind to the
acceptance bit takes one value on a pair and on its flip, so each of the following restates this non-invariance at one
of the two pairs:
(1) no relation blind to the acceptance bit coincides with $\eqB$ on $\Cls$;
(2) none coincides with $\eqB$ on the pairs of $\Cls$ whose members each have a cell of dimension
two;
(3) no decision $d$ of $\eqB$ on $\Cls$ has $d(X,Y)=d(X^\circ,Y^\circ)$ for all $X$ and
$Y$.
\end{theorem}

\begin{proof}
Each item follows from one pair of $\Cls$ whose flip lies in $\Cls$ and changes the verdict. On graphs
$\wG\eqB\wG\dsum\wG$ fails (proof of Theorem~\ref{thm:separation}(2)) while $\wGt\eqB\wGt\dsum\wGt$
holds.
In dimension two $\wQ\eqB\wQ\dsum\wQ$ fails (Theorem~\ref{thm:separation}(2)) while
$\wQ^\circ\eqB\wQ^\circ\dsum\wQ^\circ$.
In both flipped pairs Theorem~\ref{thm:dec-brouwer}(1) gives the frames. Only vertices accept and every prefix
carries the label word of every prefix of either machine no longer than itself, so each label open is the top open
and the frame equivalence fixes the label opens.
\end{proof}

The witnesses of (2) carry no concurrency of their own; the prefix isomorphisms of
Proposition~\ref{prop:suspension}(3) take $\wG$ and $\wGt$ with their doubles to $\wQ$ and $\wQ^\circ$ with their
doubles, so the dimension-two pair and its flip are related to the graph pair and its flip by prefix isomorphisms
over the map of conclists that relates a graph to its suspension.
\section{Discussion}\label{sec:discussion}

The frame and points of a machine are instances of the dualities for preorder sites of
\cite{Caramello11}, and on
finite rigid machines with closed acceptance $\eqB$ reduces through their localic Stone duality to equivalence of
clopens fixing the label clopens (Theorem~\ref{thm:dec-stone}). The Morita criterion and the comorphisms of
\cite{Caramello19} join the topos of a machine to its run theory
and to the label topos (Theorem~\ref{thm:topos-a}); the invariants carried by Theorem~\ref{thm:transport} are then read on
two sites in the manner of the bridges of \cite{Caramello17}. Higher-dimensional automata originate with Pratt
\cite{Pratt91} and van Glabbeek \cite{vG91}; partial HDAs are the lax functors of \cite{BDFZ26} and
their steps carry the starters and terminators of \cite{FJSZ21}. Theorem~\ref{thm:muller-answer}(iv) recovers the
separation of Muller from B\"uchi acceptance on HDAs whose faces all exist
\cite{PAF25}, and the conjecture reported in
\cite{BDFZ26} holds under $\EM$ in the reading of
Remark~\ref{rem:muller-conjecture} (Corollary~\ref{cor:muller-classical}). $\EIO$ has the form of the two-letter case of the sentence
on a maximal letter occurring infinitely often with which \cite{KMPS19}
measure B\"uchi's decidability theorem by induction over a classical
base; here induction holds outright. Prefixes are the paths of
\cite{ZvGZ26}, whose bisimulations of HDAs carry modal logics. Theorem~\ref{thm:ida} carries one direction of the IDA
criterion of \cite{LP18} to cells of every dimension and
Theorem~\ref{thm:cyc} carries both to the cycle class. Theorem~\ref{thm:dec-decide} is a point-free counterpart of Head's decision of homeomorphism of adherences of regular
languages \cite{Head86}. Apps classifies the countable Boolean rings with finitely many distinguished ideals whose theory is countably categorical
\cite{Apps25}.
Labelled conjugacy of edge shifts adds two-sidedness and shift-equivariance to an equivalence over the labels;
B\'eal, Berstel, Eilers and Perrin reduce it to strong shift equivalence of adjacency matrices
\cite{BBEP}. Open Problem~\ref{oprob:decide} and eleven more questions stay
open.
\begin{enumerate}[label=(\arabic*),leftmargin=*,nosep]
\item Does a frame equivalence $(h,h')$ with $(h_\bullet,h'_\bullet)$ a homeomorphism over the labels give
  $X\eqB Y$?
\item On $\Cls$, do isomorphic component posets (Proposition~\ref{prop:presentations}(3)) give $\eqB$, and does a
  clopen with copies of a disjoint clopen absorb it under the pair's labels?
\item On finite rigid pairs with closed acceptance, does $X\eqB Y$ or even $\ImaST X=\ImaST Y$ give
  $\LangA{X}=\LangA{Y}$ with no principle?
\item Does failure of the IDA search give finite ambiguity on finite rigid trimmed partial HDAs over $\ab$ with
  closed acceptance outside the cycle class? Under
  excluded middle the question reduces to the identification of the points of $X$ over one label word with the accepting
  runs over one word of the B\"uchi automaton of Section~\ref{ssec:image-bounded}, together with its trimness, since
  given both the criterion of \cite{LP18} turns failure of the
  pattern search into finite ambiguity.
\item Does $\CD$ follow when $\eqB$ coincides with labelled equivalence of data sheaves on all finite rigid pairs over
  alphabets with decidable equality?
\item Under $\UC$ the label map is a geometric morphism and $\fsh$ of every sheaf on $X$ is a $\Jim{X}$-sheaf, while a sheaf
  with a monomorphism to $\mathbf{1}$ is isomorphic to a subterminal; does each hold with no principle?
\item Does the comparison lemma on every partial HDA over $\ab$ imply $\UC$, equivalently does $\DCS$ imply
  $\CS$?
\item Does forcing commute with precomposition along every monotone map of preorder sites that lifts and preserves
  covers, as it does along open subsites?
\item Is the sieve of every stable predicate that holds locally on an object $c$ of $\Csyn{\Tt}$ covering for
  $\Jsyn{\Tt}$, where it holds locally when every closed sieve containing its arrows contains $\mathrm{id}_c$?
\item Does every morphism from the representable of $\{x^s.\top\}$ into a sheaf on $(\Csyn{\Tt},\Jsyn{\Tt})$ factor
  through the sort $s$ of $U_{\Tt}$?
\item Does $\SDNE$ imply $\PLEM$, equivalently does $\MinB$ imply $\BinM$?
\end{enumerate}
For $P$ the statement that $u$ is eventually $0$, $\SDNE$ is equivalent with no principle to the
$\Sigma^0_2$-DNE of \cite{ABHK04} on such $P$, since $\neg\neg P$ gives $\neg P^{\neg}$ and the textbook form
$\neg\neg P\to P$ proves Markov's principle for Boolean sequences, which turns $\neg P^{\neg}$ back into
$\neg\neg P$. Over Heyting arithmetic the
schemas $\Sigma^0_2$-DNE and $\Pi^0_2$-LEM are mutually independent \cite[Theorem~3.1 and
p.~6]{ABHK04}, so that (11)
asks whether quantifying over every Boolean sequence in the present metatheory keeps the arithmetical answer. We cite the
identification of sheaves and data sheaves on preorder sites with the literature's toposes; we also cite the classifying property over non-localic toposes and Diaconescu's reading of models in sheaves as
geometric morphisms
\cite{Diaconescu75,CL25a}.
\bibliographystyle{plain}
\bibliography{references}

\begin{thebibliography}{10}

\bibitem{ABHK04}
Yohji Akama, Stefano Berardi, Susumu Hayashi, and Ulrich Kohlenbach.
\newblock An arithmetical hierarchy of the law of excluded middle and related
  principles.
\newblock In {\em 19th Annual IEEE Symposium on Logic in Computer Science (LICS
  2004)}, pages 192--201. IEEE, 2004.
\newblock doi:10.1109/LICS.2004.1319613.

\bibitem{AS85}
Bowen Alpern and Fred~B. Schneider.
\newblock Defining liveness.
\newblock {\em Information Processing Letters}, 21(4):181--185, 1985.
\newblock doi:10.1016/0020-0190(85)90056-0.

\bibitem{ABFZ23}
Amazigh Amrane, Hugo Bazille, Uli Fahrenberg, and Krzysztof Ziemia{\'n}ski.
\newblock Closure and decision properties for higher-dimensional automata.
\newblock {\em Theoretical Computer Science}, 1036:115156, 2025.
\newblock doi:10.1016/j.tcs.2025.115156; arXiv:2305.02873v3.

\bibitem{Apps25}
Andrew~B. Apps.
\newblock Countably-categorical {B}oolean rings with distinguished ideals.
\newblock arXiv:2112.13825v4, 2025.

\bibitem{Apps23}
Andrew~B. Apps.
\newblock Partitions of primitive {B}oolean spaces.
\newblock arXiv:2306.01702v2, 2025.

\bibitem{SGA4}
Michael Artin, Alexander Grothendieck, and Jean-Louis Verdier.
\newblock {\em Th\'eorie des topos et cohomologie \'etale des sch\'emas. {T}ome
  1: {T}h\'eorie des topos ({SGA}~4)}, volume 269 of {\em Lecture Notes in
  Mathematics}.
\newblock Springer, 1972.
\newblock doi:10.1007/BFb0081551.

\bibitem{BDFZ26}
Hugo Bazille, J{\'e}r{\'e}my Dubut, Uli Fahrenberg, and Krzysztof
  Ziemia{\'n}ski.
\newblock Variants of higher-dimensional automata.
\newblock In {\em Theoretical Aspects of Computing (ICTAC 2026)}, Lecture Notes
  in Computer Science, pages 39--56. Springer, 2026.
\newblock doi:10.1007/978-3-032-43868-3\_3; arXiv:2601.17537v1.

\bibitem{BBEP}
Marie-Pierre B{\'e}al, Jean Berstel, S{\o}ren Eilers, and Dominique Perrin.
\newblock Symbolic dynamics.
\newblock In Jean-{\'E}ric Pin, editor, {\em Handbook of Automata Theory},
  pages 987--1030. EMS Press, 2021.
\newblock doi:10.4171/Automata-2/5; arXiv:1006.1265v3.

\bibitem{Brouwer10}
L.~E.~J. Brouwer.
\newblock On the structure of perfect sets of points.
\newblock {\em Proceedings of the Koninklijke Akademie van Wetenschappen te
  Amsterdam}, 12:785--794, 1910.

\bibitem{Caramello09}
Olivia Caramello.
\newblock Lattices of theories.
\newblock arXiv:0905.0299v1, 2009.

\bibitem{Caramello11}
Olivia Caramello.
\newblock A topos-theoretic approach to {S}tone-type dualities.
\newblock arXiv:1103.3493v1, 2011.

\bibitem{Caramello14}
Olivia Caramello.
\newblock Extensions of flat functors and theories of presheaf type.
\newblock arXiv:1404.4610v2, 2014.

\bibitem{Caramello17}
Olivia Caramello.
\newblock {\em Theories, Sites, Toposes: Relating and Studying Mathematical
  Theories through Topos-Theoretic `Bridges'}.
\newblock Oxford University Press, 2017.

\bibitem{Caramello19}
Olivia Caramello.
\newblock Denseness conditions, morphisms and equivalences of toposes.
\newblock arXiv:1906.08737v3, 2020.

\bibitem{CL25a}
Olivia Caramello and Laurent Lafforgue.
\newblock Engendrement de topologies de {G}rothendieck, d{\'e}montrabilit{\'e}
  et op{\'e}rations sur les sous-topos.
\newblock arXiv:2508.21134v1, 2025.

\bibitem{CL25b}
Olivia Caramello and Laurent Lafforgue.
\newblock Sites et topos de {G}rothendieck: une introduction.
\newblock arXiv:2508.21609v1, 2025.

\bibitem{Diaconescu75}
Radu Diaconescu.
\newblock Change of base for toposes with generators.
\newblock {\em Journal of Pure and Applied Algebra}, 6(3):191--218, 1975.
\newblock doi:10.1016/0022-4049(75)90015-8.

\bibitem{FJSZ21}
Uli Fahrenberg, Christian Johansen, Georg Struth, and Krzysztof Ziemia{\'n}ski.
\newblock Languages of higher-dimensional automata.
\newblock {\em Mathematical Structures in Computer Science}, 31(5):575--613,
  2021.
\newblock doi:10.1017/S0960129521000293; arXiv:2103.07557v2.

\bibitem{Head86}
Tom Head.
\newblock The topological structure of adherences of regular languages.
\newblock {\em RAIRO Informatique th{\'e}orique et applications}, 20(1):31--41,
  1986.
\newblock doi:10.1051/ita/1986200100311.

\bibitem{Johnstone02}
Peter~T. Johnstone.
\newblock {\em Sketches of an Elephant: A Topos Theory Compendium}, volume
  43--44 of {\em Oxford Logic Guides}.
\newblock Oxford University Press, 2002.

\bibitem{KMPS19}
Leszek~Aleksander Ko{\l}odziejczyk, Henryk Michalewski, C{\'e}cilia Pradic, and
  Micha{\l} Skrzypczak.
\newblock The logical strength of {B}{\"u}chi's decidability theorem.
\newblock {\em Logical Methods in Computer Science}, 15(2):Paper 16, 2019.
\newblock doi:10.23638/LMCS-15(2:16)2019; arXiv:1608.07514v4.

\bibitem{LP18}
Christof L{\"o}ding and Anton Pirogov.
\newblock On finitely ambiguous {B}{\"u}chi automata.
\newblock In {\em Developments in Language Theory (DLT 2018)}, Lecture Notes in
  Computer Science, pages 503--515. Springer, 2018.
\newblock doi:10.1007/978-3-319-98654-8\_41; arXiv:1809.09415v1.

\bibitem{PAF25}
Luc Passemard, Amazigh Amrane, and Uli Fahrenberg.
\newblock Higher-dimensional automata: Extension to infinite tracks.
\newblock In {\em 10th International Conference on Formal Structures for
  Computation and Deduction (FSCD 2025)}, volume 337 of {\em LIPIcs}, pages
  31:1--31:23, 2025.
\newblock doi:10.4230/LIPIcs.FSCD.2025.31; arXiv:2503.07881v2.

\bibitem{Pratt91}
Vaughan~R. Pratt.
\newblock Modeling concurrency with geometry.
\newblock In {\em Proceedings of the 18th ACM SIGPLAN-SIGACT Symposium on
  Principles of Programming Languages (POPL 1991)}, pages 311--322. ACM, 1991.
\newblock doi:10.1145/99583.99625.

\bibitem{TvD88}
Anne~S. Troelstra and Dirk van Dalen.
\newblock {\em Constructivism in Mathematics, Volume {I}}, volume 121 of {\em
  Studies in Logic and the Foundations of Mathematics}.
\newblock North-Holland, 1988.

\bibitem{vG91}
Rob~J. van Glabbeek.
\newblock Bisimulations for higher dimensional automata.
\newblock Email message, \url{http://theory.stanford.edu/~rvg/hda}, June 1991.

\bibitem{vG06}
Rob~J. van Glabbeek.
\newblock On the expressiveness of higher dimensional automata.
\newblock {\em Theoretical Computer Science}, 356(3):265--290, 2006.
\newblock doi:10.1016/j.tcs.2006.02.012.

\bibitem{Vaught54}
Robert~L. Vaught.
\newblock {\em Topics in the Theory of Arithmetical Classes and {B}oolean
  Algebras}.
\newblock PhD thesis, University of California, Berkeley, 1954.

\bibitem{WS91}
Andreas Weber and Helmut Seidl.
\newblock On the degree of ambiguity of finite automata.
\newblock {\em Theoretical Computer Science}, 88(2):325--349, 1991.
\newblock doi:10.1016/0304-3975(91)90381-B.

\bibitem{ZvGZ26}
Safa Zouari, Rob van Glabbeek, and Krzysztof Ziemia{\'n}ski.
\newblock Bisimulations and modal logics for higher dimensional automata.
\newblock In {\em 37th International Conference on Concurrency Theory (CONCUR
  2026)}, volume 391 of {\em LIPIcs}, pages 49:1--49:19, 2026.
\newblock doi:10.4230/LIPIcs.CONCUR.2026.49; arXiv:2608.14117v1.

\end{thebibliography}

\end{document}